\documentclass[10pt,aps,prd,notitlepage,preprintnumbers,eqsecnum]{revtex4-2}
\usepackage{graphicx}
\usepackage{slashed}
\usepackage{xcolor}
\usepackage{amsmath}
\usepackage{amssymb}
\usepackage{amsfonts}
\usepackage[bookmarks=false, colorlinks, linkcolor=black,citecolor=blue]{hyperref}
\newcommand{\ep}{\epsilon}
\newcommand{\la}{\lambda}
\newcommand{\de}{\delta}
\newcommand{\ga}{\gamma}
\newcommand{\al}{\alpha}
\newcommand{\be}{\beta}
\newcommand{\sig}{\sigma}
\newcommand{\s}[1]{\slashed{#1}}
\newcommand{\no}{\notag\\}

\begin{abstract}
{
We study one-loop QCD corrections to the single transverse spin asymmetry in
Drell-Yan process. The invariant mass of virtual photon and angular distributions of final lepton in Collins-Soper frame are measured. Especially, the transverse
momentum of virtual photon is integrated out. Collinear twist-3 factorization
formalism is adopted for the asymmetry. We use Feynman gauge in this work. To
eliminate dependent twist-3 distribution functions, equation of motion for quark
is used. The results satisfy both QED and QCD gauge invariance. It is confirmed that all divergences from virtual and real corrections can be removed consistently by collinear subtraction. Finite hard coefficients for the convolution of $\bar{q}(x)$ and $T_F(x_1,x_2)$ are presented.}
\end{abstract}

\begin{document}
\title{One-loop QCD corrections to SSA in unweighted Drell-Yan processes}
\author{Guang-Peng Zhang}
\email{gpzhang@ynu.edu.cn}
\affiliation{Department of physics, Yunnan University, Kunming, Yunnan 650091, China}
\maketitle

\tableofcontents

\section{Introduction}
The collinear twist-3 factorization for single transverse spin asymmetry (SSA) in pion hadronic production has been proposed for many years\cite{Efremov:1981sh,
Efremov:1984ip,Qiu:1991pp,Qiu:1998ia}. This factorization formalism can also be used
to describe SSAs in direct photon production\cite{Qiu:1991wg},
semi-inclusive deeply inelastic scattering(SIDIS)\cite{Eguchi:2006qz,Eguchi:2006mc,Ji:2006br}
and Drell-Yan process(DY)\cite{Ji:2006ub,Ji:2006vf} etc. However, a proof
for the factorization is still missing for these processes. Different
from twist-2 factorization, there are few one-loop corrections at twist-3 level  in literatures for the SSA.
So far, QCD corrections to
several weighted SSAs in DY and SIDIS are calculated to
one-loop level\cite{Vogelsang:2009pj,Kang:2012ns,
Dai:2014ala,Yoshida:2016tfh,Chen:2016dnp,Chen:2017lvx}.
In these weighted SSAs,
the transverse momentum $q_\perp$ for virtual photon in DY or
$P_{h\perp}$ for final detected hadron in SIDIS is integrated out.
It is found the twist-3 factorization really holds.
These are non-trivial checks for the factorization.
However, it does not imply the unweighted SSAs can also be factorized
at the same order of $\al_s$.
At tree level, the asymmetry for the angular distribution
of final lepton in DY with $q_\perp$ integrated out was shown to be nonzero, please see \cite{Ma:2014uma} and reference therein.
In \cite{Ma:2014uma}, the tree level tensor structure of the hadronic tensor
is made clear (see eq.(\ref{eq:w-tree}) in the following), which has two
parts: one part is
proportional to the derivative of $\de^2(q_\perp)$ and the other part is proportional to $\de^2(q_\perp)$ itself. The former is called derived part and
the latter is non-derivative part. Because the weight is proportional to $q_\perp$,
the weighted SSA receives virtual correction only from derivative part. It has been illustrated in \cite{Chen:2016dnp} that this virtual correction is the same as the correction to the usual quark form factor, which is a twist-2 quantity. The direct calculation in
\cite{Vogelsang:2009pj} confirms this. If there is
no weight, both derivative part and non-derivative part of the hadronic tensor can contribute to SSA. Thus, the examine of factorization based on
weighted SSA is incomplete. Moreover, unweighted SSA is expected to have smaller
errors than weighted SSA in experiments. One can consult \cite{COMPASS:2023vqt} for
recent experimental results for SSA in DY and \cite{Niemiec:2024zou} for preliminary weighted SSA.

In this work, we study the one-loop correction to unweighted SSA in
DY with $q_\perp$ integrated.
Explicitly, we calculate the angular distribution of final lepton.
To avoid possible ambiguity for soft-gluon-pole contribution, we take
Feynman gauge in this work. The method of diagram expansion\cite{Qiu:1998ia,Eguchi:2006mc}
is adopted here. The troubles for the expansion are mainly two
aspects: one is
there are many dependent twist-3 distribution functions. Some of these functions
contain the bad component of quark field. The other is the gauge invariant
distribution functions contain gluon field strength tensor and gauge links, but
it seems impossible to recover these two quantities completely. In order to
solve these two problems, we use at most one longitudinal gluon($G^+$)
to do collinear expansion, and then use equation of motion for quark field to
eliminate the bad component. After these treatments we get five independent
quark-gluon-quark or quark-quark correlation functions. Three of them can be
identified to $q_\partial$, $T_F$ and $T_\Delta$, which are gauge invariant.
Remaining two correlation functions are not gauge invariant and dangerous to the factorization assumption. To preserve QCD gauge invariance, the
hard coefficients before these two functions must be zero.
Really, our
calculation confirms this. This indicates our expansion scheme preserves
QCD gauge invariance. QED gauge invariance for the hadronic
tensor is also checked at one-loop level, i.e., $q^\mu W_{\mu\nu}=0$, with
$W^{\mu\nu}$ the hadronic tensor. In this work, we consider the
contribution proportional to $\bar{q}\otimes T_F$. That is, for unpolarized
hadron, only the contribution from twist-2 anti-quark distribution function
$\bar{q}(x)$ is considered and for polarized hadron, only twist-3
quark-gluon-quark distribution function $T_F$ is considered. The hard coefficients from virtual and real corrections are calculated explicitly. After collinear
subtraction, all divergences from virtual and real corrections are removed consistently.

Very recently, \cite{Rein:2025a,Rein:2025b} give the one-loop correction to a single spin asymmetry $A_{UT}$ for $E_hd^3\sig/d^3 P_h$ in lepton-hadron scattering $l+p(s_\perp)\rightarrow h+X$. The final lepton is undetected. Please see eq.(1) of \cite{Rein:2025a} for the illustration of the asymmetry. Our method for collinear
expansion is different from theirs. In \cite{Rein:2025a,Rein:2025b}, light-cone
gauge is adopted. But we really find many common features
for the one-loop corrections. For example, there are nonpole or integral contributions in virtual corrections, which come from the absorptive part of loop
integrals rather than from an on-shell propagator. Especially, these nonpole
contributions contain collinear divergences and have a different tensor structure
from tree level hadronic tensor. However, these divergences can still
be subtracted by one-loop corrections to $T_F(x,x)$. This is a new feature of twist-3 calculation. It is important for us
to extend the method of this work to lepton-hadron scattering in future.

The structure of this paper is as follows: Sec.2
is the kinematics for lepton angular distributions in Drell-Yan process;
Sec.3 is the definition of all involved twist-3
distributions and their relations resulting from equation of motion of quark field. Our expansion formalism is also presented in this section;
Sec.4 contains tree level results; Sec.5 contains one-loop virtual corrections; Sec.6 contains real corrections; Sec.7 is for the renormalization of twist-3
distribution function and collinear subtraction. The final finite hard
coefficients are also given; Sec.8 is our summary.

\section{Kinematics}
The polarized Drell-Yan process is
\begin{align}
h_A(p_a,s_\perp)+h_B(p_b)\rightarrow \ga^*[\rightarrow e^-(l)e^+(\bar{l})]+X,
\end{align}
where $h_A$ is a spin-$\frac{1}{2}$ hadron polarized transversely with $s_\perp^\mu$ the spin vector; $h_B$ is an unpolarized hadron; $X$ represents undetected hadrons;
the final lepton pair(here we take electron and positron as an example) is assumed from the decay of a virtual photon, and their momenta $l^\mu$ and $\bar{l}^\mu$ are detected. As usual we introduce $q^\mu=l^\mu+\bar{l}^\mu$ for the virtual photon.  $p_a,p_b$ are momenta of hadrons. The total energy squared is $s=(p_a+p_b)^2$.

The invariant mass squared of virtual photon $Q^2=q^2$ is a hard scale. Under Bjorken limit $Q^2\rightarrow \infty$ and $\tau=Q^2/s$ fixed, we can ignore all
masses of hadrons and leptons\cite{Kodaira:1998jn}. The angular distribution of final lepton we want to study is
\begin{align}
\frac{d\sig}{dQ^2 d\Omega}=\frac{\al_{em}^2}{4s Q^4}
\int d^nq \de(q^2-Q^2) L_{\mu\nu}W^{\mu\nu}.
\end{align}
$\Omega$ is the solid angle of final lepton with momentum $l^\mu$, defined in Collins-Soper(CS) frame\cite{Collins:1977iv}. In this work, we take dimensional regularization to regulate ultraviolet(UV) and infrared(IR) divergences. The dimension of $q$-integration has been set to $n=4-\ep$.
$L^{\mu\nu}$ and $W^{\mu\nu}$ are leptonic and hadronic tensors, respectively.
That is,
\begin{align}
L^{\mu\nu}
=&4(l^\mu \bar{l}^\nu+l^\nu \bar{l}^\mu -\frac{1}{2}Q^2 g^{\mu\nu})
= 4(-2l^\mu l^\nu+l^\mu q^\nu+l^\nu q^\mu -\frac{1}{2}Q^2 g^{\mu\nu}),\no
W^{\mu\nu}=& \int \frac{d^n x}{(2\pi)^n}
e^{iq\cdot x}
\sum_X\langle h_B,h_As_\perp|j^\nu(0)|X\rangle \langle X| j^\mu(x)|h_As_\perp,h_B\rangle.
\end{align}
$j^\mu=\bar{\psi}\ga^\mu\psi$ is electro-magnetic current. $\sum_X$ is the phase space integration for all
possible final hadrons. At parton level, the hadrons are quarks and physical gluons.
In this work we use light-cone coordinates. A four-vector $a^\mu$ is written as
$a^\mu=(a^+,a^-,a_\perp^\mu)$, with $a^\pm=\frac{1}{\sqrt{2}}(a^0\pm a^3)$. Two light-like vectors $n^\mu,\bar{n}^\mu$ are introduced, so that
\begin{align}
a^+=a\cdot n,\ a^-=a\cdot\bar{n},\ a_\perp\cdot n=a_\perp\cdot\bar{n}=0,\ n\cdot\bar{n}=1.
\end{align}
Two transverse tensors are also introduced: transverse metric $g_\perp^{\mu\nu}$
and transverse anti-symmetric tensor $\ep_\perp^{\mu\nu}$ as follows:
\begin{align}
g_\perp^{\mu\nu}=g^{\mu\nu}-n^\mu\bar{n}^\nu-n^\nu \bar{n}^\mu,\
\ep_\perp^{\mu\nu}=\ep^{-+\mu\nu}=\ep^{\rho\tau\mu\nu}\bar{n}_\rho n_\tau.
\end{align}
Note that $\ep^{0123}=+1$ and $\ep_{\perp}^{12}=+1$ in this work.

In the center of mass(CM) frame of initial hadrons, under Bjorken limit, the masses
of hadrons can be ignored, so, $p_a^\mu,p_b^\mu$ become light-like, that is,
\begin{align}
p_a^\mu=p_a^+ \bar{n}^\mu,\ p_b^\mu=p_b^- n^\mu.
\end{align}
The solid angle $\Omega$ is defined in CS frame, which is a rest frame of lepton pair and is obtained from CM frame by two boosts\cite{Collins:1977iv}.
The first boost is along $Z$ axis so that
$q^z=0$ after boost; the second boost is along the direction $\vec{q}_\perp$, so
that $\vec{q}_\perp=0$ after boost. $l^\mu$ in CS frame is parameterized as
\begin{align}
l_{cs}^\mu=(l^0,l^1,l^2,l^3)
=\frac{Q}{2}
(1,\sin\theta\cos\phi,\sin\theta\sin\phi,\cos\theta).
\end{align}
We also define
$ l_{cs}^\pm=\frac{Q}{2\sqrt{2}}(1\pm \cos\theta)$
and $l_{\perp,cs}^\mu$ for the longitudinal and transverse components of $l^\mu_{cs}$, respectively. With the two Lorentz boosts done explicitly,
the momentum $l^\mu$ in CM frame can be expressed in terms of $l_{cs}^\mu$
as follows:
\begin{align}
l^+=& \frac{q^+}{E_{t}}\Big[\frac{E_{t}}{2}
+l_{cs}^z-\frac{1}{Q} l_{\perp,cs}\cdot q_{\perp}\Big],\no
l^-=& \frac{q^-}{E_{t}}\Big[\frac{E_{t}}{2}
-l_{cs}^z-\frac{1}{Q} l_{\perp,cs}\cdot q_{\perp}\Big],\no
l_\perp^\mu=& l_{\perp,cs}^\mu
+(\frac{E_{t}}{Q}-1)\frac{l_{\perp,cs}\cdot q_{\perp}}{q_\perp^2}
q_{\perp}^\mu
+\frac{1}{2}q_{\perp}^\mu.
\label{eq:lepton-momentum}
\end{align}
As a convention, for quantities defined in CM frame, the subscription ``cm" is suppressed. The dot product is defined for four-vector, that is,
$a_\perp\cdot b_\perp=-\vec{a}_\perp\cdot\vec{b}_\perp$ and especially
$a_\perp^2=-\vec{a}_\perp\cdot\vec{a}_\perp<0$. The transverse energy $E_t$ for virtual photon is $E_t=\sqrt{Q^2-q_\perp^2}\geq Q$.
Above representation of lepton momentum in CM frame is crucial for our calculation.
In the following we do all calculations in CM frame. The spin vector $s_\perp^\mu$
is perpendicular to $p_a$ in CM frame. If we let $s_\perp^\mu=(0^+,0^-,1,0)|s_\perp|$, i.e., $\vec{s}_\perp$ defines X-axis in CM frame, then,
$\tilde{s}_\perp^\mu=\ep_\perp^{\mu\nu}s_{\perp\nu}=(0^+,0^-,0,1)|s_\perp|$,
and then,
\begin{align}
l_{\perp,cs}\cdot\tilde{s}_\perp=-\frac{Q}{2}|s_\perp|\sin\theta\sin\phi.
\end{align}
The SSA we are considering is proportional to this quantity.

In this work, we check the twist-3 factorization for two quantities $I\langle L\rangle$ and $I\langle P_7\rangle$, which are defined as
\begin{align}
I\langle L\rangle =&\int d^nq \de(q^2-Q^2) L_{\mu\nu}W^{\mu\nu},\no
I\langle P_7\rangle =&\int d^nq \de(q^2-Q^2) P_{7,\mu\nu}W^{\mu\nu},\
P_7^{\mu\nu}=\frac{1}{\tilde{s}_\perp^2}
\Big(q_\perp^\mu \tilde{s}_\perp^\nu+q_\perp^\nu \tilde{s}_\perp^\mu\Big).
\end{align}
$I\langle L\rangle$ is proportional to the differential cross section listed above, i.e.,
\begin{align}
\frac{d\sig}{dQ^2 d\Omega}=\frac{\al_{em}^2}{4s Q^4}I\langle L\rangle.
\end{align}
$I\langle P_7\rangle$ on the other hand is proportional to the weighted cross section studied in
\cite{Chen:2016dnp}. Since the weighted cross section has been shown to be factorized, $I\langle P_7\rangle$ here is used as a check of our calculation.

\section{Twist-3 distribution functions}\label{sec:def-tw3}
In this work we study the contribution from chiral-even twist-3 distribution functions. At twist-3 level, there are only two independent chiral-even
quark-gluon-quark correlation
functions,
\begin{align}
\tilde{s}_\perp^\rho T_F(x_1,x_2)
=&
g_s\int \frac{d\xi^-d\xi_1^-}{4\pi}e^{i\xi^-k^+ + i\xi_1^- k_1^+}
\langle ps|
\bar{\psi}(0)\mathcal{L}_n(0)\mathcal{L}_n^\dagger(\xi^-)
\ga^+G_\perp^{+\rho}(\xi^-)\mathcal{L}_n(\xi^-)
\mathcal{L}^\dagger_n(\xi_1^-)\psi(\xi_1^-)|ps\rangle,\no
s_\perp^\rho T_{\Delta}(x_1,x_2)
=&
g_s\int \frac{d\xi^-d\xi_1^-}{4\pi}e^{i\xi^-k^+ + i\xi_1^- k_1^+}
\langle ps|\bar{\psi}(0)\mathcal{L}_n(0)\mathcal{L}_n^\dagger(\xi^-)
(-i)\ga^+\ga_5
G_\perp^{+\rho}(\xi^-)\mathcal{L}_n(\xi^-)
\mathcal{L}^\dagger_n(\xi_1^-)\psi(\xi_1^-)|ps\rangle,
\end{align}
where $G_\perp^{+\rho}=G_{a\perp}^{+\rho}T_a$ is gluon field strength tensor and
$\mathcal{L}_n$ is the gauge link which ensures that the two distributions are gauge invariant. The definition of gauge link is
\begin{align}
\mathcal{L}_n(\xi^-)=P e^{-ig_s\int_{-\infty}^{\xi^-}d\la^- G^+(\la^-) },\
G^+=G^+_a T_a,
\end{align}
with $G_a^\mu$ gluon field and $T_a$ the generator of fundamental representation of $SU(N_c)$. $P$ is path ordering operator:
\begin{align}
PG^+(\la_1^-)G^+(\la_2^-)
=\theta(\la_1^--\la_2^-)G^+(\la_1^-)G^+(\la_2^-)
+\theta(\la_2^--\la_1^-)G^+(\la_2^-)G^+(\la_1^-).
\end{align}
In Drell-Yan process, the gauge link points to $-\infty$. For parton momenta, throughout the paper we use following notations
\begin{align}
k^+=xp_a^+,\ k_1^+=x_1p_a^+,\ k_2^+=x_2 p_a^+,\ k_b^-=x_b p_b^-,
\end{align}
and $k_2=k+k_1$, $x_2=x+x_1$. $k_b$ is the momentum of anti-quark from unpolarized
hadron.

In addition to these three-point distributions, there are three two-point
distributions as follows
\begin{align}
q_T(x)s_\perp^\rho=&p^+\int\frac{d\xi^-}{4\pi}
e^{i\xi^- xp^+}\langle ps|
\bar{\psi}(0)\mathcal{L}_n(0)\ga_\perp^\rho\ga_5\mathcal{L}_n^\dagger(
\xi^-)\psi(\xi^-)|ps\rangle,\no
-iq'_\partial(x)\tilde{s}_\perp^\rho =&\int\frac{d\xi^-}{4\pi}
e^{i\xi^- xp^+}\langle ps|
\bar{\psi}(0)\mathcal{L}_n(0)\ga^+\partial_\perp^\rho \mathcal{L}_n^\dagger(
\xi^-)\psi(\xi^-)|ps\rangle,\no
-iq_\partial(x)s_\perp^\rho =&\int\frac{d\xi^-}{4\pi}
e^{i\xi^- xp^+}\langle ps|
\bar{\psi}(0)\mathcal{L}_n(0)\ga^+\ga_5\partial_\perp^\rho \mathcal{L}_n^\dagger(
\xi^-)\psi(\xi^-)|ps\rangle.
\end{align}
However, they are not independent due to
\begin{align}
\frac{1}{2\pi}\int dx_1 P\frac{1}{x_1-x_2}\Big[T_F(x_1,x_2)+T_\Delta(x_1,x_2)
\Big]=-x_2 q_T(x_2)+q_\partial(x_2),\
T_F(x,x)=2q'_\partial(x).
\end{align}
These distribution functions and relations between them can be found for example in \cite{Boer:2003cm,Zhou:2009jm,Kanazawa:2015ajw}.
$q_T$ and $q'_\partial$ can be eliminated. Another kind of twist-3 distributions
involving covariant derivative can be expressed by the distribution functions
introduced above. So, we expect the factorized cross section
can be expressed by $T_F,T_\Delta$ and $q_\partial$. In \cite{Kanazawa:2015ajw},
$T_F,T_\Delta$ are called dynamical twist-3 distribution functions, while
$q_\partial,q'_\partial$ and $q_T$ are kinematical ones. With the notation of
\cite{Kanazawa:2015ajw}, our twist-3 distribution functions are expressed as
\begin{align}
&T_F(x_1,x_2)=-2\pi M F_{FT}^q(x_2,x_1),\ T_\Delta(x_1,x_2)=2\pi M G_{FT}^q(x_2,x_2),\no
&q_T(x)=M g_T(x),\ q_\partial(x)=M g_{1T}^{(1),q}(x),\
q'_\partial(x)=M f_{1T}^{\perp(1),q}(x),
\end{align}
where $M$ is hadron mass.
The minus sign for $T_F$ relation is from different definition of covariant derivatives. We have $D^\mu\equiv \partial^\mu +ig_s G^\mu$, but \cite{Kanazawa:2015ajw} has $D^\mu\equiv \partial^\mu -ig_s G^\mu$. That is,
the two strong couplings are opposite.
The relation
for $q'_\partial(x)$ holds if the gauge links in $q'_\partial$ and in Sivers function $f_{1T}^{\perp,q}(x,k_\perp)$ point to $-\infty^-$ or $+\infty^-$
at the same time.

It is difficult to recover $T_F,T_\Delta$ and $q_\partial$ in practical calculation, because of the gluon field strength tensor and gauge links. We do not try to
recover the complete gluon field strength tensor and gauge links in this calculation.
%
Instead, we first do collinear expansion using the matrix elements containing at most one $G^+$, which are
\begin{align}
\langle ps|\bar{\psi}\tilde{\Gamma}\psi|ps\rangle,\
\langle ps|\bar{\psi}\tilde{\Gamma}G^+ \psi|ps\rangle,\
\langle ps|\bar{\psi}\Gamma \partial_\perp^\rho \psi|ps\rangle,\
\langle ps|\bar{\psi}\Gamma (\partial_\perp^\rho G^+) \psi|ps\rangle,\
\langle ps|\bar{\psi}\Gamma G^+\partial_\perp^\rho\psi|ps\rangle,
\end{align}
with $\tilde{\Gamma}=\ga_\perp^\rho,\ga_\perp^\rho\ga_5$,
$\Gamma=\ga^+,\ga^+\ga_5$. Then, we try to eliminate the dependent matrix elements
by using equation of motion(EOM) and parity and time reversal(PT) symmetries. Such
an expansion scheme can be understood because the operators in $q_\partial$, $T_F$
and $T_\Delta$ can always be expanded into $\bar{\psi}_+$, $\psi_+$, $G^+$ and $G_\perp$, where $\bar{\psi}_+$ and $\psi_+$ are good components of fermion field(
see eq.(\ref{eq:good-fermion})). Then, if the factorization for $q_\partial$, $T_F$ and $T_\Delta$ is right, preserving the expansion to a certain power of $G^+$ is also right, in the sense that the coefficient functions of the resulting matrix elements are finite. In our scheme, we keep the expansion to $O(G^+)$ and $O((G_\perp)^0)$. Details of the expansion is given below.

We introduce following three types of correlation
functions. The first type does not contain $G^+$,
\begin{align}
\int \frac{d\xi^-}{2\pi}e^{ik^+\xi^-}
\langle ps|\bar{\psi}_j(0)\psi_i(\xi^-)|ps\rangle
=& \frac{\de_{ij}}{4N_c}\Big[
\ga^- 2q(x)+ \ga_5\ga_\perp^\rho s_{\perp\rho}M^{(0)}_{\ga_\perp\ga_5}(k^+)
+\ga_\perp^\rho \tilde{s}_{\perp\rho}M^{(0)}_{\ga_\perp}(k^+)
\Big]_{ij},\no
\int \frac{d\xi^-}{2\pi}e^{ik^+\xi^-}
\langle ps|\bar{\psi}_j(0)\partial_\perp^\rho\psi_i(\xi^-)|ps\rangle
=& \frac{\de_{ij}}{4N_c}\Big[
i\ga_5\ga^-s_{\perp}^\rho M^{(0)}_{\ga^+\ga_5,\partial_\perp\psi}(k^+)
+\ga^-\tilde{s}_{\perp}^\rho M^{(0)}_{\ga^+,\partial_\perp\psi}(k^+)
\Big]_{ij},
\end{align}
with $k^+=xp^+$.
The superscript $(0)$ implies there is no gluon. The subscript represents the
gamma matrix in the correlation function, and the fields with transverse derivative
are also shown. From PT symmetry, it can be shown
\begin{align}
M^{(0)}_{\ga_\perp}(k^+)=M^{(0)}_{\ga^+,\partial_\perp\psi}(k^+)=0.
\label{eq:M_perp}
\end{align}
So, there are only two twist-3 two-point correlation functions, which are related
to $\ga_5$. $q(x)$ is the usual unpolarized twist-2 parton distribution function(PDF) for quark.

The second type contains one $G^+$ but no $\partial_\perp$,
\begin{align}
g_s\int\frac{d\xi^-d\xi_1^-}{(2\pi)^2}
e^{ik^+\xi^-+ik_1^+\xi_1^-}
\langle ps|\bar{\psi}_j(0)G_a^+(\xi^-)\psi_i(\xi_1^-)|ps\rangle
=\frac{T^a_{ij}}{4N_c C_F}
\Big[i\ga_\perp^\rho \tilde{s}_{\perp\rho}M^{(1)}_{\ga_\perp}(k^+,k_1^+)
+\ga_5\ga_\perp^\rho s_{\perp\rho} M^{(1)}_{\ga_\perp\ga_5}(k^+,k_1^+)
\Big]_{ij};
\end{align}
The third type contains one $G^+$ and one $\partial_\perp$,
\begin{align}
&g_s\int\frac{d\xi^-d\xi_1^-}{(2\pi)^2}
e^{ik^+\xi^-+ik_1^+\xi_1^-}
\langle ps|
\bar{\psi}_j(0)G_a^+(\xi^-)\partial_\perp^\rho\psi_i(\xi_1^-)|ps\rangle\no
&=\frac{T^a_{ij}}{4N_c C_F}
\Big[\ga^- \tilde{s}_{\perp}^\rho M^{(1)}_{\ga^+,\partial_\perp\psi}(k^+,k_1^+)
+i\ga_5\ga^- s_\perp^\rho M^{(1)}_{\ga^+\ga_5,\partial_\perp\psi}(k^+,k_1^+)
\Big]_{ij};\no
&g_s\int\frac{d\xi^-d\xi_1^-}{(2\pi)^2}
e^{ik^+\xi^-+ik_1^+\xi_1^-}
\langle ps|\bar{\psi}_j(0)[\partial_\perp^\rho G_a^+(\xi^-)]\psi_i(\xi_1^-)|ps\rangle\no
&=\frac{T^a_{ij}}{4N_c C_F}
\Big[\ga^- \tilde{s}_{\perp}^\rho M^{(1)}_{\ga^+,\partial_\perp G^+}(k^+,k_1^+)
+i\ga_5\ga^- s_\perp^\rho M^{(1)}_{\ga^+\ga_5,\partial_\perp G^+}(k^+,k_1^+)
\Big]_{ij}.
\end{align}
From PT symmetry of QCD, all $M^{(i)}$ are real. Note that all $M^{(1)}$
are proportional to $g_s$.

For fermion field, the good and bad components are defined as
\begin{align}
\psi_+=\frac{\ga^-\ga^+}{2}\psi,\ \psi_-=\frac{\ga^+\ga^-}{2}\psi.
\label{eq:good-fermion}
\end{align}
In collinear expansion, the bad component of fermion field $\psi_-$ is
power suppressed relative to the good component $\psi_+$. From EOM of
fermion $\s{D}\psi=0$ with $D^\mu=\partial^\mu +ig_s G^\mu$, we have
\cite{Ma:2014uma}
\begin{align}
\psi_-(\xi^-)=-\frac{1}{2}\mathcal{L}_n(\xi^-)
\int_{-\infty}^{\xi^-}d\la^- \mathcal{L}_n^\dagger(\la^-)
\ga^+ \ga_\perp\cdot D_\perp(\la^-)\psi_+(\la^-).
\label{eq:bad-fermion}
\end{align}
The suppression is caused by covariant derivative $D_\perp^\rho
=\partial_\perp^\rho+ig_s G_\perp^\rho$.
This relation enables us to eliminate the bad component. After eliminating
the bad component we get following useful relations between $M^{(i)}$,
\begin{align}
M^{(1)}_{\ga_\perp\ga_5}(k^+,k_1^+)\doteq&-\frac{1}{2}P\Big(
\frac{1}{k_1^+}+\frac{1}{k_2^+}
\Big)M^{(1)}_{\ga^+\ga_5,\partial_\perp\psi}(k^+,k_1^+)
-\frac{1}{2}P\Big(
\frac{1}{k_1^+}-\frac{1}{k_2^+}
\Big)M^{(1)}_{\ga^+,\partial_\perp\psi}(k^+,k_1^+)\no
&-\frac{1}{2k_2^+}\Big(
M^{(1)}_{\ga^+\ga_5,\partial_\perp G^+}(k^+,k_1^+) - M^{(1)}_{\ga^+,\partial_\perp G^+}(k^+,k_1^+)\Big),\no
M^{(1)}_{\ga_\perp}(k^+,k_1^+)\doteq&\frac{1}{2}P\Big(
\frac{1}{k_1^+}+\frac{1}{k_2^+}
\Big)M^{(1)}_{\ga^+,\partial_\perp\psi}(k^+,k_1^+)
+\frac{1}{2}P\Big(
\frac{1}{k_1^+}-\frac{1}{k_2^+}
\Big)M^{(1)}_{\ga^+\ga_5,\partial_\perp\psi}(k^+,k_1^+)\no
&+\frac{1}{2k_2^+}\Big(
M^{(1)}_{\ga^+,\partial_\perp G^+}(k^+,k_1^+)
-M^{(1)}_{\ga^+\ga_5,\partial_\perp G^+}(k^+,k_1^+)
\Big),\no
\
M^{(0)}_{\ga_\perp\ga_5}(k_2^+)\doteq&-\frac{1}{k_2^+}
\int dk^+ P\frac{1}{k_1^+}
\Big(
M^{(1)}_{\ga^+\ga_5,\partial_\perp\psi}(k^+,k_1^+)
+M^{(1)}_{\ga^+\partial_\perp\psi}(k^+,k_1^+)\Big)
-\frac{1}{k_2^+}M^{(0)}_{\ga^+\ga_5,\partial_\perp\psi}(k_2^+).
\label{eq:EOMs}
\end{align}
The $\doteq$ implies $G_\perp$ and higher order of $G^+$ such as $(G^+)^2$ terms
are ignored on right hand side. P means principal value(PV), and $k_2^+=k^++k_1^+$.
To one-loop level, a consistent treatment of $\ga_5$ is important.
We adopt HVBM scheme\cite{tHooft:1972tcz,Breitenlohner:1977hr}
in this work. In this scheme $\ga_5$ is defined as a four dimensional
quantity. In addition, spin vectors $s_\perp^\mu$ and $\tilde{s}_\perp^\mu$ are
also defined as four dimensional quantities. Thus following identities can be applied
\begin{align}
\ga^+\s{s}_\perp(i\ga_5\ga^-)=-i\ga_5 \s{s}_\perp -\s{\tilde{s}}_\perp,\
\ga^+\s{\tilde{s}}_\perp\ga^-=-\s{\tilde{s}}_\perp-i\ga_5\s{s}_\perp,
\end{align}
which reduce the number of gamma matrices. Our conventions for $\ga_5$ are listed
in Appendix.\ref{sec:ga5}.

Now our calculation scheme is clear. First, we use all possible twist-3
matrix elements $M^{(i)}$ to do collinear expansion and get all corresponding
hard coefficients. For hadronic tensor, the result is
\begin{align}
8N_c^2 C_F W^{\mu\nu}=& \int dk_b^- dk_2^+
\bar{q}(k_b^-)
\Big[
H_0^{\mu\nu}M_{\ga_\perp\ga_5}^{(0)}(k_2^+)
+\tilde{H}_0^{\mu\nu}M^{(0)}_{\ga^+\ga_5,\partial_\perp\psi}(k_2^+)
\Big]\no
&+\int dk_b^- dk^+ dk_1^+ \bar{q}(k_b^-)
\Big[
H_1^{\mu\nu}M^{(1)}_{\ga_\perp\ga_5}(k^+,k_1^+)
+H_2^{\mu\nu}M^{(1)}_{\ga_\perp}(k^+,k_1^+)\no
&+H_3^{\mu\nu}M^{(1)}_{\ga^+\ga_5,\partial_\perp\psi}(k^+,k_1^+)
+H_4^{\mu\nu}M^{(1)}_{\ga^+,\partial_\perp\psi}(k^+,k_1^+)
+H_5^{\mu\nu}M^{(1)}_{\ga^+\ga_5,\partial_\perp G^+}(k^+,k_1^+)
+H_6^{\mu\nu}M^{(1)}_{\ga^+,\partial_\perp G^+}(k^+,k_1^+)
\Big].
\end{align}
The hard coefficients are
\begin{align}
H^{\mu\nu}_0=& \text{Tr}\Big[C_F H^{\mu\nu}\otimes \ga_5\s{s}_\perp\otimes \ga^+\Big],\no
\tilde{H}^{\mu\nu}_0=&
\text{Tr}\Big[C_F i\frac{\partial H^{\mu\nu}}{\partial k_{2\perp}^\tau}
\otimes i\ga_5\ga^-s_\perp^\tau \otimes \ga^+\Big],\no
H^{\mu\nu}_1=& \text{Tr}\Big[H^{\mu\nu}\otimes T^a\ga_5\s{s}_\perp\otimes \ga^+\Big],\no
H^{\mu\nu}_2=& \text{Tr}\Big[H^{\mu\nu}\otimes  i T^a\s{\tilde{s}}_\perp\otimes \ga^+\Big],\no
H^{\mu\nu}_3=& \text{Tr}\Big[i\frac{\partial H^{\mu\nu}}{\partial k_{1\perp\tau}}
\otimes iT^a\ga_5\ga^-s_\perp^\tau\otimes \ga^+\Big],\no
H^{\mu\nu}_4=& \text{Tr}\Big[i\frac{\partial H^{\mu\nu}}{\partial k_{1\perp\tau}}
\otimes T^a\ga^- \tilde{s}_{\perp\tau}\otimes \ga^+\Big],\no
H^{\mu\nu}_5=& \text{Tr}\Big[i\frac{\partial H^{\mu\nu}}{\partial k_{\perp\tau}}
\otimes iT^a\ga^5\ga^-s_\perp^\tau\otimes \ga^+\Big],\no
H^{\mu\nu}_6=& \text{Tr}\Big[i\frac{\partial H^{\mu\nu}}{\partial k_{\perp\tau}}
\otimes T^a\ga^-\tilde{s}_\perp^\tau\otimes \ga^+\Big].
\end{align}
$\otimes$ is the product in Dirac and color space.

Then we use EOM relations eq.(\ref{eq:EOMs}) to eliminate dependent correlation functions to get
\begin{align}
8N_c^2 C_F W^{\mu\nu}=&
\int dk_b^-\bar{q}(x_b)\int dk_2^+
g_0^{\mu\nu}\times M^{(0)}_{\ga^+\ga_5,\partial_\perp\psi}(k_2^+)\no
&+\int dk_b^-\bar{q}(x_b)\int dk^+ dk_1^+
\Big[g_1^{\mu\nu}\times M^{(1)}_{\ga^+,\partial_\perp\psi}
+g_2^{\mu\nu}\times M^{(1)}_{\ga^+\ga_5,\partial_\perp\psi}
+g_3^{\mu\nu}\times M^{(1)}_{\ga^+,\partial_\perp G^+}
+g_4^{\mu\nu}\times M^{(1)}_{\ga^+\ga_5,\partial_\perp G^+}\Big],
\end{align}
with
\begin{align}
g_0=& -\frac{1}{k_2^+}H_0+\tilde{H}_0,\no
k_2^+ g_1=& -P\frac{1}{k_1^+}H_0-P\frac{k^+}{2 k_1^+}H_1
+P\frac{k_1^+ + k_2^+}{2k_1^+}H_2 + k_2^+H_4,\no
k_2^+ g_2=&-P\frac{1}{k_1^+}H_0-P\frac{k_1^+ + k_2^+}{2 k_1^+}H_1
+P\frac{k^+}{2 k_1^+}H_2+k_2^+H_3,\no
k_2^+ g_3=&\frac{1}{2}H_1+\frac{1}{2}H_2 + k_2^+ H_6,\no
k_2^+ g_4=&-\frac{1}{2}H_1-\frac{1}{2}H_2+k_2^+ H_5.
\label{eq:hi2gi}
\end{align}
$\mu,\nu$ indices in $H_i$ and $g_i$ are suppressed. Since all $M^{(i)}$ are real,
the coefficients $g_i$ are also real for symmetric $\mu,\nu$.

Besides EOM relations, there is one more important relation if $k^+=0$, i.e.,
\begin{align}
M^{(1)}_{\ga^+,\partial_\perp\psi}(k^+,k_1^+)\Big|_{k^+=0}
=M^{(1)}_{\ga^+,\partial_\perp\bar{\psi}}(k^+,k_1^+)\Big|_{k^+=0}
=-\frac{1}{2}M^{(1)}_{\ga^+,\partial_\perp G^+}(k^+,k_1^+)\Big|_{k^+=0}.
\label{eq:SGP-relation}
\end{align}
This is equivalent to the relation between $q'_\partial$ and $T_F$, and can be
derived from PT symmetry. It is possible
that the coefficient $g_1$ contains a soft-gluon-pole(SGP) part, that is,
\begin{align}
g_1(k^+,k_1^+)=\tilde{g}_1(k^+,k_1^+)+\de(k^+)g_1^{SGP}(k_1^+),
\end{align}
where $\tilde{g}_1(k^+,k_1^+)$ is finite at $k^+=0$. If this is the case, then
\begin{align}
&\int dk^+ dk_1^+\Big[g_1(k^+,k_1^+)M^{(1)}_{\ga^+,\partial_\perp\psi}(k^+,k_1^+)
+g_3(k^+,k_1^+)M^{(1)}_{\ga^+,\partial_\perp G^+}(k^+,k_1^+)
\Big]\no
=&\int dk^+ dk_1^+\Big[\tilde{g}_1(k^+,k_1^+)M^{(1)}_{\ga^+,\partial_\perp\psi}(k^+,k_1^+)
+\Big(g_3(k^+,k_1^+)-\frac{1}{2}\de(k^+)g_1^{SGP}(k_1^+)\Big)
M^{(1)}_{\ga^+,\partial_\perp G^+}(k^+,k_1^+)\Big].
\end{align}
On the other hand, since $q_\partial$ contains gauge link $\mathcal{L}_n$ while $M^{(0)}_{\ga^+\ga_5,\partial_\perp\psi}$ does not, we should extract
some parts of $g_1,g_3$ to produce the gauge link $\mathcal{L}_n$, if we
want to write $M^{(0)}_{\ga^+\ga_5,\partial_\perp\psi}$ into $q_\partial$.
This is not difficult if we notice that
\begin{align}
\int dk^+
\frac{e^{ik^+ (\xi^--\xi_1^-)}}{k^+ +i\ep}G^+(\xi^-)
=-2\pi i \theta(\xi_1^- -\xi^- )G^+(\xi^-),
\end{align}
and
\begin{align}
\int dk^+P\frac{1}{k^+}M^{(1)}_{\ga^+\ga_5,\partial_\perp\psi}
=&\frac{1}{2}\int dk^+ (\frac{1}{k^+ +i\ep}+\frac{1}{k^+ -i\ep})M^{(1)}_{\ga^+\ga_5,\partial_\perp\psi}
=\frac{1}{2}\int dk^+ \frac{1}{k^+ +i\ep} M^{(1)}_{\ga^+\ga_5,\partial_\perp\psi}+c.c.
\end{align}
Then,
\begin{align}
&2s_\perp^\rho\int dk^+ P\frac{1}{k^+}\Big(
M^{(1)}_{\ga^+\ga_5,\partial_\perp\psi}(k^+,k_1^+)+
M^{(1)}_{\ga^+\ga_5,\partial_\perp G^+}(k^+,k_1^+)
\Big)\no
=&s_\perp^\rho M^{(0)}_{\ga^+\ga_5,\partial_\perp\psi}(k_2^+)
-\int\frac{d\xi^-}{2\pi}e^{i\xi^- k_2^+}
\langle ps|\bar{\psi}(0)\mathcal{L}_n(0)(-i\ga^+\ga_5)
\partial_\perp^\rho \Big(\mathcal{L}_n^\dagger(\xi^-)\psi(\xi^-)\Big)|ps\rangle
+O((G^+)^2).
\end{align}
Or equivalently,
\begin{align}
M^{(0)}_{\ga^+\ga_5,\partial_\perp\psi}(k_2^+)-2\int dk^+ P\frac{1}{k^+}\Big(
M^{(1)}_{\ga^+\ga_5,\partial_\perp\psi}(k^+,k_1^+)+
M^{(1)}_{\ga^+\ga_5,\partial_\perp G^+}(k^+,k_1^+)
\Big)=-2 q_\partial(x_2)+O((G^+)^2).
\end{align}
Then,
\begin{align}
&\int dk_2^+ g_0(k_2^+)M^{(0)}_{\ga^+\ga_5,\partial_\perp\psi}(k_2^+)
+\int dk^+ dk_1^+ [
g_2(k^+,k_1^+)M^{(1)}_{\ga^+\ga_5,\partial_\perp\psi}
+g_4(k^+,k_1^+)M^{(1)}_{\ga^+\ga_5,\partial_\perp G^+}
]\no
=& -2\int dk_2^+ g_0(k_2^+)q_\partial(x_2)\no
&+\int dk^+ dk_1^+ \Big[
\Big(g_2(k^+,k_1^+)+P\frac{2g_0(k_2^+)}{k^+}\Big)
M^{(1)}_{\ga^+\ga_5,\partial_\perp\psi}
+\Big(g_4(k^+,k_1^+)+P\frac{2g_0(k_2^+)}{k^+}
\Big)M^{(1)}_{\ga^+\ga_5,\partial_\perp G^+}\Big].
\end{align}
After gauge link is recovered, remaining $G^+$ can be viewed as a part of gluon
field strength tensor. We replace
$\partial_\perp^\rho G^+$ in $M^{(1)}_{\ga^+,\partial_\perp G^+}$ and
$M^{(1)}_{\ga^+\ga_5,\partial_\perp G^+}$ into $-G_\perp^{+\rho}$ as done in \cite{
Qiu:1998ia} or
\begin{align}
M^{(1)}_{\ga^+,\partial_\perp G^+}(k^+,k_1^+)
\rightarrow -\frac{1}{\pi}T_F(x_1,x_2),\
M^{(1)}_{\ga^+\ga_5,\partial_\perp G^+}(k^+,k_1^+)
\rightarrow -\frac{1}{\pi}T_\Delta(x_1,x_2),
\label{eq:M2TF}
\end{align}
with $x_2=x+x_1$, $k^+=xp_a^+$, $k_1^+=x_1 p_a^+$.
The final formula is
\begin{align}
8N_c^2 C_F W^{\mu\nu}=&
\int dk_b^-\bar{q}(x_b)\int dk_2^+
\tilde{g}_0^{\mu\nu}(k_2^+) q_\partial(x_2)\no
&+\int dk_b^-\bar{q}(x_b)\int dk^+ dk_1^+
\Big[\tilde{g}_1^{\mu\nu}(k^+,k_1^+)M^{(1)}_{\ga^+,\partial_\perp\psi}(k^+,k_1^+)
+\tilde{g}_2^{\mu\nu}(k^+,k_1^+) M^{(1)}_{\ga^+\ga_5,\partial_\perp\psi}(k^+,k_1^+)\no
&-\frac{1}{\pi}\tilde{g}_3^{\mu\nu}(k^+,k_1^+) T_F(x_1,x_2)
-\frac{1}{\pi}\tilde{g}_4^{\mu\nu}(k^+,k_1^+) T_\Delta(x_1,x_2)\Big],
\label{eq:main-W}
\end{align}
with
\begin{align}
\tilde{g}_0(k_2^+)=&-2g_0(k_2^+),\no
\tilde{g}_1(k^+,k_1^+)=& g_1(k^+,k_1^+)-\de(k^+)g_1^{SGP}(k_1^+),\no
\tilde{g}_2(k^+,k_1^+)=& g_2(k^+,k_1^+)+P\frac{2g_0(k_2^+)}{k^+},\no
\tilde{g}_3(k^+,k_1^+)=& g_3(k^+,k_1^+)-\frac{1}{2}\de(k^+)g_1^{SGP}(k_1^+),\no
\tilde{g}_4(k^+,k_1^+)=& g_4(k^+,k_1^+)+P\frac{2g_0(k_2^+)}{k^+}.
\label{eq:gi_tilde}
\end{align}
This is our main formula for calculation. Now it is clear that if $\tilde{g}_1$ or $\tilde{g}_2$ is nonzero, the collinear
expansion we employed does not preserve QCD gauge invariance. This is an important check of our calculation.

\section{Tree level result}
\begin{figure}
\begin{minipage}{0.8\textwidth}
\begin{center}
\begin{minipage}{0.45\textwidth}
\includegraphics[scale=0.4]{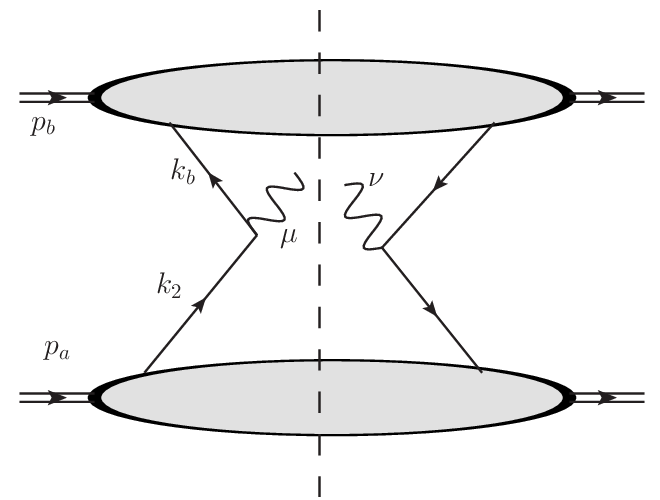}\\
(a)
\end{minipage}
\begin{minipage}{0.45\textwidth}
\includegraphics[scale=0.4]{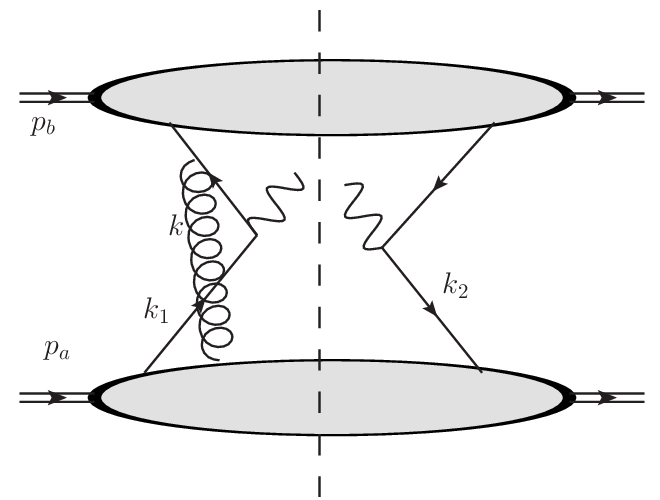}\\
(b)
\end{minipage}
\end{center}
\caption{Tree diagrams contributing to $W^{\mu\nu}$. The conjugated diagram of (b)
is not shown, but included in the calculation.}
\label{fig:tree}
\end{minipage}
\end{figure}

According to our expansion scheme, at most one $G^+$ appears. There are only
two diagrams at tree level, as shown in Fig.\ref{fig:tree}. The contribution of
conjugated diagram of Fig.\ref{fig:tree}(b) is included but not shown.
Suppose $\vec{p}_a$
is along $+Z$ axis in CM frame. $p_a^\mu\simeq(p_a^+,0,0_\perp)$, $p_b^\mu\simeq (0,p_b^-,0_\perp)$.  Under collinear limit,
the partons connecting the hard part and hadrons are collinear, whose momenta are
\begin{align}
k_2^\mu=(k_2^+,k_2^-,k_{2\perp})\sim Q(1,\la^2,\la),\
k_b^\mu=(k_b^+,k_b^-,k_{b\perp})\sim Q(\la^2,1,\la),
\end{align}
with $\la$ a small quantity. $k$ and $k_1$ are also collinear to $p_a$, like $k_2$.
At twist-3 level, the hard part should be expanded to $O(\la)$. One can find
the details about collinear expansion in \cite{Qiu:1998ia,Eguchi:2006mc,Chen:2016dnp}.
Because the hard coefficients $H_i$ contain the delta function for
momentum conservation $\de^n(k_2+k_b-q)$, we also need to do power expansion
for it, that is,
\begin{align}
\de^{n-2}(k_{2\perp}-q_\perp)=\de^{(n-2)}(q_\perp)-\frac{\de^{(n-2)}(q_\perp)}{
\partial q_\perp^\rho}k_{2\perp}^\rho+O(k_{2\perp}^2).
\end{align}
$\partial \de^{n-2}(q_\perp)/\partial q_\perp^\rho$ gives the derivative part and $\de^{(n-2)}(q_\perp)$ gives the non-derivative part. Such an expansion is necessary because we have to integrate out $q_\perp$ in the cross section. For example,
considering a test function $t(q_\perp)$, we have
\begin{align}
\int d^{n-2}q_\perp t(q_\perp)\de^{(n-2)}(q_\perp-k_{2\perp})
=t(k_{2\perp}).
\end{align}
Then, collinear expansion gives
\begin{align}
t(k_{2\perp})=t(0)+\frac{\partial t(k_{2\perp})}{\partial k_{2\perp}^\rho}\Big|_{k_{2\perp}=0}k_{2\perp}^\rho+O(k_{2\perp}^2).
\end{align}
Now the first and the second terms on RHS can be reproduced by the non-derivative
and derivative terms of the expansion of $\de^{(n-2)}(q_\perp-k_{2\perp})$.
Next we calculate these two parts separately.

Because final leptons are unpolarized, the leptonic tensor $L^{\mu\nu}$ is symmetric in $\mu,\nu$. So, we just need the symmetric part of the hadronic tensor.
For non-derivative part, only $H_{5,6}^{\mu\nu}$ can have a symmetric part. We have
 \begin{align}
H_5^{\mu\nu}|_{\text{tree}}^{\text{Fig.1b}}
=&H_6^{\mu\nu}|_{\text{tree}}^{\text{Fig.1b}}
= -2i N_cC_F\frac{1}{k_b\cdot k+i\ep}\frac{1}{p_a^+}\de(k_2^+-q^+)\de(k_b^--q^-)\de^{n-2}(q_\perp)
(p_a^\mu\tilde{s}_\perp^\nu +p_a^\nu \tilde{s}_\perp^\mu).
\end{align}
We just need the real part. Because
\begin{align}
\frac{1}{k_b\cdot k+i\ep}=P\frac{1}{k_b\cdot k}-i\pi\de(k_b\cdot k),
\end{align}
only the delta function gives a real part. This is called soft-gluon-pole(SGP)
contribution, since the delta function forces the gluon momentum
$k^+$ to be zero. We have
 \begin{align}
H_5^{\mu\nu}|_{\text{tree}}^{\text{Fig.1b+c.c}}
=&H_6^{\mu\nu}|_{\text{tree}}^{\text{Fig.1b+c.c}}
= -4\pi N_cC_F\de(k_b\cdot k)
\frac{1}{p_a^+}\de(k_2^+-q^+)\de(k_b^--q^-)\de^{n-2}(q_\perp)
(p_a^\mu\tilde{s}_\perp^\nu +p_a^\nu \tilde{s}_\perp^\mu).
\end{align}
However, from PT symmetry,
\begin{align}
M^{(1)}_{\ga^+\ga_5,\partial_\perp G^+}(k^+,k_1^+)\Big|_{k^+=0}=0.
\end{align}
So, $H_5$ can be ignored. We have
\begin{align}
8N_c^2 C_F W^{\mu\nu}_{non-de}
=&-4\pi N_c C_F \int dk_b^-\int dk_2^+ \de(k_2^+ -q^+)\de(k_b^--q^-)
\de^{n-2}(q_\perp)
\Big[\bar{q}(x_b)M^{(1)}_{\ga^+,\partial_\perp G^+}(0,k_2^+)
\frac{(p_a^\mu\tilde{s}_\perp^\nu +p_a^\nu\tilde{s}_\perp^\mu)}{p_a\cdot q}
\Big].
\end{align}

For the derivative part, $H_{0,1,2}^{\mu\nu}$ do not appear, because corresponding
matrix elements contain a bad component of fermion field. The contributions are of
twist-4 at least. So, only $\tilde{H}_0^{\mu\nu}$ and $H_{3-6}^{\mu\nu}$ may contribute. However, $\tilde{H}_0^{\mu\nu}$ and $H_{3,5}^{\mu\nu}$ contain one
$\ga_5$. After taking the trace, they are proportional to $\ep_\perp^{\mu\nu}$,
which is anti-symmetric in $\mu,\nu$. Thus, $\tilde{H}_0^{\mu\nu}$ and $H_{3,5}^{\mu\nu}$ do not contribute. As a result, we just need to calculate $H_4^{\mu\nu}$ and $H_6^{\mu\nu}$, which are
\begin{align}
H_4^{\mu\nu}=&i\frac{\partial \de^n(k+k_1+k_b-q)}{\partial k_{1\perp}^\rho}\tilde{s}_\perp^\rho Tr[H^{\mu\nu}\otimes T^a \ga^-\otimes \ga^+],\no
H_6^{\mu\nu}=&i\frac{\partial \de^n(k+k_1+k_b-q)}{\partial k_\perp^\rho}\tilde{s}_\perp^\rho Tr[H^{\mu\nu}\otimes T^a \ga^-\otimes \ga^+].
\end{align}
From Fig.\ref{fig:tree} without conjugated diagrams, we get
\begin{align}
p_a^+ H_4^{\mu\nu}|_{\text{tree}}^{\text{Fig.1b}}
=p_a^+ H_6^{\mu\nu}|_{\text{tree}}^{\text{Fig.1b}}
=-i\frac{\partial \de^n(k+k_1+k_b-q)}{\partial q_\perp^\rho}\tilde{s}_\perp^\rho 4N_cC_F \frac{1}{x+i\ep}g_\perp^{\mu\nu}.
\end{align}
With conjugated diagrams taken into account,
only real part contributes, which is proportional to $\de(x)$.
With the help of eq.(\ref{eq:SGP-relation}), $\partial_\perp\psi$ can be converted
to $\partial_\perp G^+$. Then, we have
\begin{align}
8N_c^2 C_F  W^{\mu\nu}_{de} =& \int dk_b^- \int dk^+ dk_1^+ \bar{q}(k_b^-)
\Big(-\frac{1}{2}H_4^{\mu\nu}+H_6^{\mu\nu}\Big)M^{(1)}_{\ga^+,\partial G^+}\no
=& -4\pi N_c C_F \int dk_b^-\int dk_2^+ \de(k_2^+ -q^+)\de(k_b^--q^-)
\bar{q}(x_b)M^{(1)}_{\ga^+,\partial G^+}(0,k_2^+)
\frac{\partial \de^{n-2}(q_\perp)}{\partial q_\perp^\rho}\tilde{s}_\perp^\rho
g_\perp^{\mu\nu}.
\end{align}
Thus, the total tree level result is
\begin{align}
W_{tree}^{\mu\nu}=& W^{\mu\nu}_{non-de}+W^{\mu\nu}_{de}\no
=&\frac{-\pi}{2N_c}\int dk_b^-\int dk_2^+ \de(k_2^+ -q^+)\de(k_b^--q^-)
\bar{q}(x_b)M^{(1)}_{\ga^+,\partial G^+}(0,k_2^+)
\Big[\frac{\de^{n-2}(q_\perp)}{p_a\cdot q}(p_a^\mu\tilde{s}_\perp^\nu +p_a^\nu\tilde{s}_\perp^\mu)
+\tilde{s}_\perp^\rho\frac{\partial \de^{n-2}(q_\perp)}{\partial q_\perp^\rho}
g_\perp^{\mu\nu}\Big]\no
=&\frac{1}{2N_c}\int \frac{dx_b}{x_b}\frac{dx_2}{x_2}
\de(1-\hat{x}_2)\de(1-\hat{x}_b)
\bar{q}(x_b)T_F(x_2,x_2)
\Big[\de^{n-2}(q_\perp)
\frac{1}{p_a\cdot q}(p_a^\mu\tilde{s}_\perp^\nu +p_a^\nu\tilde{s}_\perp^\mu)
+\tilde{s}_\perp^\rho\frac{\partial \de^{n-2}(q_\perp)}{\partial q_\perp^\rho}
g_\perp^{\mu\nu}\Big].
\label{eq:w-tree}
\end{align}
with
\begin{align}
\xi\equiv \frac{q^+}{p_a^+},\ \tau_\xi\equiv \frac{\tau}{\xi}=\frac{Q^2}{s\xi},\
\hat{x}_2\equiv \frac{\xi}{x_2},\ \hat{x}_b\equiv\frac{\tau_\xi}{x_b}.
\end{align}
In the last equality we have used eq.(\ref{eq:M2TF}) to change $M^{(1)}_{\ga^+,\partial_\perp G^+}$ to $T_F$. Above $W^{\mu\nu}_{tree}$ agrees with known result in \cite{Ma:2014uma}. It has been pointed out in \cite{Ma:2014uma}, such a structure of $W^{\mu\nu}$
with a derivative in $\de^{(n-2)}(q_\perp)$ satisfies QED gauge invariance in
the sense of distribution, i.e.,
\begin{align}
\int d^{n-2}q_\perp t(q_\perp)q_\mu W_{tree}^{\mu\nu}=0,
\end{align}
with $t(q_\perp)$ a test function which is normal at $q_\perp=0$. With explicit
expression of lepton momentum $l^\mu$ given in eq.(\ref{eq:lepton-momentum}), we
get
\begin{align}
(p_a^\mu\tilde{s}_\perp^\nu+p_a^\nu\tilde{s}_\perp^\mu)L_{\mu\nu}\Big|_{q_\perp=0}
=\frac{4s_2}{x_2}\cos\theta
\tilde{s}_\perp\cdot l_{\perp,cs},\quad
g_\perp^{\mu\nu}\frac{\partial L_{\mu\nu}}{\partial q_\perp^\rho}
\Big|_{q_\perp=0}=0.
\end{align}
The second equation indicates that derivative term does not contribute to SSA. We thus get the asymmetry
\begin{align}
A_{UT}\equiv & \frac{\frac{d\sig(s_\perp)}{dQ^2 d\Omega}
-\frac{d\sig(-s_\perp)}{dQ^2 d\Omega}
}{\frac{d\sig(s_\perp)}{dQ^2 d\Omega}+\frac{d\sig(-s_\perp)}{dQ^2 d\Omega}}
= -\frac{\sin 2\theta \sin\phi}{2Q(1+\cos^2\theta)}
\frac{\int\frac{d\xi}{\xi}\frac{dx_b}{x_b}\frac{dx_2}{x_2}
\de(1-\hat{x}_2)\de(1-\hat{x}_b)
\bar{q}(x_b)T_F(x_2,x_2)}
{\int \frac{d\xi}{\xi}\frac{dx_b}{x_b}\frac{dx_2}{x_2}
\de(1-\hat{x}_2)\de(1-\hat{x}_b)\bar{q}(x_b)q(x_2)}.
\end{align}
Chiral-Odd contribution is not included. As illustrated in \cite{Ma:2014uma},
this is the same as \cite{Zhou:2010ui}, but only $1/2$ of \cite{Boer:1997bw,
Boer:1999si,Lu:2011th}, \cite{Boer:2001tx}. We also notice
that if one ignores the time-reversal odd function in
\cite{Boer:1997bw,Boer:1999si,Lu:2011th}, i.e., $f_T(x)=0$ in their notation, the
result becomes the same. $f_T(x)=0$ is equivalent to our $M^{(0)}_{\gamma_\perp}=0$
as indicated by our eq.(\ref{eq:M_perp}). This equation is a consequence of PT
symmetry. Since we work in Feynman gauge, for which gauge potentials vanish at infinity, we think PT symmetry is preserved and this equation is reasonable.
We have used this equation in our calculation. One can find further discussions about this discrepancy in \cite{Ma:2014uma}.

\section{Virtual corrections}
In this section we present our results for one-loop virtual corrections. We
first give the corrections to hadronic tensor $W^{\mu\nu}$. Same as tree level
$W^{\mu\nu}$, the virtual correction contains nonderivative part and derivative part, which are calculated separately.
Direct calculation of the
one-loop integrals is complicated because a lot of tensor integrals
are involved. A better method is to use FIRE\cite{Smirnov:2008iw} to reduce these
tensor integrals to standard scalar integrals. The reduced integrals are
very simple: only standard two-point integrals remain. In the calculation, both
UV and IR divergences are regulated by dimensional regularization, and we do not
distinguish UV and IR divergences. UV divergences will be cancelled by
counter term contributions discussed in Sec.\ref{sec:reno}. After corrections to $W^{\mu\nu}$
are obtained, we give the result for $I\langle L\rangle$ and  $I\langle P_7\rangle$.

\begin{figure}
\begin{minipage}{0.8\textwidth}
\begin{flushleft}
\includegraphics[scale=0.4]{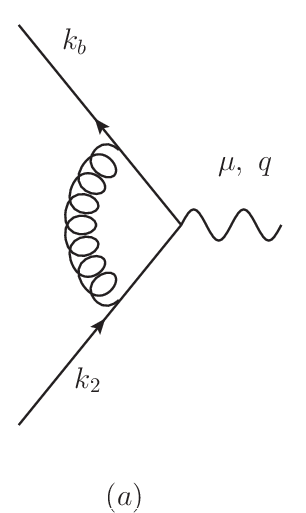}\hfill
\includegraphics[scale=0.4]{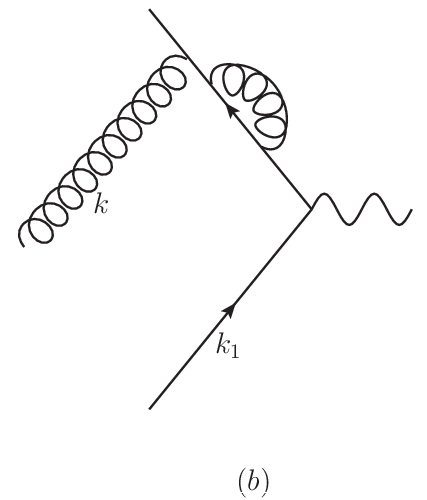}\hfill
\includegraphics[scale=0.4]{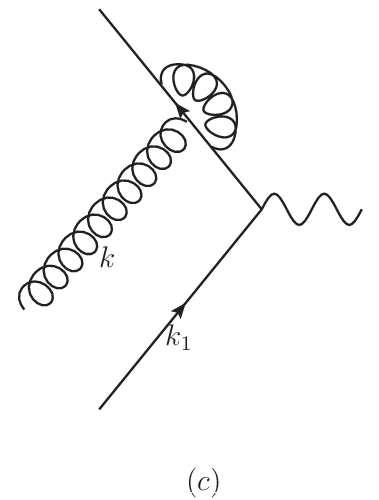}\hfill
\includegraphics[scale=0.4]{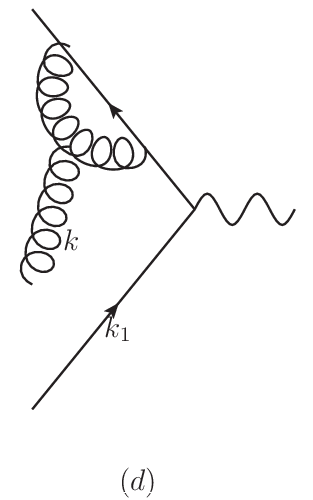}\hfill
\includegraphics[scale=0.4]{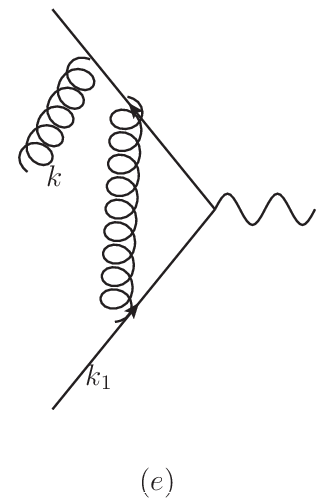}\\
\includegraphics[scale=0.4]{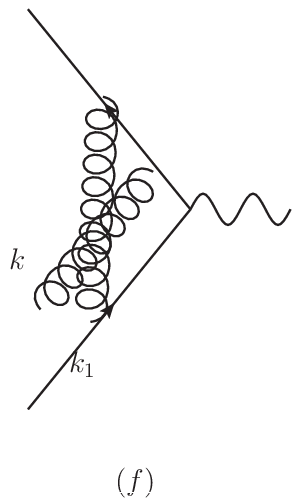}\hfill
\includegraphics[scale=0.4]{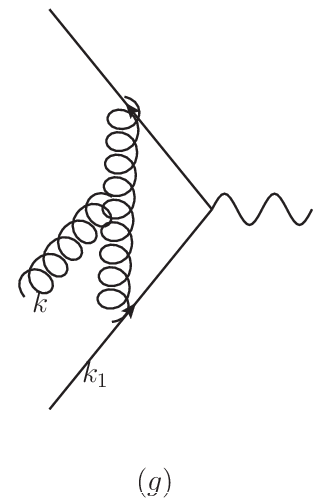}\hfill
\includegraphics[scale=0.4]{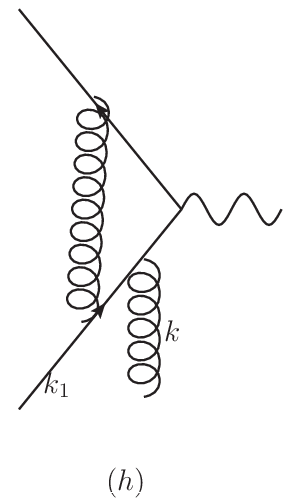}\hfill
\includegraphics[scale=0.4]{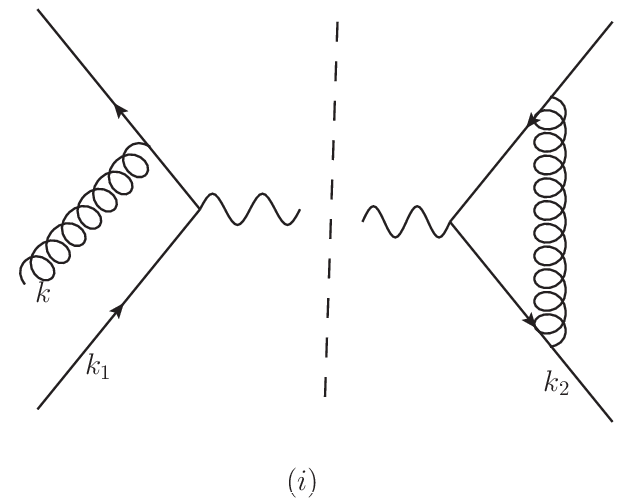}\hfill
\end{flushleft}
\caption{Diagrams for the hard part of one-loop virtual correction to $W^{\mu\nu}$. The right part
of (a-h) is not drawn, which is a tree level photon-quark vertex. The last diagram
(i) contains both left part and right part.
All conjugated diagrams are not shown, but included in the calculation. }
\label{fig:vir}
\end{minipage}
\end{figure}

The diagrams we consider in this part are shown in Fig.\ref{fig:vir}.
In order to get
the real part of $W^{\mu\nu}$ in physical region, we have to make clear
the analyticity of the amplitude about $s_0$ and $s_1$, where
\begin{align}
s_0=(k+k_b)^2=2k\cdot k_b,\ s_1=(k_1+k_b)^2=2k_1\cdot k_b.
\end{align}
We also define $s_2=s_0+s_1=2k_2\cdot k_b$, but do not use it to eliminate
$s_0$ or $s_1$ in the amplitude. The elimination will break the analytic property
about $s_0$ or $s_1$. Taking $s_0,s_1$ as variables is crucial for the extraction
of real part of $W^{\mu\nu}$.

By using Feynman parameters, it can be shown that for the diagrams in Fig.\ref{fig:vir},
the hard part of $W^{\mu\nu}$ is analytic on the upper half planes of $s_0$ and $s_1$, respectively. For example, one of the scalar integrals
appearing for Fig.\ref{fig:vir}(f) is
\begin{align}
I=\int \frac{d^n k_g}{(2\pi)^n} \frac{1}{
[(k_b+k_g)^2+i\ep][(k_b+k+k_g)^2+i\ep][(k_1-k_g)^2+i\ep]
[k_g^2+i\ep]}.
\label{eq:int-I}
\end{align}
With Feynman parameters and momentum shift, $k_g$ can be integrated out. Then,
\begin{align}
I\sim \int_0^1 \prod dx_i \de(1-x_1-x_2-x_3-x_4)
\Delta^{\frac{n}{2}-4},
\end{align}
with
\begin{align}
\Delta=(x_1 k_b+x_2(k_b+k)-x_3 k)^2-x_2(k_b+k)^2-i\ep
=-x_2(1-x_1-x_2)s_0-x_3(x_1+x_2) s_1-i\ep.
\end{align}
Since $0\leq x_i\leq 1$ and $1-x_1-x_2=x_3+x_4>0$, the integral is well defined
if $s_0$ and $s_1$ have positive imaginary parts. It can be checked that all
integrals appearing in Fig.\ref{fig:vir} have such a feature. So, we conclude that
the hard part is analytic on the upper half planes of $s_0,s_1$. Further,
there are only three massive quantities in $W^{\mu\nu}$, i.e., $s_0,s_1,s_0+s_1$,
any scalar integral $I$ can be written into following form
\begin{align}
I=\frac{1}{(s_0+i\ep)^\al(s_1+i\ep)^\be (s_0+s_1+i\ep)^\ga}f(s_0,s_1),
\end{align}
with $f(s_0,s_1)$ a polynomial of $s_0,s_1$, and $\al,\be,\ga$ some constants
depending on $\ep=4-n$. After this is clear, following calculation is straightforward. Reduced by FIRE,
all $H_i$ can be expressed by three two-point integrals:
\begin{align}
B(s_0)=&\mu^{\ep}\int\frac{d^n l}{(2\pi)^n}\frac{1}{[l^2+i\ep][(l+k_{b0})^2+i\ep]},\no
B(s_1)=&\mu^{\ep}\int\frac{d^n l}{(2\pi)^n}\frac{1}{[l^2+i\ep][(l+k_{b1})^2+i\ep]},\no
B(s_2)=&\mu^{\ep}\int\frac{d^n l}{(2\pi)^n}\frac{1}{[l^2+i\ep][(l+k_{b2})^2+i\ep]},
\end{align}
with $k_{b0}=k_b+k$, $k_{b1}=k_b+k_1$, $k_{b2}=k_b+k_2$.
Moreover, the complex conjugate of $B(s_2)$ is denoted by $B_c(s_2)$, i.e.,
$B_c(s_2)=B^*(s_2)$. The expression of $B(u)$ for a general $u$ is
\begin{align}
B(u)=&i\frac{(4\pi\mu^2)^{\ep/2}(-u-i\ep)^{-\ep/2}}{16\pi^2}
\Gamma(1+\frac{\ep}{2})\frac{2}{\ep}B(1-\frac{\ep}{2},1-\frac{\ep}{2}).
\end{align}
If $u<0$, $B(u)$ is purely imaginary. Expansion in $\ep$ gives
\begin{align}
B(u)=&i\frac{1}{16\pi^2}\Big(\frac{4\pi\mu^2}{|u|}\Big)^{\ep/2}
\Gamma(1+\frac{\ep}{2})
\Big[
\frac{2}{\ep}+2+(2-\frac{\pi^2}{12})\ep + \theta(u)\Big(-\frac{\pi^2\ep}{4}\Big)
+i\pi (1+\ep)\theta(u)+O(\ep^2)\Big].
\label{eq:Bu}
\end{align}
Now, the real part of $B(u)$ is made explicit. The real part
is finite.

We list nonderivative and derivative contributions
separately in the following.

\subsection{ Corrections to nonderivative part of $W^{\mu\nu}$}
Define $h_i^{\mu\nu}$ as
\begin{align}
H_i^{\mu\nu}=\de(k_2^+-q^+)\de(k_b^--q^-)\de^{n-2}(q_\perp)
h_i^{\mu\nu},\ i=0,1,\cdots,6.
\end{align}
$\tilde{h}_0^{\mu\nu}$ is defined similarly. By definition, $q_\perp=0$ in $h_i$. There is only one transverse vector in $h_i$, that is $\tilde{s}_\perp$. So,
the tensor structure for $h_i$ is,
\begin{align}
h_i^{\mu\nu}=&2g_s^2\text{Re}\Big\{A_i t_a^{\mu\nu}+B_i t_b^{\mu\nu}
\Big\},\ i=0,\cdots,6;\no
t_a^{\mu\nu}=& p_a^\mu\tilde{s}_\perp^\nu +p_a^\nu\tilde{s}_\perp^\mu,\
t_b^{\mu\nu}=p_b^\mu\tilde{s}_\perp^\nu +p_b^\nu\tilde{s}_\perp^\mu.
\end{align}
Only real part is needed, as indicated by $\text{Re}\{\cdots\}$. The overall
factor 2 is due to conjugated diagrams. The coefficients $A_i,B_i$
are obtained directly from the diagrams in Fig.\ref{fig:vir}. As stated before,
they are analytic on the upper half-planes of $s_0,s_1$.

For $h_0$, we have
\begin{align}
\frac{1}{N_cC_F}p_b^- A_0=& -\frac{x_2 (\epsilon -2)
   C_F}{x_b}B\left(s_2\right),\no
\frac{1}{N_cC_F}p_b^- B_0=& -\frac{\left(\epsilon ^2+8\right)
    C_F}{\epsilon }B\left(s_2\right).
\label{eq:H0}
\end{align}
For $\tilde{h}_0$, we have
\begin{align}
\frac{1}{N_cC_F}\tilde{A}_0=& -\frac{2 x_2 (\epsilon -2)
   C_F}{s_2}B\left(s_2\right),\no
\frac{1}{N_cC_F}\tilde{B}_0=& -\frac{2 x_b \left(\epsilon ^2+8\right)
   C_F}{s_2 \epsilon }B\left(s_2\right).
\end{align}
For $h_1$, we have
\begin{align}
\frac{1}{N_c C_F}A_1=&\frac{2x_2}{s_2 (x+i\ep)}\frac{-2+\ep}{\ep}
\Big[x C_A (B(s_0)-B(s_2))-\ep x_2 C_F (B(s_2)-B_c(s_2))\Big],\no
\frac{1}{N_c C_F}B_1=&\frac{2x_b}{s_2 (x+i\ep)}\frac{1}{\ep}\Big[
C_A(-2+\ep)\Big(x_2 B(s_0)-x B(s_2)\Big)
+C_F(8+\ep^2)x_2 \Big(B(s_2)+B_c(s_2)\Big)
\Big].
\end{align}
For $h_2$, we have
\begin{align}
\frac{1}{N_c C_F}A_2=&\frac{2x_2}{s_2 (x+i\ep)}\frac{-2+\ep}{\ep}
\Big[
C_A x \Big(B(s_0)-B(s_2)\Big)-C_F x_2 \ep \Big(B(s_2)-B_c(s_2)\Big)
\Big],\no
\frac{1}{N_c C_F}B_2=&\frac{2x_b}{s_2 (x+i\ep)}\frac{1}{\ep}\Big[
C_A (-2+\ep)\Big(x B(s_2)-x_2 B(s_0)\Big)
+C_F (8-4\ep +\ep^2) x_2 \Big(B(s_2)-B_c(s_2)\Big)
\Big].
\end{align}
For $h_3$, we have
\begin{align}
\frac{1}{N_cC_F}p_a^+A_3=&-\frac{2x_2}{s_2 (x+i\ep) (x_1+i\ep)}\frac{-2+\ep}{\ep}
\Big[
C_A x \Big( B(s_0)-B(s_2)\Big)
-\ep C_F  x_1 \Big(B(s_2)+B_c(s_2)\Big)
\Big],\no
\frac{1}{N_cC_F}p_a^+B_3=&\frac{2x_b}{s_2 (x+i\ep) (x_1+i\ep)}\frac{1}{\ep}\Big[
C_A(-2+\ep)\Big(x_2 B(s_0)-x B(s_2)\Big)
+C_F(8+\ep^2) x_1 \Big( B(s_2)+B_c(s_2) \Big)
\Big].
\end{align}
For $h_4$, we have
\begin{align}
\frac{1}{N_cC_F}p_a^+A_4=&-\frac{2x_2(-2+\ep)}{s_2 (x+i\ep) (x_1+i\ep)}\frac{1}{\ep}
\Big[C_A x \Big(B(s_0)-B(s_2)\Big)
-C_F \ep x_1 \Big(B(s_2)-B_c(s_2)\Big)\Big],\no
\frac{1}{N_cC_F}p_a^+B_4=&\frac{2x_b}{s_2 (x+i\ep)(x_1+i\ep)}\frac{1}{\ep}\Big[
C_A (-2+\ep)\Big(x_2 B(s_0)-x B(s_2)\Big)
+C_F (8+\ep^2) x_1 \Big( B(s_2)-B_c(s_2)\Big)
\Big].
\end{align}
For $h_5$, we have
\begin{align}
\frac{1}{N_c C_F}p_a^+ A_5=&
-\frac{2x_2(2C_A+\ep C_F)}{s_2 (x_1+i\ep)}\Big(B(s_0)-B(s_2)\Big)\no
&+\frac{2(C_A-2C_F)x_2^2(-4+\ep^2)}{s_2 (x+i\ep)^2 \ep}\Big(B(s_1)-B(s_2)\Big)\no
&-\frac{2x_2}{s_2 (x+i\ep) \ep}\Big[
\ep(2C_A +\ep C_F) B(s_0)
-2\Big(C_A \ep +C_F(4-3\ep+\ep^2)\Big)B(s_2)
+C_F (8+\ep^2) B_c(s_2)\Big],\no
\frac{1}{N_c C_F}p_a^+B_5=&
\frac{2x_b (2C_A+\ep C_F)}{s_2 (x_1+i\ep)}\Big(B(s_0)-B(s_2)\Big)\no
&-\frac{2(C_A-2C_F)x_2 x_b}{s_2 (x+i\ep)^2 \ep}\Big(B(s_1)-B(s_2)\Big)\no
&+\frac{2x_b}{s_2 (x+i\ep) \ep}\Big(
\ep(2C_A+\ep C_F)B(s_0)-2\ep (C_A-2C_F)B(s_2)+(8+\ep^2)C_F B_c(s_2)
\Big).
\end{align}
For $h_6$, we have
\begin{align}
\frac{1}{N_cC_F}p_a^+(A_5-A_6)=& \frac{4C_F x_2 (-2+\ep)}{s_2 (x+i\ep)}B_c(s_2),\no
\frac{1}{N_cC_F}p_a^+(B_5-B_6)=& \frac{4C_F x_b(8+\ep^2)}{s_2 (x+i\ep) \ep}B_c(s_2).
\end{align}
By using the formula
\begin{align}
\frac{1}{x+i\ep}=P\frac{1}{x}-i\pi\de(x),
\end{align}
$P\frac{1}{x}$ or $\de(x)$ may contribute to the real parts of $A_i,B_i$, since
$B(s_i)$ is complex for general $s_i$. The contribution proportional to $\de(x)$
(or $P\frac{1}{x}$) is called pole contribution (or non-pole contribution).
Pole contribution may be proportional to $\de(x)$ or $\de(x_1)$. The former is called soft-gluon-pole(SGP)
contribution, and the latter is called soft-fermion-pole(SFP) contribution.
Before proceeding, we should show that SFP does not contribute to $h_i$, otherwise,
the PV in eq.(\ref{eq:EOMs}) for $x_1$ is ill-defined.

From the explicit results of $h_i$, $h_{3,4,5,6}$ may contain SFP contributions, which are given by following combination
\begin{align}
\frac{B(s_0)-B(s_2)}{x_1+i\ep}
,\ \text{or}\ \frac{x_2 B(s_0)-x B(s_2)}{x_1+i\ep}.
\end{align}
Under the limit $x_1\rightarrow 0$, we have $x=x_2$, $s_0=s_2$, so,
\begin{align}
\de(x_1)(B(s_0)-B(s_2))=0,\ \de(x_1)(x_2 B(s_0)-x B(s_2))=0.
\end{align}
Thus, SFP contribution vanishes, and all $h_i$'s are well defined at $x_1=0$ and
PV in eq.(\ref{eq:EOMs}) for $x_1$ is well defined.

Similarly, $g_i^{\mu\nu}$ is decomposed as
\begin{align}
g_i^{\mu\nu}=2g_s^2 \text{Re}\Big[
C_i t_a^{\mu\nu}+D_i t_b^{\mu\nu}
\Big]\de^n(k_2+k_b-q),
\end{align}
with $i=0,1,\cdots,4$. Using eq.(\ref{eq:hi2gi}),
we get the coefficients $C_i,D_i$ as follows.

For $g_0$, we have
\begin{align}
C_0=D_0=0.
\end{align}

For $g_1$, we have
\begin{align}
\frac{1}{N_cC_F}p_a^+ C_1=& 0,\no
\frac{1}{N_cC_F}p_a^+ D_1=& \frac{4C_F(2x_2-x)x_b}{s_2 x_1}\frac{4-\ep+\ep^2}{\ep}
\frac{1}{x+i\ep}\Big(B(s_2)-B_c(s_2)\Big).
\end{align}
Because $B(s_2)-B_c(s_2)$ is purely imaginary, only SGP gives nonzero contribution.
$1/x_1$ is a PV, which is introduced by EOM relations.
For $g_2$, we have
\begin{align}
\frac{1}{N_cC_F}p_a^+ C_2=& 0,\no
\frac{1}{N_cC_F}p_a^+ D_2=& \frac{4C_F x_b}{s_2 x_1}\frac{4-\ep+\ep^2}{\ep}
\Big(B(s_2)-B_c(s_2)\Big).
\end{align}
$D_2$ has no pole contribution. Because $B(s_2)-B_c(s_2)$ is purely imaginary, we have $\text{Re}D_2=0$. As a result, $g_2=0$.

For $g_3$, we have
\begin{align}
\frac{1}{N_cC_F}p_a^+ C_3=& \frac{2x_2}{s_2}
\Big[
-\frac{2C_A+\ep C_F}{x_1}\Big(B(s_0)-B(s_2)\Big)
+\frac{(C_A-2C_F)x_2}{(x+i\ep)^2}\frac{\ep^2-4}{\ep}\Big(B(s_1)-B(s_2)\Big)\no
&+\frac{1}{x+i\ep}\Big(
-(2C_A+\ep C_F)B(s_0)
+2(C_A+C_F\frac{4-3\ep+\ep^2}{\ep})B(s_2)
+2C_F\frac{-4+\ep-\ep^2}{\ep}B_c(s_2)
\Big)
\Big],\no
\frac{1}{N_cC_F}p_a^+ D_3=&
\frac{2x_b}{s_2}\Big[
\frac{2C_A+\ep C_F}{x_1}\Big(B(s_0)-B(s_2)\Big)
-\frac{(C_A-2C_F)x_2}{(x+i\ep)^2}\frac{\ep^2-4}{\ep}\Big(B(s_1)-B(s_2)\Big)\no
&+\frac{1}{x+i\ep}\Big(
(2C_A+\ep C_F)B(s_0)+2(-C_A+C_F\frac{4+\ep+\ep^2}{\ep})B(s_2)
+2C_F\frac{-4+\ep-\ep^2}{\ep}B_c(s_2)\Big)
\Big].
\end{align}

For $g_4$, we notice that $g_3-g_4$ is very simple,
\begin{align}
\frac{1}{N_c C_F}p_a^+ (C_3-C_4)=& 0,\no
\frac{1}{N_c C_F}p_a^+ (D_3-D_4)=&\frac{8C_F x_b}{s_2 (x+i\ep)}\frac{4-\ep+\ep^2}{\ep}
\Big(B(s_2)-B_c(s_2)\Big).
\label{eq:g3-g4}
\end{align}
For convenience, we define $\Delta g^{\mu\nu}\equiv g_3^{\mu\nu}-g_4^{\mu\nu}$.
Its expression is shown above. As can be seen, $\Delta g$ contains only SGP contribution.

As a summary, we find $g_0=g_2=0$. $g_1$ is nonzero, but contains only SGP contribution. Considering eq.(\ref{eq:gi_tilde}), we have $\tilde{g}_1=0$.
Thus, QCD gauge invariance is preserved. $g_3,g_4$ contain SGP and non-pole contributions. Next, we present the results separately.

\subsubsection{SGP contribution}
The limit $x\rightarrow 0$ in $g_i$ gives SGP contribution. What is special
in dimensional regularization is $B(s_0)=0$ if $s_0=0$. In addition,
$B(s_1)$ should be expanded near $s_2$ since in physical region $s_1=s_2-s_0$.
\begin{align}
B(s_1)-B(s_2)=\frac{\ep}{2}\frac{s_0}{s_2}B(s_2)+\frac{\ep}{4}(1+\frac{\ep}{2})
\frac{s_0^2}{s_2^2}B(s_2)+O(s_0^3).
\end{align}
$s_0$ on right hand side eliminates the double pole $1/(x+i\ep)^2$ in $C_3,D_4$. After this we get SGP contribution by the replacement $1/(x+i\ep)\rightarrow-i\pi\de(x)$. The results are
\begin{align}
\frac{1}{N_cC_F}p_a^+ C_1^{SGP}=&0,\no
\frac{1}{N_cC_F}p_a^+ D_1^{SGP}=&\de(x)\frac{16\pi C_F x_b}{s_2}\frac{4-\ep+\ep^2}{\ep}
\text{Im}\Big(B(s_2)\Big),\no
\frac{1}{N_cC_F}p_a^+ C_3^{SGP}=&\de(x) \frac{8\pi C_F x_2 }{s_2 }\frac{4-\ep+\ep^2}{\ep}
\text{Im}\Big(B(s_2)\Big),\no
\frac{1}{N_cC_F}p_a^+ D_3^{SGP}=&\de(x) \frac{8\pi C_F x_b }{s_2 }\frac{4-\ep+\ep^2}{\ep}
\text{Im}\Big(B(s_2)\Big).
\end{align}
So,
\begin{align}
C_3^{SGP}/x_2=D_3^{SGP}/x_b,\ D_1^{SGP}=2D_3^{SGP}.
\end{align}
These two relations are important.

As discussed in Sec.\ref{sec:def-tw3},
it is $-\frac{1}{2}g_1+g_3$ that gives the final contribution
from $T_F(x_2,x_2)$. From above result, we have
\begin{align}
\frac{1}{N_cC_F}p_a^+ (-\frac{1}{2}C_1^{SGP}+C_3^{SGP})
=&\frac{1}{N_cC_F}p_a^+ C_3^{SGP},\no
\frac{1}{N_cC_F}p_a^+ (-\frac{1}{2}D_1^{SGP}+D_3^{SGP})
=&0.
\end{align}
The vanishing of the second equation indicates SGP part has the same
tensor structure as tree level.

For $g_4$, because $T_{\Delta}(x_1,x_2)=-T_{\Delta}(x_2,x_1)$,
the SGP contribution does not exist.

In summary, SGP contribution to the nonderivative part of $W^{\mu\nu}$ is
\begin{align}
8N_c^2 C_F W^{\mu\nu}_{SGP}
=&-\frac{2g_s^2}{\pi} t_a^{\mu\nu}
\int dk_b^- \int dk^+ dk_1^+ \bar{q}(x_b)\de^n(k+k_1+k_b-q)
C_{3}^{SGP}
T_F(x_2,x_2).
\label{eq:vir-sgp}
\end{align}

\subsubsection{Non-pole contribution}
For non-pole contribution, $x\neq 0$, only the real part of $B(s_i)$ contributes to $g_i$.
The combination $B(s_2)-B_c(s_2)$ is purely imaginary and does not
contribute. So, $g_{0,1,2}$ are zero.
Moreover, the difference between $D_3$ and $D_4$ is proportional to $B(s_2)-B_c(s_2)$, thus the non-pole part of $g_3$ and $g_4$
is the same, i.e.,
\begin{align}
C_4^{NP}=C_3^{NP},\ D_4^{NP}=D_3^{NP}.
\end{align}
For $C_3^{NP}$, we have
\begin{align}
\frac{1}{N_cC_F}p_a^+ C^{NP}_3=&
-\frac{4x_2 C_A}{x_1 s_2}\text{Re}\Big(B(s_0)-B(s_2)\Big)
-\frac{4x_2}{s_2 (x)_p }\text{Re}\Big(C_A B(s_0)-(C_A-2C_F)B(s_2)\Big)\no
&-\frac{8(C_A-2C_F)}{\ep}\frac{x_2^2}{s_2 (x^2)_p}\text{Re}\Big(B(s_1)-B(s_2)\Big).
\end{align}
$1/(x)_p$ is for the principal value of $1/x$. $1/(x^2)_p$ is well defined because
$B(s_1)-B(s_2)=0$ when $x=0$.
Because $s_0,s_1$ can be negative or positive, $C_3^{NP}$ is non-zero.
Note that
\begin{align}
\frac{1}{N_cC_F}p_a^+ \Big[\frac{1}{x_2}C_3+\frac{1}{x_b}D_3\Big]
=& \frac{8C_F}{s_2 (x+i\ep)}\frac{4-\ep+\ep^2}{\ep}(B(s_2)-B_c(s_2))
\end{align}
does not contain non-pole contribution, because $B(s_2)-B_c(s_2)$ is purely imaginary. So, the non-pole parts of $C_3$ and $D_3$ satisfy
\begin{align}
\frac{1}{x_2}C_3^{NP}+\frac{1}{x_b}D_3^{NP}=0.
\label{eq:cd-NP}
\end{align}
Due to principal value, the expansion of $C_3^{NP}$ in $\ep$ is subtle. When we
calculate $B(s_0)/x_p$, we encounter following quantity
\begin{align}
I_0(x)=\frac{(-x-i\ep)^{-\ep/2}}{x_p}.
\end{align}
To extract its real part and imaginary part, we write $I_0(x)$ as
\begin{align}
I_0(x)
=& \frac{|x|^{-\ep/2}}{x_p}[(-1-i\ep)^{-\ep/2}\theta(x)+\theta(-x)]\no
=& \frac{|x|^{-\ep/2}}{x_p}[\theta(x)+\theta(-x)+i\frac{\pi\ep}{2}\theta(x)
+O(\ep^2)]\no
=& \frac{|x|^{-\ep/2}}{x_p}[1+i\frac{\pi\ep}{2}\theta(x)+O(\ep^2)].
\end{align}
For the imaginary part, the expansion $|x|^{-\ep/2}=1-\frac{\ep}{2}\ln |x|+\cdots$ is not allowed, because we will encounter $\theta(x)/x_p$, which is meaningless. Actually,
$\theta(x)x^{-\ep}/x_p$ should be understood as $\theta(x)x^{-1-\ep}$, which
can be expanded as follows
\begin{align}
\theta(x)x^{-1-\ep/2}=\theta(x)\Big[-\frac{2}{\ep}\de(x)+\frac{1}{(x)_+}
-\frac{\ep}{2}\Big(\frac{\ln x}{x}\Big)_++O(\ep^2)\Big].
\end{align}
Now, the singularity at $x=0$ is extracted. Plus functions in the above is standard and  defined in the region $x\in (0,1)$.
On the other hand, the real part can be expanded directly,
\begin{align}
\frac{|x|^{-\ep/2}}{x_p}=\frac{1}{x_p}(1-\frac{\ep}{2}\ln |x|+O(\ep^2)).
\end{align}
Without $\theta(x)$, $1/x_p$ and $\ln|x|/x_p$ are well defined. The final result is
\begin{align}
I_0(x)
=& \frac{1}{x_p}\Big(1-\frac{\ep}{2}\ln |x|\Big)
+i\pi\theta(x)\Big(-\de(x)+\frac{\ep}{2}\frac{1}{(x)_+}\Big)
+O(\ep^2).
\end{align}
For definiteness, we let $\theta(0)=1$. Compared with the real part, the imaginary
part of $I_0(x)$ is not suppressed by $\ep$.

After this point is clear, $C_3^{NP}$ can be obtained, i.e.,
\begin{align}
\frac{p_a^+}{N_c C_F}C_3^{NP}
=& \frac{x_2}{4\pi s_2\Gamma(1-\frac{\ep}{2})}
\Big(\frac{4\pi\mu^2}{s_2}\Big)^{\ep/2}\Big\{
C_A \Big[
-\frac{\theta(-x)}{x_2-x}
+\de(x)\Big(-\frac{2}{\ep}-2-\ln x_2 \Big)
+\frac{\theta(x)}{x_+}\Big]\no
&+(C_A-2C_F)\Big[
-\frac{1}{x_p}-\frac{x_2\theta(x-x_2)}{x^2}\Big(\frac{2}{\ep}+2\Big)
-\frac{x_2\theta(x_2-x)}{(x^2)_p}\ln\frac{x_2-x}{x_2}
\Big]
\Big\}.
\label{eq:c3-np}
\end{align}
$x_2$ or $s_2$ is always positive.

In summary, for non-derivative part of $W^{\mu\nu}$, the non-pole contribution is
\begin{align}
8N_c^2 C_F W^{\mu\nu}_{NP}= &-\frac{2g_s^2}{\pi}
\int dk_b^- \int dk^+ dk_1^+ \bar{q}(x_b)
\de^n(k_b+k+k_1-q)
\Big(C_3^{NP}t_a^{\mu\nu}+D_3^{NP}t_b^{\mu\nu}\Big)
\Big[T_F(x_1,x_2)+T_{\Delta}(x_1,x_2)\Big].
\label{eq:vir-np}
\end{align}
$D_3^{NP}$ can be obtained from $C_3^{NP}$ from eq.(\ref{eq:cd-NP}).
Both $t_a^{\mu\nu}$ and $t_b^{\mu\nu}$ appear in this part, while only $t_a^{\mu\nu}$ appears in SGP part and in tree level result. Moreover,
the coefficients here are divergent. Different from twist-2 cases, the new
tensor structure with divergent coefficients does not imply a breaking of
factorization, as we will explain later.

\subsection{Corrections to derivative part of $W^{\mu\nu}$}
Similar to tree level contribution, for this part, only $H_4,H_6$ are nonzero and
only SGP contribution is possible. We have
\begin{align}
8N_c^2 C_F W_{de}^{\mu\nu}=& \int dk_b^- \int dk^+ dk_1^+
\de(q^- -k_b^-)\de(k^++k_1^+-q^+)
(H_6^{\mu\nu}-\frac{1}{2}H_4^{\mu\nu})M^{(1)}_{\ga^+,\partial G^+},
\end{align}
and
\begin{align}
H_6^{\mu\nu}=H_4^{\mu\nu}
=-i\frac{\partial \de^{n-2}(q_\perp)}{\partial q_\perp^\rho}\tilde{s}_\perp^\rho
Tr[H^{\mu\nu}\otimes T^a\ga^-\otimes \ga^+].
\end{align}
Because $k_\perp=k_{1\perp}=0$ in the trace, the trace is an on-shell quantity.
After calculation, we have
\begin{align}
8N_c^2 C_F W^{\mu\nu}_{de}=&-\frac{2g_s^2}{\pi}\int dk_b^- \int dk^+ dk_1^+
\de(q^- -k_b^-)\de(k^++k_1^+-q^+)
T_F(x_2,x_2) E g_\perp^{\mu\nu}
\frac{\partial \de^{n-2}(q_\perp)}{\partial q_\perp^\rho}\tilde{s}_\perp^\rho.
\label{eq:vir-de}
\end{align}
Both $\mu,\nu$ are transverse. The coefficient $E$ is related to $C_3^{SGP}$ as follows
\begin{align}
g_s^2 E=g_s^2 p_a\cdot q C_3^{SGP}
=& \al_s\frac{N_cC_F^2}{x_2 p_a^+}
\de(x)
\Big(\frac{4\pi\mu^2}{Q^2}\Big)^{\ep/2}\frac{1}{\Gamma(1-\frac{\ep}{2})}
\Big[\frac{8}{\ep^2}+\frac{6}{\ep}+8-\pi^2+O(\ep)\Big],
\end{align}
which is just the correction to quark form factor, as pointed out in \cite{Chen:2016dnp}. Note that $s_2=Q^2$ for virtual correction.

\subsection{Total virtual corrections}
Now, the complete virtual correction to hadronic tensor is the sum of eqs.(\ref{eq:vir-sgp},\ref{eq:vir-np},\ref{eq:vir-de}), that is,
\begin{align}
8N_c^2 C_F W^{\mu\nu}=&8N_c^2 C_F[W^{\mu\nu}_{SGP}+W^{\mu\nu}_{NP}+W^{\mu\nu}_{de}]\no
=&-\frac{2g_s^2}{\pi}
\int dk_b^-\bar{q}(x_b)\int dk_2^+\de(k_b^- - q^-)\de(k_2^+-q^+)
\Big\{\no
&\int dk^+ T_F(x_2,x_2)
\Big[
\de^{n-2}(q_\perp) t_a^{\mu\nu}
+\tilde{s}_\perp^\rho \frac{\partial \de^{n-2}(q_\perp)}{\partial q_\perp^\rho}
g_\perp^{\mu\nu} p_a\cdot q\Big]C_{3}^{SGP}
\no
&+\int dk^+  \de^{n-2}(q_\perp) \Big(
C_{3}^{NP}t_a^{\mu\nu}+D_{3}^{NP}t_b^{\mu\nu}\Big)
\Big(T_F(x_1,x_2)+T_\Delta(x_1,x_2)\Big)\Big\}.
\end{align}
If there is no non-pole contribution, the virtual correction has the
same structure as tree level hadronic tensor. The pole contribution, the last second
line of above result, satisfies QED gauge invariance, as shown for tree level hadronic tensor. The
correction to derivative part is the same as quark form factor, in agreement with
the conclusion of \cite{Chen:2016dnp}. In \cite{Chen:2016dnp}, such correction is
inferred from Ward identity for longitudinal gluon $G^+$. Here we recover the result
by direct calculation. The non-pole contribution, the last line, also satisfies such invariance: with $q_\perp=0$,
\begin{align}
q_\mu \Big(
C_{3,NP}t_a^{\mu\nu}+D_{3,NP}t_b^{\mu\nu}\Big)
=& \tilde{s}_\perp^\nu p_a\cdot p_b \Big(
x_b C_{3,NP}  +x_2 D_{3,NP}  \Big).
\end{align}
Due to eq.(\ref{eq:cd-NP}), this is zero. Thus, QED gauge invariance is satisfied.

Using eq.(\ref{eq:lepton-momentum}), we have
\begin{align}
t_a^{\mu\nu}L_{\mu\nu}\Big|_{q_\perp=0}=\frac{4s_2}{x_2}\cos\theta
\tilde{s}_\perp\cdot l_{\perp,cs},\quad
t_b^{\mu\nu}L_{\mu\nu}\Big|_{q_\perp=0}=-\frac{4s_2}{x_b}\cos\theta
\tilde{s}_\perp\cdot l_{\perp,cs},\quad
g_\perp^{\mu\nu}\frac{\partial L_{\mu\nu}}{\partial q_\perp^\rho}
\Big|_{q_\perp=0}=0.
\end{align}
Then, $I\langle L\rangle$ and $I\langle P_7\rangle$ can be obtained easily, which
are written as
\begin{align}
I\langle L\rangle\Big|_{v}
=&I\langle L\rangle\Big|_v^{SGP}+I\langle L\rangle\Big|_v^{NP},\no
I\langle P_7\rangle\Big|_{v}
=&I\langle P_7\rangle\Big|_v^{SGP}.
\end{align}
For SGP part, we have
\begin{align}
I\langle L\rangle \Big|^{SGP}_{v}
=&2\times l_{\perp,cs}\cdot\tilde{s}_{\perp}\cos\theta
\frac{\al_s}{128\pi N_c^2}\bar{A}_\ep
\int\frac{d\xi}{\xi}\frac{dx_b}{x_b}\frac{dx_2}{x_2}
\bar{q}(x_b)T_F(x_2,x_2)\de(1-\hat{x}_2)\de(1-\hat{x}_b)\no
&\times (-32)(N_c^2-1)\Big(
\frac{8}{\ep^2}+\frac{6}{\ep}+8-\pi^2
\Big),\no
I\langle P_7\rangle\Big|^{SGP}_{v} =& -\frac{1}{4l_{\perp,cs}\cdot\tilde{s}_{\perp}\cos\theta}
I\langle L\rangle \Big|^{SGP}_{v}.
\label{eq:vir-complete}
\end{align}
The overall factor $2$ in the first equation comes from the contribution of conjugated diagrams.

Only $I\langle L\rangle$ receives non-pole contribution:
\begin{align}
I\langle L\rangle \Big|^{NP}_{v}
=&2\times l_{\perp,cs}\cdot\tilde{s}_{\perp}\cos\theta
\frac{\al_s}{128\pi N_c^2}\bar{A}_\ep
\int\frac{d\xi}{\xi}\frac{dx_b}{x_b}\de(1-\hat{x}_b)
\bar{q}(x_b)
\int_{x^*_2-1}^1 dx_2 \Big[\no
&N_c^2(\frac{128}{\ep}+128+64\ln x^*_2)\de(x_2)T_F(x^*_2-x_2,x^*_2)\no
&+64T_F(x^*_2-x_2,x^*_2)
\Big(
(\frac{2}{\ep}+2)\frac{x^*_2\theta(x_2-x^*_2)}{x_2^2}
+\frac{N_c^2\theta(-x_2)}{x^*_2-x_2}
-\frac{N_c^2\theta(x_2)}{(x_2)_+}+\frac{1}{(x_2)_p}
+\frac{x^*_2 \theta(x^*_2-x_2)}{(x_2^2)_p}\ln\frac{x^*_2-x_2}{x^*_2}
\Big)
\Big].
\label{eq:vir-nonpole}
\end{align}
As stated before, non-pole part is divergent.
In these expressions,
\begin{align}
\bar{A}_\ep=\frac{1}{\Gamma(1-\frac{\ep}{2})}\Big(
\frac{4\pi\mu^2}{Q^2}
\Big)^{\ep/2},
\end{align}
and
\begin{align}
\hat{x}_2\equiv \frac{x_2^*}{x_2},\ \hat{x}_b\equiv \frac{x_b^*}{x_b},\
x_2^*=\xi,\ x_b^*=\tau_\xi,\ \xi=\frac{q^+}{p_a^+},\ \tau=\frac{Q^2}{2p_a\cdot p_b},\ \tau_\xi=\frac{\tau}{\xi}.
\end{align}
For SGP contribution, $x_2^*$ and $x_b^*$ serve as the lower bounds of integration about $x_2$, $x_b$, respectively. For non-pole contribution, the bounds for
$x_2$ are determined by the theta functions and the support of $T_F$.

Recently, \cite{Rein:2025b} gives NLO corrections to SSA in lepton-hadron
collision. For virtual corrections, their diagrams are the same as ours. Their result (eq.(37) of \cite{Rein:2025b}) contains both SGP contribution and non-pole
contribution. Our SGP correction is the same as theirs, up to a $\pi^2$, which
is due to the continuation of $Q^2$. Their result contains two kinds of non-pole contributions. In their notation, the first one is for the region $x'>x$, which
corresponds to the first line of our eq.(\ref{eq:c3-np}). The second nonpole contribution is
for the region $x'<0$, which corresponds to the second line of our eq.(\ref{eq:c3-np}). They
found that the divergent parts of these two non-pole contributions can be subtracted out by the evolution kernel of twist-3 distribution function. This is also the case
of our results. We will show this later. It should be noted that \cite{Rein:2025b}
performs the calculation in light-cone gauge, but we use Feynman gauge. So, they
use $G_\perp$ to do collinear expansion, but we use $G^+$. Moreover,
the integration order is different. \cite{Rein:2025b} first integrates out
the loop momentum and then do expansion in $k_T$, with $k_T$ the small transverse
momentum of initial quark. But we do expansion in transverse momentum first, and then reduce the resulting integrals. Reproducing
the structure of the virtual correction of \cite{Rein:2025b} is a support to our calculation scheme.

\section{Real corrections}
For real corrections, $q_\perp\neq 0$ is assumed. $W^{\mu\nu}$ thus depends on
$p_a^\mu,p_b^\mu,q_\perp^\mu$ and $\tilde{s}_\perp^\mu$. Because of QED gauge invariance
$q_\mu W^{\mu\nu}=0$, there are six independent tensor structures for transversely polarized case\cite{Arnold:2008kf}. For our convenience, the tensors can be chosen as
\begin{align}
P_{1,\ldots,4}^{\mu\nu}=
& \tilde{s}_\perp\cdot q_\perp \{
g_\perp^{\mu\nu}-\frac{q_\perp^\mu q_\perp^\nu}{q_\perp^2},\
\tilde{p}_a^\mu \tilde{p}_a^\nu,
\tilde{p}_b^\mu \tilde{p}_b^\nu,
\tilde{p}_a^\mu \tilde{p}_b^\nu+\tilde{p}_b^\mu \tilde{p}_a^\nu
\},\no
P_{5,6}^{\mu\nu}=&\{
\tilde{p}_a^\mu \Big(\tilde{s}_\perp^\nu-\frac{\tilde{s}_\perp\cdot q_\perp}{q_\perp^2}q_\perp^\nu
\Big)+(\mu\leftrightarrow \nu),\
\tilde{p}_b^\mu \Big(\tilde{s}_\perp^\nu-\frac{\tilde{s}_\perp\cdot q_\perp}{q_\perp^2}q_\perp^\nu
\Big)+(\mu\leftrightarrow \nu)
\},
\end{align}
with $\tilde{p}_i^\mu=p_i^\mu-p_i\cdot q q^\mu/q^2$, so that $q\cdot\tilde{p}_i=0$.

$W^{\mu\nu}$ is decomposed as
\begin{align}
W^{\mu\nu}=\sum_{i=1}^6 W_i P_i^{\mu\nu},
\label{eq:w-tensor}
\end{align}
with $W_i$ a scalar function of $p_a,p_b,q$.
To solve $W_i$, we first contract both sides of eq.(\ref{eq:w-tensor})
with $P_j^{\mu\nu}$ to get
\begin{align}
P_{j}\cdot W=\sum_{i=1}^6 W_i P_i\cdot P_j.
\label{eq:pij}
\end{align}
The dot product represents the contraction of Lorentz indices, e.g., $P_j\cdot W=
P_{j,\mu\nu}W^{\mu\nu}$. Eq.(\ref{eq:pij}) can be solved directly. But the
coefficients before $P_i\cdot W$ may depend on spin vector $\tilde{s}_\perp$ in a very complicated
way, since $P_i\cdot P_j$ depends on $\tilde{s}_\perp$. Noticing that
$W_i$ does not depend on the direction of $q_\perp$, we can integrate out
the angles of $q_\perp$ on both sides of eq.(\ref{eq:pij})
and then solve the obtained equations. According
to this treatment, we successfully get
\begin{align}
W_i=\frac{2-\ep}{\Omega_{n-2}\tilde{s}_\perp^2}\sum_j
C_{ij}\xi_j,\ \xi_j\equiv \int d\Omega_{n-2} P_j\cdot W,
\label{eq:xi-i}
\end{align}
where $d\Omega_{n-2}$ is for $q_\perp$.
After integration, $\xi_j$ depends on $\tilde{s}_\perp^2,
q_\perp^2,q\cdot p_a,q\cdot p_b$.
The coefficients $C_{ij}$ do not depend on $\tilde{s}_\perp$,
whose expressions are
\begin{align}
C_{11}=& \frac{1}{(\epsilon -1) q_t^2},\no
C_{22}=& -\frac{\left(q\cdot p_b\right){}^4}{q_t^6 \left(p_a\cdot p_b\right){}^4},\no
C_{23}=& -\frac{\left(q_t^2 p_a\cdot p_b-q\cdot p_a q\cdot p_b\right){}^2}{q_t^6 \left(p_a\cdot p_b\right){}^4},\no
C_{24}=& \frac{\left(q\cdot p_b\right){}^2 \left(q_t^2 p_a\cdot p_b-q\cdot p_a q\cdot p_b\right)}{q_t^6
   \left(p_a\cdot p_b\right){}^4},\no
C_{33}=& -\frac{\left(q\cdot p_a\right){}^4}{q_t^6 \left(p_a\cdot p_b\right){}^4},\no
C_{34}=& \frac{\left(q\cdot p_a\right){}^2 \left(q_t^2 p_a\cdot p_b-q\cdot p_a q\cdot p_b\right)}{q_t^6
   \left(p_a\cdot p_b\right){}^4},\no
C_{44}=& -\frac{q_t^4 \left(p_a\cdot p_b\right){}^2-2 q_t^2 p_a\cdot p_b q\cdot p_a q\cdot p_b+2 \left(q\cdot
   p_a\right){}^2 \left(q\cdot p_b\right){}^2}{2 q_t^6 \left(p_a\cdot p_b\right){}^4},\no
C_{55}=& \frac{\left(q\cdot p_b\right){}^2}{2 (\epsilon -1) q_t^2 \left(p_a\cdot p_b\right){}^2},\no
C_{56}=& \frac{q\cdot p_a q\cdot p_b-q_t^2 p_a\cdot p_b}{2 (\epsilon -1) q_t^2 \left(p_a\cdot p_b\right){}^2},\no
C_{66}=& \frac{\left(q\cdot p_a\right){}^2}{2 (\epsilon -1) q_t^2 \left(p_a\cdot p_b\right){}^2}.
\label{eq:cij}
\end{align}
Notice that $C_{ij}$ is symmetric, $C_{ij}=C_{ji}$ and $q_t=|\vec{q}_\perp|$.

Then,
\begin{align}
I\langle L\rangle
=& \int d^n q \de(q^2-Q^2)L^{\mu\nu}W_{\mu\nu}\no
=& \sum_{i,j} \int d^n q \de(q^2-Q^2) L\cdot P_i \frac{2-\ep}{\Omega_{n-2}\tilde{s}_\perp^2}
C_{ij} \xi_j.
\end{align}
Since $C_{ij}$ and $\xi_j$ are functions of $q_\perp^2$, the angle of $q_\perp$
is just contained in $L\cdot P_i$. Again, we integrate out the angle of $q_\perp$
first,
\begin{align}
\int d^n q \de(q^2-Q^2) L\cdot P_i \frac{2-\ep}{\Omega_{n-2}\tilde{s}_\perp^2}
C_{ij} \xi_j
=&\int\frac{dq^+}{2q^+} \int dq_t q_t^{n-3} \Big(\int d\Omega_{n-2}L\cdot P_i \Big)
\frac{2-\ep}{\Omega_{n-2}\tilde{s}_\perp^2}C_{ij} \xi_j.
\end{align}
There is a transverse momentum $l_{\perp,cs}$ in $L^{\mu\nu}$, with
lepton momentum $l^\mu$ given in eq.(\ref{eq:lepton-momentum}). Since
$P_i$ contains only one $\tilde{s}_\perp$, we must have
\begin{align}
\Big(\int d\Omega_{n-2}L\cdot P_i \Big)\propto
l_{\perp,cs}\cdot \tilde{s}_{\perp}\Omega_{n-2}.
\end{align}
To obtain the coefficients is easy and we have
\begin{align}
\sum_{i,j}\Big(\int d\Omega_{n-2}L\cdot P_i \Big)
\frac{2-\ep}{\Omega_{n-2}\tilde{s}_\perp^2}C_{ij} \xi_j
=l_{\perp,cs}\cdot\tilde{s}_\perp \sum_i a_i \xi_i\frac{1}{\tilde{s}_\perp^2}.
\end{align}
Then,
\begin{align}
I\langle L\rangle
=& \frac{l_{\perp,cs}\cdot\tilde{s}_\perp}{\tilde{s}_\perp^2}
\sum_i \int \frac{dq^+}{2q^+}\int dq_t q_t^{n-3} a_i \xi_i.
\end{align}

The coefficients $a_i$ are
\begin{align}
a_1=&0,\no
a_2=& -\frac{4 Q E_t^2 l_{cs}^z}{q_t^2 \left(q\cdot
   p_a\right){}^2},\no
a_3=& \frac{16 Q l_{cs}^z \left(q\cdot
   p_a\right){}^2}{E_t^2 q_t^2 \left(p_a\cdot
   p_b\right){}^2},\no
a_4=& 0,\no
a_5=& -\frac{4 E_t l_{cs}^z}{q\cdot p_a},\no
a_6=& \frac{8 l_{cs}^z q\cdot p_a}{E_t p_a\cdot p_b},
\end{align}
with $E_t=\sqrt{Q^2-q_\perp^2}=\sqrt{Q^2+q_t^2}$, $l_{cs}^z=\frac{Q}{2}\cos\theta$.
A nontrivial feature is all $a_i$ are proportional to $l^z_{cs}$ or $\cos\theta$.

The formula can be further simplified. With $\xi_i$ given in eq.(\ref{eq:xi-i}),
we have
\begin{align}
I\langle L\rangle
=&\frac{l_{\perp,cs}\cdot\tilde{s}_\perp}{\tilde{s}_\perp^2}\sum_i
\int \frac{dq^+}{2q^+}\int dq_t q_t^{n-3} a_i \int d\Omega_{n-2} P_i\cdot W\no
=&\frac{l_{\perp,cs}\cdot\tilde{s}_\perp}{\tilde{s}_\perp^2}\sum_i
\int \frac{dq^+}{2q^+}\int dq_t q_t^{n-3}\int d\Omega_{n-2}  a_i  P_i\cdot W\no
=&\frac{l_{\perp,cs}\cdot\tilde{s}_\perp}{\tilde{s}_\perp^2}\sum_i
\int d^n q\de(q^2-Q^2) a_i  P_i\cdot W.
\end{align}
Because
\begin{align}
P_i\cdot W=\tilde{s}_\perp^{\tau}\tilde{s}_{\perp\rho}P_{i,\mu\nu\tau}W^{\mu\nu,\rho},
\end{align}
and
\begin{align}
\int d^n q \de(q^2-Q^2)a_i P_{i,\mu\nu\tau}W^{\mu\nu,\rho}=\tilde{W}_i g_{\perp\tau}^{\rho},\
\tilde{W}_i=\frac{1}{n-2}\int d^n q \de(q^2-Q^2) a_i P_{i,\mu\nu\tau}W^{\mu\nu,\rho}
g_{\perp\rho}^\tau,
\label{eq:wtilde}
\end{align}
we have
\begin{align}
I\langle L\rangle=l_{\perp,cs}\cdot\tilde{s}_\perp\sum_{i=1}^6 \tilde{W}_i.
\end{align}
This is our main formula for real corrections. There is no $\tilde{s}_\perp$ or
$l_{\perp,cs}$ in $\tilde{W}_i$ now. With $W^{\mu\nu}$ replaced
by $\tilde{W}_i$, the formula eq.(\ref{eq:main-W}) can be applied.
Further calculation is the same as that for real correction of weighted
cross section studied in \cite{Chen:2016dnp}.

\begin{figure}
\begin{center}
\begin{minipage}{0.3\textwidth}
\includegraphics[scale=0.3]{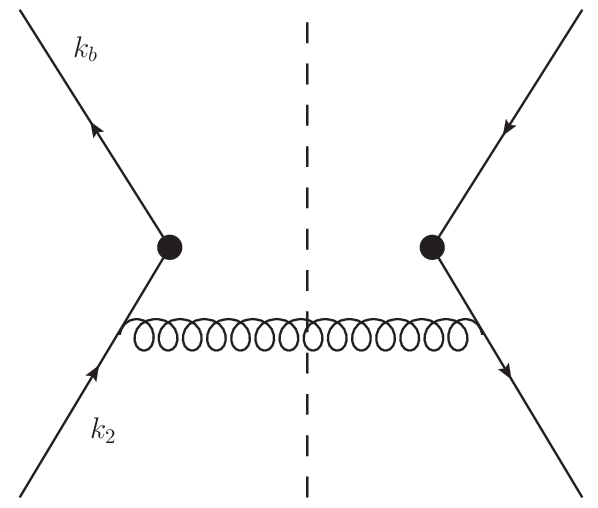}\\
(a)
\end{minipage}
\begin{minipage}{0.3\textwidth}
\includegraphics[scale=0.3]{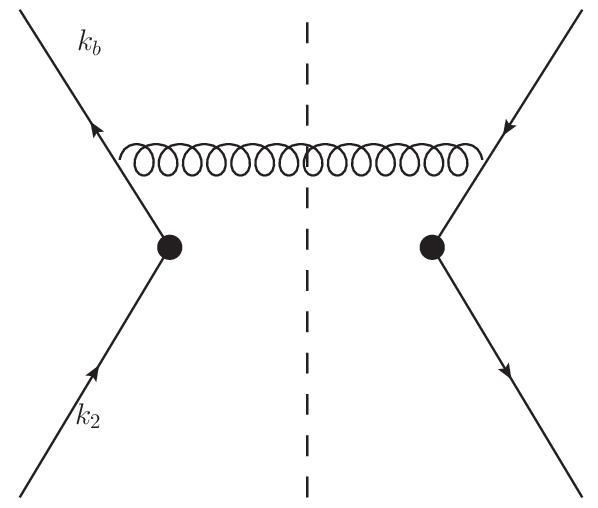}\\
(b)
\end{minipage}
\begin{minipage}{0.3\textwidth}
\includegraphics[scale=0.3]{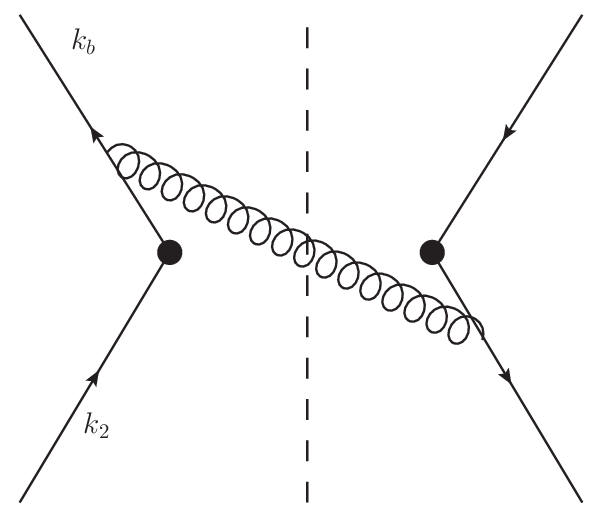}\\
(c)
\end{minipage}
\end{center}
\caption{The diagrams for the hard part of real corrections related to twist-3 two-point correlation functions. The conjugated diagram of (c) is not shown. The black dot represents quark-photon interaction.}
\label{fig:two-point-real}
\end{figure}

Same as weighted cross section, there is no contribution from two-point correlation
functions, i.e.,Fig.\ref{fig:two-point-real}. The amplitudes in these diagrams do not contain any absorptive part. So, $M^{(0)}_{\ga_\perp\ga_5}$ does not contribute
and we just need to consider three-point distribution functions. Still, three types of poles contribute: SGP, SFP and hard pole(HP). The details can be found in \cite{Chen:2016dnp}, here we just present the final result for each pole contribution. We note that for HP and SFP, the momentum fraction of initial
gluon $x\neq 0$, and it can be shown easily that the collinear expansion
based on $G^+$ or $G_\perp$ leads to the same hard coefficients\cite{Eguchi:2006mc}. We have checked this by direct calculation. In App.\ref{sec:hp-c}, we present some
details for the calculation of HP contribution based on $G_\perp$ expansion. The procedure based on $G^+$ expansion is too lengthy and not shown.
For SGP contribution, we use $G^+$ to do collinear expansion only. Especially,
for Fig.\ref{fig:sgp}, the contribution from $M^{(1)}_{\ga_\perp\ga_5}$ vanishes,
because corresponding $W^{\mu\nu}$ projected by $\ga_5\ga_\perp$ is anti-symmetric
in $\mu,\nu$. $M^{(1)}_{\ga_\perp}$ does not contribute because
\begin{align}
M^{(1)}_{\ga_\perp}(k^+,k_1^+)\Big|_{k^+=0}=0,
\end{align}
which can be shown from PT symmetry or from the second equation of eq.(\ref{eq:EOMs}) and eq.(\ref{eq:SGP-relation}). The SGP contribution from
$M^{(1)}_{\ga^+\ga_5,\partial_\perp\psi}(k^+,k_1^+)$ also vanishes,
although $M^{(1)}_{\ga^+\ga_5,\partial_\perp\psi}(k^+,k_1^+)$ may be nonzero
at $k^+=0$. In this work, we consider
only the contribution relevant to $\bar{q}\otimes T_F$. Because
$M^{(1)}_{\ga^+\ga_5,\partial_\perp G^+}$ is related to $T_\Delta$ rather than $T_F$, we do not consider the
contribution from this matrix elements in the following.
Our SGP contributions
are given by $M^{(1)}_{\ga^+,\partial_\perp\psi}$ and $M^{(1)}_{\ga^+,\partial_\perp G^+}$. As illustrated in Sec.\ref{sec:def-tw3}, SGP contributions from these two
matrix elements can be expressed by $T_F$. In addition, we also use $L^{\mu\nu}$
instead of the projection operators $P_i^{\mu\nu}$ to calculate. The same $I\langle L\rangle$ is obtained. This is a check of our calculation. In the calculation, we
first integrate out $k_g$, then do expansion in $k_\perp$ and $k_{2\perp}$. After
this step, $q_\perp$ is integrated out. It is easy to see if we first do expansion
in $k_\perp$ and $k_{2\perp}$, the result is the same.
Such an order of expansion and integration is different from the order taken
in \cite{Rein:2025b}, where $k_{\perp}$ or $k_{2\perp}$ expansion is done after
phase space integration is completed.

\subsection{SGP contributions}
\begin{figure}
\begin{center}
\begin{minipage}{0.15\textwidth}
\begin{center}
\includegraphics[scale=0.3]{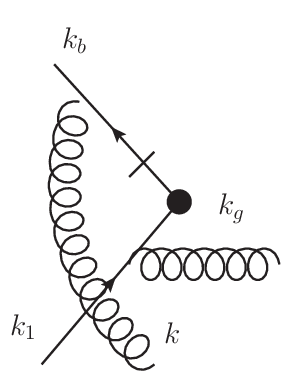}\\
(a)
\end{center}
\end{minipage}
\begin{minipage}{0.15\textwidth}
\begin{center}
\includegraphics[scale=0.3]{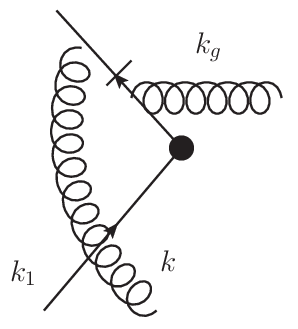}\\
(b)
\end{center}
\end{minipage}
\begin{minipage}{0.15\textwidth}
\begin{center}
\includegraphics[scale=0.3]{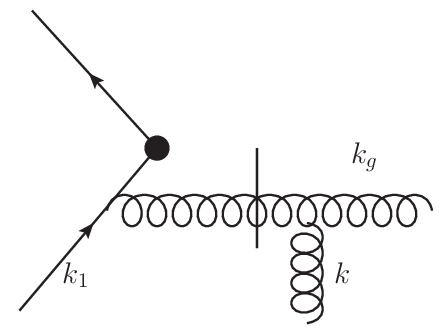}\\
(c)
\end{center}
\end{minipage}
\begin{minipage}{0.15\textwidth}
\begin{center}
\includegraphics[scale=0.3]{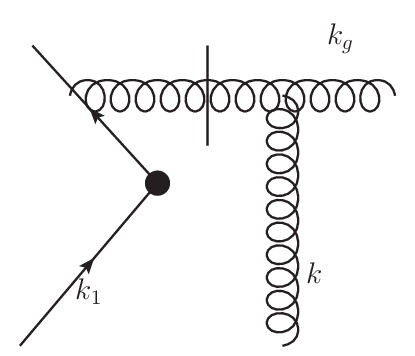}\\
(d)
\end{center}
\end{minipage}
\begin{minipage}{0.2\textwidth}
\begin{center}
\includegraphics[scale=0.3]{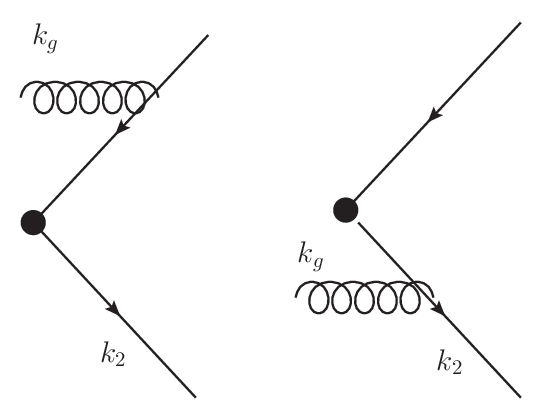}
\end{center}
\end{minipage}
\end{center}
\caption{Diagrams for the hard part of SGP contributions in $\bar{q}+qg$ channel.(a-d) are of the left parts and the last
two diagrams are of the right part. The black dot represents photon-quark interaction. The propagator with short bar is on-shell. Conjugated diagrams are not shown. }
\label{fig:sgp}
\end{figure}

SGP contribution is given by Fig.\ref{fig:sgp}. The short bar indicates the propagator is on-shell. Fig.\ref{fig:sgp}(c,d) are mirror diagrams\cite{Eguchi:2006mc},
which do not contribute. Our calculation confirms this.
The results from Fig.\ref{fig:sgp}(a,b) are written as
\begin{align}
\{I\langle P_7\rangle,I\langle L\rangle\}
=&2\times \frac{\al_s\{1,l_{\perp,cs}\cdot\tilde{s}_\perp\cos\theta\}}{128\pi N_c^2}\bar{A}_\ep
\int_r \bar{q}(x_b)T_F(x_2,x_2)
\Big[\no
&g_0\de(1-\hat{x}_2)\de(1-\hat{x}_b)
+g_1\de(1-\hat{x}_b)+g_2\de(1-\hat{x}_2)+\frac{g_3}{(1-\hat{x}_2)_+
(1-\hat{x}_b)_+}
\Big],
\end{align}
with
\begin{align}
\int_r\equiv \int_\tau^1\frac{d\xi}{\xi}\int_\xi^1\frac{dx_2}{x_2}
\int_{\tau_\xi}^1\frac{dx_b}{x_b}.
\label{eq:int-r}
\end{align}
The integration bounds are determined by $k_g^\pm \geq 0$ with $k_g$ the momentum of final gluon.

For $I\langle P_7\rangle$, the result is
\begin{align}
g_0=& \frac{64}{\epsilon ^2}+O\left(\epsilon^1\right),\no
g_1=&-\frac{16
   \left(\hat{x}_2^2+1\right)}{\left(1-\hat{x}_2
   \right)_+ \epsilon }+8
   \left(\left(L_2-1\right)
   \hat{x}_2^2+L_2-\hat{x}_2\right)+O\left(\epsilon ^1\right),\no
g_2=&-\frac{16
   \left(\hat{x}_b^3+\hat{x}_b\right)}{\epsilon
   \left(1-\hat{x}_b\right)_+}+8 \left(L_1\hat{x}_b^3+\left(L_1+1\right)
   \hat{x}_b+1\right)+O\left(\epsilon^1\right),\no
g_3=&8 \Big(\hat{x}_2^2 \left(2
   \hat{z}-1\right)+\hat{x}_2 \left(2
   \hat{z}^2-3 \hat{z}+1\right)+\hat{z}
   \left(\hat{z}^2-2 \hat{z}+2\right)\Big),
\label{eq:p7-sgp}
\end{align}
with
\begin{align}
\hat{z}\equiv 1-\hat{x}_2(1-\hat{x}_b)\text{ or }
\hat{x}_b=\frac{\hat{x}_2+\hat{z}-1}{\hat{x}_2}.
\end{align}

For $I\langle L\rangle$, the result is
\begin{align}
g_0=& -\frac{256}{\epsilon ^2}+O\left(\epsilon^1\right),\no
g_1=&\frac{64
   \left(\hat{x}_2^2+1\right)}{\left(1-\hat{x}_2
   \right)_+ \epsilon }-32
   \left(\left(L_2+1\right)
   \hat{x}_2^2+L_2-\hat{x}_2\right)+O\left(\epsilon ^1\right),\no
g_2=&\frac{64
   \left(\hat{x}_b^3+\hat{x}_b\right)}{\epsilon
   \left(1-\hat{x}_b\right)_+}-32 \left(L_1
   \hat{x}_b^3+\left(L_1+1\right)
   \hat{x}_b-1\right)+O\left(\epsilon^1\right),\no
g_3=& -\frac{32 \left(\hat{x}_2+\hat{z}-1\right)
   \left(\hat{x}_2^2 \left(\hat{z}
   \left(E_t-Q\right)+Q\right)+\hat{x}_2
   \left(\hat{z}-1\right) \hat{z} \left(E_t+Q
   \hat{z}\right)+Q \hat{x}_2^3
   \left(\hat{z}-1\right)+Q \hat{z}^2
   \left(\hat{z}^2-\hat{z}+1\right)\right)}{Q
   \hat{x}_2 \hat{z}^2},
\label{eq:IL-sgp}
\end{align}
with
\begin{align}
L_1=\Big(\frac{\ln(1-\hat{x}_b)}{1-\hat{x}_b}\Big)_+
-\frac{\ln\hat{x}_b}{1-\hat{x}_b},\
L_2=\Big(\frac{\ln(1-\hat{x}_2)}{1-\hat{x}_2}\Big)_+.
\end{align}

\subsection{HP contributions}
\begin{figure}
\begin{center}
\begin{minipage}{0.15\textwidth}
\begin{center}
\includegraphics[scale=0.3]{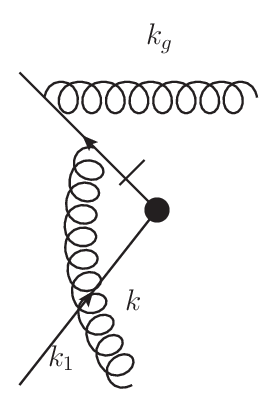}\\
(a)
\end{center}
\end{minipage}
\begin{minipage}{0.15\textwidth}
\begin{center}
\includegraphics[scale=0.3]{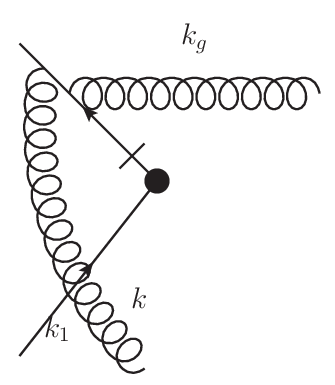}\\
(b)
\end{center}
\end{minipage}
\begin{minipage}{0.15\textwidth}
\begin{center}
\includegraphics[scale=0.3]{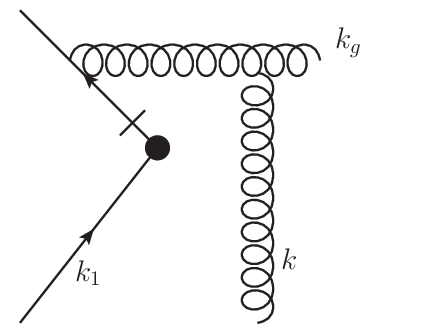}\\
(c)
\end{center}
\end{minipage}
\begin{minipage}{0.2\textwidth}
\begin{center}
\includegraphics[scale=0.3]{fig/qg-hr}\\
\end{center}
\end{minipage}
\end{center}
\caption{Diagrams for the hard part of hard pole contributions from $\bar{q}+qg$ channel. (a,b,c) are of the left part and the last two diagrams are of the right part. Conjugated diagrams are not shown.}
\label{fig:hp1}
\end{figure}

The hard pole contribution is given by Figs.\ref{fig:hp1},\ref{fig:hp2}.
There are two channels which
correspond to two different subprocesses: $\bar{q}+qg\rightarrow g+\ga^*$
and $\bar{q}+q\bar{q}\rightarrow q+\ga^*$. We next give their results separately.
\subsubsection{$\bar{q}+qg$ channel}
The diagrams are given in Fig.\ref{fig:hp1}.
The pole condition is $(k_1-q)^2=0$ or
\begin{align}
x_1=x_1^*=\frac{Q^2}{2p_a\cdot q}.
\end{align}
The results are summarized as
\begin{align}
\{I\langle P_7\rangle,I\langle L\rangle\}
=&2\times \frac{\al_s\{1,l_{\perp,cs}\cdot\tilde{s}_\perp\cos\theta\}}{128\pi N_c^2}\bar{A}_\ep
\int_r \bar{q}(x_b)T_F(x_1^*,x_2)
\Big[\no
&g_0\de(1-\hat{x}_2)\de(1-\hat{x}_b)
+g_1\de(1-\hat{x}_b)+g_2\de(1-\hat{x}_2)+\frac{g_3}{(1-\hat{x}_2)_+(1-\hat{x}_b)_+}
\Big].
\end{align}
For $I\langle P_7\rangle$, the result is
\begin{align}
g_0=& -\frac{64 N_c{}^2}{\epsilon ^2}-\frac{32
   N_c{}^2}{\epsilon }-16
   N_c{}^2+O\left(\epsilon ^1\right),\no
g_1=& \frac{16 \left(\left(\hat{x}_2+1\right)
   N_c{}^2\right)}{\left(1-\hat{x}_2\right)_+
   \epsilon }+\frac{8
   \left(\left(\hat{x}_2+1\right) \left(L_2
   \hat{x}_2-L_2+1\right)
   N_c{}^2\right)}{(1-\hat{x}_2)_+}+O\left(\epsilon ^1\right),\no
g_2=& \frac{16 \left(\left(\hat{x}_b^2+1\right)
   \left(\hat{x}_b+N_c{}^2-1\right)\right)}{\epsilon  \left(1-\hat{x}_b\right)_+}+\frac{8
   \left(\left(L_1 \hat{x}_b^3-L_1
   \hat{x}_b^2+\left(L_1+2\right)
   \hat{x}_b-L_1\right)
   \left(\hat{x}_b+N_c{}^2-1\right)\right)}{(1-\hat{x}_b)_+}+O\left(\epsilon ^1\right),\no
g_3=& -8 \left(\left(\hat{x}_2+\hat{z}^2\right)
   \left(N_c{}^2+\hat{z}-1\right)\right)+O\left(\epsilon ^1\right).
   \label{eq:p7-hp1}
\end{align}
For $I\langle L\rangle$, the result is
\begin{align}
g_0=&\frac{256 N_c{}^2}{\epsilon
   ^2}+O\left(\epsilon^1\right),\no
g_1=&-\frac{64 \left(\hat{x}_2+1\right)
   N_c{}^2}{\left(1-\hat{x}_2\right)_+ \epsilon
   }+32 \left(L_2 \hat{x}_2+L_2-1\right)
   N_c{}^2+O\left(\epsilon ^1\right),\no
g_2=&-\frac{64 \left(\hat{x}_b^2+1\right)
   \left(\hat{x}_b+N_c{}^2-1\right)}{\epsilon
   \left(1-\hat{x}_b\right)_+}+32 \left(L_1
   \hat{x}_b^2+L_1+2\right)
   \left(\hat{x}_b+N_c{}^2-1\right)+O\left(\epsilon ^1\right),\no
g_3=&\frac{32 \left(\hat{x}_2 \hat{z}
   E_t+\left(\hat{z}-1\right) \hat{z} E_t+Q
   \hat{x}_2^2\right)
   \left(N_c{}^2+\hat{z}-1\right)}{\hat{x}_2
   E_t}+O\left(\epsilon ^1\right).
\label{eq:hp1}
\end{align}
Except for $g_0$, the divergent part of $I\langle L\rangle $ is ``$-4$" times of
the divergent part of $I\langle P_7\rangle$.

\subsubsection{$\bar{q}+q\bar{q}$ channel}
\begin{figure}
\begin{center}
\begin{minipage}{0.15\textwidth}
\begin{center}
\includegraphics[scale=0.3]{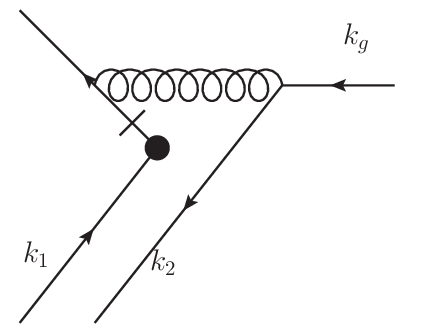}\\
(a)
\end{center}
\end{minipage}
\begin{minipage}{0.15\textwidth}
\begin{center}
\includegraphics[scale=0.3]{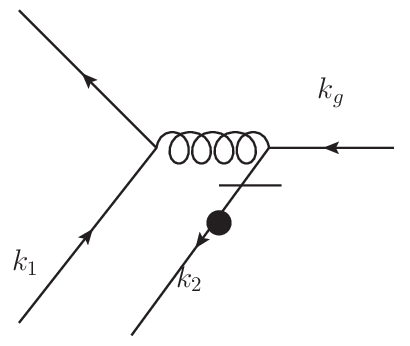}\\
(b)
\end{center}
\end{minipage}
\begin{minipage}{0.15\textwidth}
\begin{center}
\includegraphics[scale=0.3]{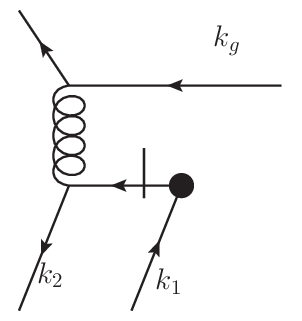}\\
(c)
\end{center}
\end{minipage}
\begin{minipage}{0.15\textwidth}
\begin{center}
\includegraphics[scale=0.3]{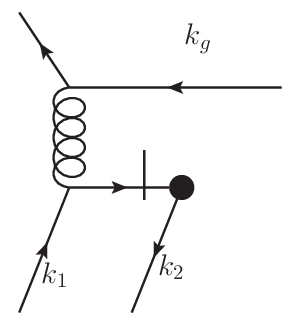}\\
(d)
\end{center}
\end{minipage}
\begin{minipage}{0.2\textwidth}
\begin{center}
\includegraphics[scale=0.25]{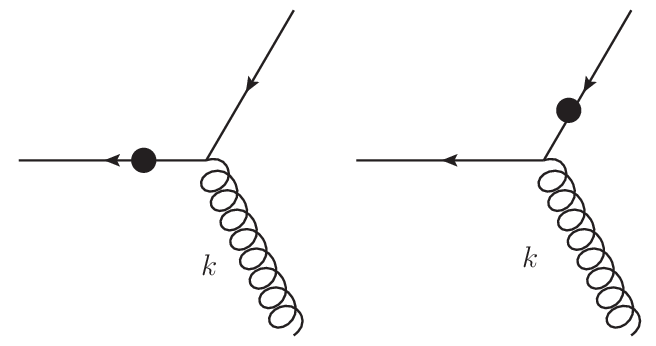}\\
\end{center}
\end{minipage}
\end{center}
\caption{Diagrams for the hard part of hard pole contributions from $\bar{q}+q\bar{q}$ channel. (a,b,c) are of the left part and the last two diagrams
are of the right part. Conjugated diagrams are not shown. }
\label{fig:hp2}
\end{figure}

This type of hard pole contribution is given by Fig.\ref{fig:hp2}.
The result is written as
\begin{align}
\{I\langle P_7\rangle,I\langle L\rangle\}
=&2\times \frac{\al_s\{1,l_{\perp,cs}\cdot\tilde{s}_\perp\cos\theta\}}{128\pi N_c^2}\bar{A}_\ep
\int_r \bar{q}(x_b)\no
&\Big\{
T_F(x_1^*,x_1^*-x_2)
\Big[
g_0\de(1-\hat{x}_2)\de(1-\hat{x}_b)
+g_1\de(1-\hat{x}_b)+g_2\de(1-\hat{x}_2)+\frac{g_3}{(1-\hat{x}_2)_+(1-\hat{x}_b)_+}
\Big]\no
&+T_F(x_2-x_1^*,-x_1^*) \frac{g_4}{(1-\hat{x}_2)_+(1-\hat{x}_b)_+}\Big\}.
\end{align}
Fig.\ref{fig:hp2}(a,c) gives $T_F(x_1^*,x_1^*-x_2)$, while
Fig.\ref{fig:hp2}(b,d) gives
$T_F(x_2-x_1^*,-x_1^*)$. For the latter, there is no contribution
proportional to delta functions.

For $I\langle P_7\rangle$, the result is
\begin{align}
g_0=& 0,\no
g_1=&-\frac{32 \hat{x}_2-16}{\epsilon }+8
   \left(\left(2 \hat{x}_2-1\right) \left(\log
   \left(1-\hat{x}_2\right)-1\right)\right),\no
g_2=& 0,\no
g_3=&-\frac{8(1-\hat{x}_2)}{\hat{z}}
\Big[
-N_c\left(\hat{z}-1\right) \left(-2 \hat{x}_2
   \left(\hat{z}-2\right)+\hat{z}^2-4
   \hat{z}+2\right)
   +\hat{z} \left(-2 \hat{x}_2+2 \hat{z}-1\right)
\Big],\no
g_4=& -\frac{8(1-\hat{x}_2)(1-\hat{x}_b)}{\hat{z}}
\Big[
\left(2 \hat{x}_2
   \left(\hat{z}-2\right)-\hat{z}^2+4
   \hat{z}-2\right) N_c+\left(\hat{z}-1\right)
   \hat{z} \left(2 \hat{x}_2-\hat{z}+1\right)
\Big].
\label{eq:p7-hp2}
\end{align}
For $I\langle L\rangle$, the result is
\begin{align}
g_0=&0 ,\no
g_1=&-\frac{64}{\epsilon }-\left(32-32 \log
   \left(1-\hat{x}_2\right)\right),\no
g_2=&0 ,\no
g_3=&-\frac{32 Q^3 \left(\hat{x}_2-1\right)
   \hat{x}_2 \left(\hat{z}-1\right) \left(2
   \hat{z} N_c-2 N_c+\hat{z}\right)}{E_t^3
   \left(\hat{x}_2+\hat{z}-1\right)}+\frac{32
   \left(\hat{x}_2-1\right)
   \left(\hat{x}_2+\hat{z}-1\right)
   \left(\hat{z}^2 N_c-\hat{z} \left(3
   N_c+1\right)+2 N_c\right)}{\hat{x}_2
   \hat{z}},\no
g_4=&-\frac{32 \left(\hat{x}_2-1\right)
   \left(\hat{z}-1\right)
   \left(\hat{x}_2+\hat{z}-1\right)
   \left(\hat{z} \left(N_c-1\right)-2
   N_c+\hat{z}^2\right)}{\hat{x}_2
   \hat{z}}+\frac{32 Q^3
   \left(\hat{x}_2-1\right) \hat{x}_2
   \left(\hat{z}-1\right)^2 \left(2
   N_c+\hat{z}\right)}{E_t^3
   \left(\hat{x}_2+\hat{z}-1\right)}.
\label{eq:hp2}
\end{align}
It is noted that the divergent part of $g_1$ is very different for the two observables.
\subsection{SFP contributions}
\begin{figure}
\begin{center}
\begin{minipage}{0.15\textwidth}
\begin{center}
\includegraphics[scale=0.3]{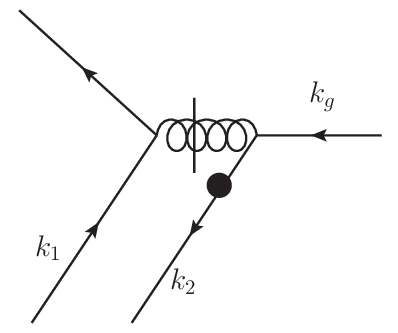}\\
(a)
\end{center}
\end{minipage}
\begin{minipage}{0.15\textwidth}
\begin{center}
\includegraphics[scale=0.3]{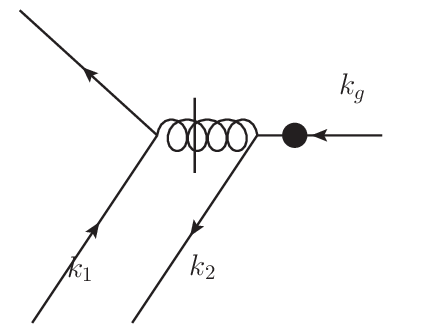}\\
(b)
\end{center}
\end{minipage}
\begin{minipage}{0.2\textwidth}
\begin{center}
\includegraphics[scale=0.3]{fig/qqbar-hr}\\
\end{center}
\end{minipage}

\end{center}
\caption{Diagrams for the hard part of SFP contribution from $\bar{q}+qg$ channel. (a,b) are of the left part and the last two diagrams are of the right part. Conjugated diagrams are not shown.}
\label{fig:sfp}
\end{figure}

In this work, we concentrate on the contribution of $\bar{q}\otimes T_F$. For this
case, the diagrams giving SFP are shown in Fig.\ref{fig:sfp}. Mirror diagrams are not shown, which do not contribute\cite{Eguchi:2006mc}.
The result is finite.
\begin{align}
\{I\langle P_7\rangle,I\langle L\rangle\}
=&2\times \frac{\al_s\{1,l_{\perp,cs}\cdot\tilde{s}_\perp\cos\theta\}}{128\pi N_c^2}\bar{A}_\ep \int_r \bar{q}(x_b)
T_F(0,-x_2)
\frac{g_3}{(1-\hat{x}_2)_+ (1-\hat{x}_b)_+}.
\end{align}
For $I\langle P_7\rangle$, the result is
\begin{align}
g_3=&-8 \left(\left(\hat{x}_2-1\right)
   \left(\hat{z}-1\right) \left(-4 \hat{x}_2
   \hat{z}+2
   \hat{x}_2^2+\hat{x}_2+\hat{z}^2\right)\right);
\label{eq:p7-sfp}
\end{align}
For $I\langle L\rangle$, the result is
\begin{align}
g_3=&-\frac{32 \left(\left(\hat{x}_2-1\right)
   \left(\hat{z}-1\right)
   \left(\hat{x}_2+\hat{z}-1\right)
   \left(\hat{x}_2 E_t+2 Q
   \hat{x}_2^2-Q \hat{z}^2\right)\right)}{Q
   \hat{x}_2 \hat{z}}.
\label{eq:sfp}
\end{align}
So far, we have listed all unsubtracted  hard coefficients relevant to $\bar{q}\otimes T_F$. For $I\langle P_7\rangle$, all results are the same as our
previous result\cite{Chen:2016dnp}.

\section{Renormalization and Subtraction}\label{sec:reno}
The counter terms and renormalization constants are given in Appendix.\ref{sec:cts}.
Here we
present some details for the renormalization of twist-3 distribution functions.
It is noted that in the definitions of $T_F$ and $M^{(1)}$
there is one explicit $g_s$.
We should be careful about the renormalization of this coupling. Consider
the renormalization for $W^{\mu\nu}$. First, we use bare quantities to write
it as
\begin{align}
W^{\mu\nu}=\Gamma_B\otimes \bar{M}^{(1)}_B\otimes \bar{q}_B,
\end{align}
where $\bar{M}^{(1)}_B$ does not include $g_s^B$ explicitly.
Then, the renormalization for $\Gamma_B$ is
\begin{align}
\Gamma_B=Z_2^{-4/2}Z_3^{-1/2}\Gamma_R.
\end{align}
Now, we extract one $g_s^B$ from $\Gamma_B$ to define $\tilde{\Gamma}_B$ as follows
\begin{align}
\Gamma_B=\tilde{\Gamma}_Bg_s^B.
\end{align}
Renormalized $\Gamma_R$ and $\tilde{\Gamma}_R$ are defined similarly,
\begin{align}
\Gamma_R=\tilde{\Gamma}_Rg_s.
\end{align}
Then,
\begin{align}
\tilde{\Gamma}_B=Z_2^{-4/2}Z_3^{-1/2}\frac{g_s}{g_s^B}\tilde{\Gamma}_R.
\end{align}
Using
\begin{align}
g_s^B=Z_g g_s,\ Z_g=Z_{1F}Z_2^{-1}Z_3^{-1/2},
\end{align}
we get
\begin{align}
\tilde{\Gamma}_B=Z_2^{-1}Z_{1F}^{-1}\tilde{\Gamma}_R.
\end{align}
It is for $g_s^B \bar{M}_B^{(1)}$ that is renormalized as a whole quantity,
\begin{align}
M_B^{(1)}=g_s^B \bar{M}^{(1)}_B =Z'_{pdf}\otimes g_s \bar{M}_R^{(1)}=Z'_{pdf}\otimes M_R^{(1)}
\end{align}
$Z'_{pdf}$ is the renormalizaition constant for twist-3 PDF, which is related to
the evolution kernel. For twist-2 PDF,
\begin{align}
\bar{q}_B =Z_{pdf}\otimes \bar{q}_R.
\end{align}
So,
\begin{align}
W^{\mu\nu}=&\tilde{\Gamma}_B\otimes g_s^B \bar{M}^{(1)}_B \otimes \bar{q}_B\no
=&\tilde{\Gamma}_R Z_2^{-1}Z_{1F}^{-1}\otimes [Z'_{pdf}\otimes M^{(1)}_R]
\otimes [Z_{pdf}\otimes \bar{q}_R].
\label{eq:w-renor2}
\end{align}
$\tilde{\Gamma}_R$ is calculated by using counter terms. At one loop level, it is
\begin{align}
\tilde{\Gamma}_R=\tilde{\Gamma}_{vir}+\tilde{\Gamma}_{tree}(1+2\de z_1^{\ga}
+\de z_{1F}-\de z_2).
\end{align}
We ignore real corrections to $\tilde{\Gamma}$ here. The last term $-\de z_2$ in
$(\cdots)$ is from the counter term contribution to Fig.\ref{fig:vir}(b).
Considering eq.(\ref{eq:w-renor2}), we have
\begin{align}
W^{\mu\nu}=&
[\tilde{\Gamma}_{vir}+\tilde{\Gamma}_{tree}(1+2\de z_1^{\ga}
-2\de z_2)]\otimes [Z'_{pdf}\otimes M^{(1)}_R]
\otimes [Z_{pdf}\otimes \bar{q}_R].
\end{align}
Note that $\de z_1^\ga=\de z_2$. So,
\begin{align}
W^{\mu\nu}=&
[\tilde{\Gamma}_{vir}+\tilde{\Gamma}_{tree}]\otimes [Z'_{pdf}\otimes M^{(1)}_R]
\otimes [Z_{pdf}\otimes \bar{q}_R].
\label{eq:w-renor}
\end{align}
That is, wave function renormalization and counter term contributions cancel each
other. As a net effect, the hard part is not affected by counter terms and self-energy corrections to external legs. Only distribution functions need renormalization. Divergence in hard part now is of IR type.

Next, we should make a collinear subtraction to remove
the collinear divergence in the hard part\cite{Collins:2008sg}.
The subtraction term is
obtained by following replacement in tree level cross section,
\begin{align}
\bar{q}(x)\rightarrow \bar{q}(x)-\Delta\bar{q}(x),\
T_F(x,x)\rightarrow T_F(x,x)-\Delta T_F(x,x).
\end{align}
$\Delta\bar{q}$ and $\Delta T_F$ are obtained from $Z_{pdf}$ and $Z'_{pdf}$
in eq.(\ref{eq:w-renor}). They are also related to the evolution kernels of $\bar{q}(x)$
and $T_F(x_1,x_2)$. In $\overline{\text{MS}}$ scheme, their expressions are
\begin{align}
\Delta \bar{q}(x)=&-\frac{\al_s}{2\pi}
\Big(\frac{2}{\ep}-\ga_E+\ln 4\pi\Big)
\int_x^1\frac{dz}{z}P_{qq}(z)\bar{q}(\frac{x}{z})
\equiv -\frac{\al_s}{2\pi}
\Big(\frac{2}{\ep}-\ga_E+\ln 4\pi\Big) P_{qq}\otimes \bar{q}(x),\no
\Delta T_F(x,x)=& -\frac{\al_s}{2\pi}\Big(\frac{2}{\ep}-\ga_E+\ln 4\pi\Big)
\mathcal{F}_q\otimes T_F(x).
\end{align}
$P_{qq}(z)$ is the standard DGLAP kernel,
\begin{align}
P_{qq}(z)=C_F\Big[\frac{1+z^2}{(1-z)_+}+\frac{3}{2}\de(1-z)\Big].
\end{align}
The kernel for $T_F$ is a little complicated\cite{Kang:2008ey,Braun:2009mi,
Zhou:2009jm,Ma:2012xn,Schafer:2012ra,Kang:2012em}, which reads
\begin{align}
\mathcal{F}_q\otimes T_F(x)
=&-N_cT_F(x,x)+\int_x^1 \frac{dz}{z}\Big[
P_{qq}(z)T_F(\xi,\xi)+\frac{N_c}{2}\Big(
T_{\Delta }(x,\xi)+\frac{
(1+z)T_F(x,\xi)-(1+z^2)T_F(\xi,\xi)}{1-z}\Big)\no
&+\frac{1}{2N_c}\Big(
(1-2z)T_F(x,x-\xi)+T_{\Delta }(x,x-\xi)
\Big)
-\frac{1}{2}\frac{(1-z)^2+z^2}{\xi}T_{G+}(\xi,\xi)
\Big],
\end{align}
with $\xi=x/z$. $T_{G+}$ is a pure gluon twist-3 distribution function, which
is ignored in the following calculation, since we are interested in quark contribution only.

From eq.(\ref{eq:w-tree}), the tree level $I\langle P_7\rangle$ and $I\langle L\rangle$ can be obtained,
\begin{align}
I\langle P_7\rangle \Big|_{tree}=& -\frac{1}{2N_c}
\int_\tau^1\frac{d\xi}{\xi}
\bar{q}(x_b^*)T_F(x_2^*,x_2^*),\no
I\langle L\rangle \Big|_{tree}=& \tilde{s}_\perp\cdot l_{\perp,cs}
\cos\theta \frac{2}{N_c}\int_\tau^1\frac{d\xi}{\xi}
\bar{q}(x_b^*)T_F(x_2^*,x_2^*),
\end{align}
with $x_b^*=\tau_\xi$ and $x_2^*=\xi$. The two tree level results differ by an
$\ep$ independent constant factor. Their subtraction terms have the same relation.
If the factorization is right, we must have
\begin{align}
I\langle w\rangle \Big|_{r+v}-I\langle w\rangle \Big|_{sub}=\text{finite.}
\end{align}
for $w=P_7,L$. We consider the subtraction for $I\langle P_7\rangle$ first.

\subsection{$I\langle P_7\rangle$ subtraction}
The subtraction term is
\begin{align}
I\langle P_7\rangle\Big|_{sub}=& -\frac{1}{2N_c}\int\frac{d\xi}{\xi}
\Big[(\Delta \bar{q}(x_b^*))T_F(x_2^*,x_2^*)
+\bar{q}(x_b^*)\Delta T_F(x_2^*,x_2^*)\Big].
\end{align}
With some transformations, it can be written into the standard form
\begin{align}
I\langle P_7\rangle\Big|_{sub}=&2\times \frac{\al_s}{128\pi N_c^2}\bar{A}_\ep
\int_r \bar{q}(x_b)\Big\{
T_F(x_2,x_2)\Big[
\de(1-\hat{x}_2)\de(1-\hat{x}_b) g_0
+\de(1-\hat{x}_b)g_1^{(a)}(\hat{x}_2)
+\de(1-\hat{x}_2)g_2(\hat{x}_b)
\Big]\no
&+T_F(x_2^*,x_2)\Big[\de(1-\hat{x}_b)g_1^{(b)}(\hat{x}_2)\Big]
+T_F(x_2^*,x_2^*-x_2)\Big[
\de(1-\hat{x}_b)g_1^{(c)}(\hat{x}_2)\Big]
\Big\},
\label{eq:sub}
\end{align}
with
\begin{align}
g_0=& \frac{-48+16N_c^2}{\ep}+(24-8N_c^2)\ln\frac{\mu^2}{Q^2},\no
g_1^{(a)}(\hat{x}_2)=&-\frac{16(1+\hat{x}_2^2)}{\ep (1-\hat{x}_2)_+}
+\frac{8(1+\hat{x}_2^2)}{(1-\hat{x}_2)_+}\ln\frac{\mu^2}{Q^2},\no
g_1^{(b)}(\hat{x}_2)=&\frac{16N_c^2(1+\hat{x}_2)}{\ep (1-\hat{x}_2)_+}
-\frac{8N_c^2(1+\hat{x}_2)}{(1-\hat{x}_2)_+}\ln\frac{\mu^2}{Q^2},\no
g_1^{(c)}(\hat{x}_2)=&\frac{16-32\hat{x}_2}{\ep}
+(16\hat{x}_2-8)\ln\frac{\mu^2}{Q^2},\no
g_2(\hat{x}_b)=&(N_c^2-1)\frac{16(1+\hat{x}_b^2)}{\ep (1-\hat{x}_b)_+}- (N_c^2-1)\frac{8(1+\hat{x}_b^2)}{(1-\hat{x}_b)_+}
\ln\frac{\mu^2}{Q^2}.
\end{align}
On the other hand, the complete virtual corrections is
\begin{align}
I\langle P_7\rangle\Big|_{v}=&2\times \frac{\al_s}{128\pi N_c^2}\bar{A}_\ep
\int_r \bar{q}(x_b)T_F(x_2,x_2)
\Big[\de(1-\hat{x}_2)\de(1-\hat{x}_b)g_0\Big],\no
g_0=& 8(N_c^2-1)\Big(
\frac{8}{\ep^2}+\frac{6}{\ep}+8-\pi^2
\Big).
\label{eq:div-vir}
\end{align}
Complete real corrections can be read from eqs.(\ref{eq:p7-sgp},\ref{eq:p7-hp1},\ref{eq:p7-hp2},\ref{eq:p7-sfp}).
Here we list their divergent parts for convenience.
\begin{align}
I\langle P_7\rangle\Big|_{r}\doteq &2\times \frac{\al_s}{128\pi N_c^2}\bar{A}_\ep\int_r \Big\{
\bar{q}(x_b)T_F(x_2,x_2)\Big[
\de(1-\hat{x}_2)\de(1-\hat{x}_b)g_0^{(a)}
+\de(1-\hat{x}_b)g_1^{(a)}+\de(1-\hat{x}_2)g_2^{(a)}\Big]\no
+& \bar{q}(x_b)T_F(x^*_1,x_2)
\Big[
\de(1-\hat{x}_2)\de(1-\hat{x}_b)g_0^{(b)}
+\de(1-\hat{x}_b)g_1^{(b)}+\de(1-\hat{x}_2)g_2^{(b)}\Big]\no
+& \bar{q}(x_b)T_F(x^*_1,x^*_1-x_2)
\Big[
\de(1-\hat{x}_2)\de(1-\hat{x}_b)g_0^{(c)}
+\de(1-\hat{x}_b)g_1^{(c)}+\de(1-\hat{x}_2)g_2^{(c)}\Big]
\Big\},
\label{eq:div-real}
\end{align}
with
\begin{align}
g_0^{(a)}=& \frac{64}{\ep^2},\no
g_1^{(a)}=& -\frac{16(\hat{x}_2^2+1)}{\ep (1-\hat{x}_2)_+},\no
g_2^{(a)}=& -\frac{16(\hat{x}_b+\hat{x}_b^3)}{\ep (1-\hat{x}_b)_+};
\end{align}
and
\begin{align}
g_0^{(b)}=& \frac{-64N_c^2}{\ep^2}-\frac{32N_c^2}{\ep},\no
g_1^{(b)}=& \frac{16N_c^2(\hat{x}_2+1)}{\ep (1-\hat{x}_2)_+},\no
g_2^{(b)}=& \frac{16(\hat{x}_b^2+1)(\hat{x}_b+N_c^2-1)}{\ep (1-\hat{x}_b)_+};
\end{align}
and
\begin{align}
g_0^{(c)}=&0,\no
g_1^{(c)}=&-\frac{32\hat{x}_2-16}{\ep},\no
g_2^{(c)}=&0.
\end{align}
For $g_0$, the sum of virtual and real corrections is
\begin{align}
g_0|_{v+r}=g_0|_{v}+g_0|_r^{(a+b+c)}
=(N_c^2-1)(\frac{64}{\ep^2}+\frac{48}{\ep})+\frac{64}{\ep^2}
-\frac{64}{\ep^2}N_c^2-\frac{32}{\ep}N_c^2
=\frac{16N_c^2-48}{\ep}.
\end{align}
This is the same as $g_0|_{sub}$, given in eq.(\ref{eq:sub}).
For $I\langle P_7\rangle$,
virtual correction contains vanishing $g_1$ and $g_2$, while the divergence of
$g_1,g_2$ from real correction is the same as that of the subtraction term, which
can be seen from eq.(\ref{eq:div-real}) and eq.(\ref{eq:sub}).

Now it is clear that
\begin{align}
I\langle P_7\rangle\Big|_{r}+I\langle P_7\rangle\Big|_{v}-
I\langle P_7\rangle\Big|_{sub}=\text{finite.}
\end{align}
So, $I\langle P_7\rangle$ can be factorized.

\subsection{$I\langle L\rangle$ subtraction}
Since
\begin{align}
I\langle L\rangle|_{tree}=-4l_{\perp,cs}\cdot\tilde{s}_\perp \cos\theta
I\langle P_7\rangle|_{tree},
\end{align}
the subtraction terms have the same relation, i.e.,
\begin{align}
I\langle L\rangle|_{sub}=-4l_{\perp,cs}\cdot\tilde{s}_\perp \cos\theta
I\langle P_7\rangle|_{sub}.
\end{align}
However, virtual corrections do not have the same relation. The divergent
part is
\begin{align}
I\langle L\rangle\Big|_{v}^{div}=&
2\times l_{\perp,cs}\cdot\tilde{s}_\perp \cos\theta
\frac{\al_s}{128\pi N_c^2}\bar{A}_\ep\int_\tau^1\frac{d\xi}{\xi}
\int_{\tau_\xi}^1\frac{dx_b}{x_b}\int \frac{dx_2}{x_2}
\Big\{
\bar{q}(x_b)T_F(x_2,x_2)\Big[\de(1-\hat{x}_2)\de(1-\hat{x}_b)g_0\Big]\no
&+\frac{dx_2}{x_2}\bar{q}(x_b)T_F(x_2^*,x_2^*-x_2)\de(1-\hat{x}_b)\Big[
g_1^{(c)}\theta(x_2-x_2^*)\Big]\Big\},
\end{align}
with
\begin{align}
g_0=& -32(N_c^2-1)\Big(\frac{8}{\ep^2}+\frac{6}{\ep}\Big)+\frac{128N_c^2}{\ep},\no
g_1^{(c)}=&\frac{128\hat{x}_2}{\ep}.
\end{align}
The last term of $g_0$ comes from $\de(x)$ part of $I\langle L\rangle|_v^{NP}$ in
eq.(\ref{eq:vir-complete}), and another term comes from SGP contribution.
$g_1^{(c)}$ is given by nonpole contributions. For $g_1^{(c)}$ term, the boundary
of $x_2$ should be made explicit.
Generally, the support of $T_F(x_1,x_2)$ is $|x_{1,2}|<1$, $|x_1-x_2|<1$. For
$T_F(x_2^*,x_2^*-x_2)$, because $0<x_2^*<1$ the allowed region of $x_2$ is
$x_2^*-1<x_2<1$. Considering $\theta(x_2-x_2^*)$, we finally have $x_2^*<x_2<1$,
which is precisely the same as that of hard pole contribution.
Thus,
\begin{align}
I\langle L\rangle\Big|_{v}^{div}= &
2\times l_{\perp,cs}\cdot\tilde{s}_\perp \cos\theta
\frac{\al_s}{128\pi N_c^2}\bar{A}_\ep\int_r
\Big\{
\bar{q}(x_b)T_F(x_2,x_2)\Big[\de(1-\hat{x}_2)\de(1-\hat{x}_b)g_0\Big]\no
&+
\bar{q}(x_b)T_F(x_2^*,x_2^*-x_2)\de(1-\hat{x}_b)g_1^{(c)}\Big\},
\end{align}
where $\int_r$ is defined by eq.(\ref{eq:int-r}).

The real corrections of $I\langle L\rangle$ are given by eqs.(\ref{eq:IL-sgp},
\ref{eq:hp1},\ref{eq:hp2},\ref{eq:sfp}).
The divergent part is summarized as follows:
\begin{align}
I\langle L\rangle\Big|_{r}^{div}= &2\times
l_{\perp,cs}\cdot\tilde{s}_\perp \cos\theta
\frac{\al_s}{128\pi N_c^2}\bar{A}_\ep\int_r\Big\{
\bar{q}(x_b)T_F(x_2,x_2)\Big[
\de(1-\hat{x}_2)\de(1-\hat{x}_b)g_0^{(a)}
+\de(1-\hat{x}_b)g_1^{(a)}+\de(1-\hat{x}_2)g_2^{(a)}\Big]\no
+& \bar{q}(x_b)T_F(x^*_1,x_2)
\Big[
\de(1-\hat{x}_2)\de(1-\hat{x}_b)g_0^{(b)}
+\de(1-\hat{x}_b)g_1^{(b)}+\de(1-\hat{x}_2)g_2^{(b)}\Big]\no
+& \bar{q}(x_b)T_F(x^*_1,x^*_1-x_2)
\Big[
\de(1-\hat{x}_2)\de(1-\hat{x}_b)g_0^{(c)}
+\de(1-\hat{x}_b)g_1^{(c)}+\de(1-\hat{x}_2)g_2^{(c)}\Big]
\Big\},
\end{align}
with
\begin{align}
g_0^{(a)}=& -\frac{256}{\ep^2},\no
g_0^{(b)}=& \frac{256N_c^2}{\ep^2},\no
g_0^{(c)}=& 0;
\end{align}
and
\begin{align}
g_1^{(a)}=& \frac{64(\hat{x}_2^2+1)}{\ep (1-\hat{x}_2)_+},\no
g_1^{(b)}=&-\frac{64N_c^2(1+\hat{x}_2)}{\ep(1-\hat{x}_2)_+},\no
g_1^{(c)}=& -\frac{64}{\ep};
\end{align}
and
\begin{align}
g_2^{(a)}=& \frac{64(\hat{x}_b^3+\hat{x}_b)}{\ep (1-\hat{x}_b)_+},\no
g_2^{(b)}=&-\frac{64(1+\hat{x}_b^2)(\hat{x}_b+N_c^2-1)}{\ep(1-\hat{x}_b)_+},\no
g_2^{(c)}=& 0.
\end{align}
Now, the sum of virtual and real corrections give
\begin{align}
g_0|_{r+v}=& g_0|_v+g_0|_r^{a+b+c}=-\frac{64N_c^2-192}{\ep}.
\end{align}
This is the same as the corresponding subtraction term, which is
\begin{align}
g_0|_{sub}=&-\frac{-192+64 N_c^2}{\ep}+32(N_c^2-1)(\pi^2-8).
\end{align}
Thus $g_0|_{r+v}-g_0|_{sub}$ is finite.

For $g_1^{(c)}$ parts in both virtual and real corrections, we notice that
they are proportional to $\de(\hat{x}_b-1)$. This implies
\begin{align}
q_t=\sqrt{\frac{Q^2}{\hat{x}_b}(1-\hat{x}_2)(1-\hat{x}_b)}=0,\
q^-=\frac{Q^2+q_t^2}{2q^+}=\frac{Q^2}{2\xi p_a^+},
\end{align}
and
\begin{align}
x_1^*=\frac{Q^2}{2p_a^+ q^-}=\xi=x_2^*.
\end{align}
Thus, $g_1^{(c)}$ part in virtual correction can be understood as a hard pole
contribution. This point is noticed also by \cite{Rein:2025b}.
Because of $T_F(x_1,x_2)=T_F(x_2,x_1)$, these two $g_1^{(c)}$ can
be added together. The result is
\begin{align}
g_1^{(c)}|_v+g_1^{(c)}|_r=\frac{128\hat{x}_2-64}{\ep}.
\end{align}
This is the same as the divergence in corresponding subtraction term $g_1^{(c)}|_{sub}$. Thus, $g_1^{(c)}|_{r+v}-g_1^{(c)}|_{sub}$ is finite.

Other divergences in real correction, including $g_1^{(a)}|_r$, $g_1^{(b)}|_r$
and $g_2|_r$, are also checked. They are the same as
the divergences of corresponding subtraction terms.

In summary, for $I\langle L\rangle$, all divergences in real and virtual
corrections can be subtracted out by subtraction terms constructed from
evolution equation of $T_F(x,x)$. Thus, twist-3 factorization holds at
one-loop level for SSA of DY. A new feature appearing here is
the divergent non-pole part of virtual correction has the same form as
hard pole contribution of real corrections. This is highly nontrivial.
The calculation performed in coordinate space at twist-3 level \cite{Scimemi:2019gge,Vladimirov:2021hdn} should be helpful to
understand how the factorization appears. We will use this method to do
calculation again in future.

\subsection{Final results}
Because we have checked that all divergences from real and virtual corrections can
be removed consistently by renormalization and subtraction, the final one-loop
correction is finite and can be obtained from
\begin{align}
I\langle L\rangle |_{final}=I\langle L\rangle |_{r+v}-I\langle L\rangle |_{sub},
\end{align}
by ignoring all divergences and setting $\bar{A}_\ep=1$. Results for $I\langle P_7\rangle $ are obtained in the same way and are the same as \cite{Chen:2016dnp}. For simplicity, we just show the result of $I\langle L\rangle$ here. It is
organized as
\begin{align}
I\langle L\rangle |_{final}=&
l_{\perp,cs}\cdot\tilde{s}_\perp \cos\theta
\frac{\al_s}{\pi N_c^2}\int\frac{d\xi}{\xi}
\Big\{-N_c\ln\frac{\mu^2}{Q^2}\Big[
\Big(P_{qq}\otimes \bar{q}(x_b^*)\Big)T_F(x_2^*,x_2^*)
+\bar{q}(x_b^*)\Big(\mathcal{F}_{q}\otimes T_F(x_2^*)\Big)
\Big]\no
&+\bar{q}(x_b^*)T_F(x_2^*,x_2^*)(N_c^2-1)(\frac{\pi^2}{2}-4)
+\bar{q}(x_b^*)\int_{x_2^*-1}^1 dx_2 T_F(x_2^*-x_2,x_2^*)
\Big[
N_c^2(2+\ln x_2^*)\de(x_2)\no
&+\frac{2x_2^*}{x_2^2}\theta(x_2-x_2^*)
+\frac{N_c^2\theta(-x_2)}{x_2^*-x_2}-\frac{N_c^2\theta(x_2)}{(x_2)_+}
+\frac{1}{(x_2)_p}+\frac{x_2^*\theta(x_2^*-x_2)}{(x_2^2)_p}
\ln\frac{x_2^*-x_2}{x_2^*}\Big]\Big\}\no
&+I\langle L\rangle\Big|_{r}^{SGP}
+I\langle L\rangle\Big|_{r}^{HP-1}
+I\langle L\rangle\Big|_{r}^{HP-2}
+I\langle L\rangle\Big|_{r}^{HP-3}
+I\langle L\rangle\Big|_{r}^{SFP}.
\end{align}
The last line is the reminder part of real corrections after subtraction. Their
explicit expressions are given below.

SGP part is:
\begin{align}
I\langle L\rangle |_{r}^{SGP}=&
l_{\perp,cs}\cdot\tilde{s}_\perp \cos\theta
\frac{\al_s}{\pi N_c^2}\int_r
\bar{q}(x_b)T_F(x_2,x_2)
\Big[g_1^{(a)}\de(1-\hat{x}_b)+g_2^{(a)}\de(1-\hat{x}_2)
+\frac{g_3^{(a)}}{(1-\hat{x}_2)_+(1-\hat{x}_b)_+}\Big],\no
g_1^{(a)}=& -\frac{1}{2}\Big[L_2(1+\hat{x}_2^2)-\hat{x}_2+\hat{x}_2^2\Big],\no
g_2^{(a)}=& -\frac{1}{2}\Big[L_1\hat{x}_b
(1+\hat{x}_b^2)+\hat{x}_b-1\Big],\no
g_3^{(a)}=& \frac{\hat{x}_2+\hat{z}-1}{2Q^2 \hat{x}_2 \hat{z}^2}
\Big[Q \hat{x}_2^3
   \left(\hat{z}-1\right)+
\hat{x}_2^2 \left(\hat{z}
   \left(E_t-Q\right)+Q\right)+\hat{x}_2
   \left(\hat{z}-1\right) \hat{z} \left(E_t+Q
   \hat{z}\right)+Q \hat{z}^2
   \left(\hat{z}^2-\hat{z}+1\right)
\Big].
\end{align}
There are three parts of hard pole contributions, which correspond to different
$T_F$. The first part is:
\begin{align}
I\langle L\rangle |_{r}^{HP-1}=&
l_{\perp,cs}\cdot\tilde{s}_\perp \cos\theta
\frac{\al_s}{\pi N_c^2}\int_r
\bar{q}(x_b)T_F(x_1^*,x_2)
\Big[g_1^{(b)}\de(1-\hat{x}_b)+g_2^{(b)}\de(1-\hat{x}_2)
+\frac{g_3^{(b)}}{(1-\hat{x}_2)_+(1-\hat{x}_b)_+}\Big],\no
g_1^{(b)}=& \frac{N_c^2}{2}\Big[L_2 (1+\hat{x}_2)-1\Big],\no
g_2^{(b)}=& \frac{\hat{x}_b+N_c^2-1}{2}\Big[L_1(1+\hat{x}_b^2)+2\Big],\no
g_3^{(b)}=& \frac{N_c^2+\hat{z}-1}{2\hat{x}_2 E_t}
\Big[\hat{x}_2 \hat{z}
   E_t+\left(\hat{z}-1\right) \hat{z} E_t+Q
   \hat{x}_2^2\Big];
\end{align}
The second part is:
\begin{align}
I\langle L\rangle |_{r}^{HP-2}=&
l_{\perp,cs}\cdot\tilde{s}_\perp \cos\theta
\frac{\al_s}{\pi N_c^2}\int_r
\bar{q}(x_b)T_F(x_1^*,x_1^*-x_2)
\Big[g_1^{(c)}\de(1-\hat{x}_b)
+\frac{g_3^{(c)}}{(1-\hat{x}_2)_+(1-\hat{x}_b)_+}\Big],\no
g_1^{(c)}=& \frac{1}{2}\Big[\ln(1-\hat{x}_2)-1\Big],\no
g_3^{(c)}=& -\frac{Q^3\hat{x}_2(\hat{x}_2-1)(\hat{z}-1)
(2N_c(\hat{z}-1)+\hat{z})}{2E_t^3(\hat{x}_2+\hat{z}-1)}
+\frac{(\hat{x}_2-1)(\hat{x}_2+\hat{z}-1)}{2\hat{x}_2\hat{z}}
\Big[N_c(\hat{z}-1)(\hat{z}-2)-\hat{z}\Big].
\end{align}
The third part is:
\begin{align}
I\langle L\rangle |_{r}^{HP-3}=&
l_{\perp,cs}\cdot\tilde{s}_\perp \cos\theta
\frac{\al_s}{\pi N_c^2}\int_r
\bar{q}(x_b)T_F(x_2-x_1^*,-x_1^*)
\frac{g_3^{(d)}}{(1-\hat{x}_2)_+(1-\hat{x}_b)_+},\no
g_3^{(d)}=&-\frac{(\hat{x}_2-1)(\hat{z}-1)(\hat{x}_2+\hat{z}-1)}{2\hat{x}_2\hat{z}}
\Big[N_c(\hat{z}-2)-\hat{z}+\hat{z}^2\Big]
+\frac{Q^3\hat{x}_2(\hat{x}_2-1)(\hat{z}-1)^2(2N_c+\hat{z})}{2E_t^3(\hat{x}_2
+\hat{z}-1)}.
\end{align}
The last part, SFP contribution, is:
\begin{align}
I\langle L\rangle |_{r}^{SFP}=&
l_{\perp,cs}\cdot\tilde{s}_\perp \cos\theta
\frac{\al_s}{\pi N_c^2}\int_r
\bar{q}(x_b)T_F(0,-x_2)
\frac{g_3^{(e)}}{(1-\hat{x}_2)_+(1-\hat{x}_b)_+},\no
g_3^{(e)}=&-\frac{(\hat{x}_2-1)(\hat{z}-1)(\hat{x}_2+\hat{z}-1)}{2Q \hat{x}_2
\hat{z}}\Big[\hat{x}_2 E_t +2 Q\hat{x}_2^2-Q\hat{z}^2\Big].
\end{align}
Note that all explicit $\ln\mu$ dependence is provided by subtraction in our formalism. With tree level result taken into account, our result is $\mu$ independent to $O(\al_s)$.

\section{Summary}
In this work, we have calculated the one-loop correction to unweighted SSA in
DY for the lepton angular distribution. We choose to work in Feynman gauge.
As a comparison, we also calculate
a weighted observable $I\langle P_7\rangle$ with the same method. We use
at most one $G^+$ to do collinear expansion, and then use EOM for fermion
to eliminate the bad component of fermion field. After this, the hadronic
tensor or cross section is expressed by independent twist-3 distribution
functions. Three of these functions are $q_\partial$, $T_F$ and $T_{\Delta}$,
which are gauge invariant. The other two distribution functions are not gauge
invariant and their hard coefficients are found to be zero.
To calculate virtual
correction, we first illustrate that the hard part is analytic on the upper
half planes of $s_0,s_1$. After this we use FIRE to reduce the
various tensor integrals to standard scalar integrals.
Due to the simple kinematics, the reduced scalar integrals are three bubble
integrals $B(s_0)$, $B(s_1)$ and $B(s_2)$. According to the analyticity of
the amplitude on $s_0,s_1$, we successfully extract
pole part and non-pole part of virtual corrections.
Both are divergent, and for the pole part only soft gluon pole contributes.
Virtual corrections to $W^{\mu\nu}$ can be written into a form consistent
with QED gauge invariance.
For real corrections, only three kinds of
pole contributions are possible, that is, SGP, HP and SFP contributions.
There is no contribution from two-point twist-3 distribution functions.
Then, we perform collinear subtraction.
As found in \cite{Rein:2025a,Rein:2025b}, the divergence of nonpole part of virtual
correction actually
can be written into the same form as hard pole contribution of real corrections.
In the sum of virtual and real corrections, all divergences can be subtracted
out by the one-loop corrections to twist-2 and twist-3 distribution functions.
Thus, the twist-3 factorization still holds for polarized DY at one-loop level.
We further give all subtracted finite hard coefficients. Only the coefficients
of the convolution $\bar{q}\otimes T_F$ are considered in this work. In future,
we will also extend the calculation to include chiral-odd and pure gluon
twist-3 distribution functions. Moreover, some NLO corrections to SSA in
lepton-hadron scattering are worked out in \cite{Rein:2025a,Rein:2025b}.
The factorization formula there has a more complicated form, e.g., the derivative
of $T_F(x,x)$ appears. We will consider this process in future to check
our formalism further.

\section*{Acknowledgements}
The author would like to thank Prof.~Jian-Ping~Ma for very helpful discussions.
The hospitality of ITP, CAS(Beijing) during the finalization of this paper is appreciated.

\appendix
\section{Counter terms of QCD}\label{sec:cts}
The counter terms we need in this work are
\begin{align}
\mathcal{L}_{QCD}\supset \de z_2 \bar{\psi}i\s{\partial}\psi
-\de z_{1F}\bar{\psi}g_s\s{G}\psi-\de z_1^\ga \bar{\psi}e\s{A}\psi.
\end{align}
All fields and coupling constants are renormalized ones. $G^\mu=G_a^\mu T^a$
is gluon field, and $A^\mu$ is photon field.
The relation between bare and renormalized quantities is
\begin{align}
\psi_B=Z_2^{1/2}\psi,\ G_{B,a}^\mu=Z_3^{1/2}G_a^\mu,\ g_s^B=\frac{Z_{1F}}{Z_2 Z_3^{1/2}}g_s.
\end{align}
In $\overline{MS}$ scheme, the values of these constants are well known, and can be
found in e.g.,\cite{Muta:1998vi}. For convenience, we list the values in Feynman gauge as follows:
\begin{align}
\de z_3=&Z_3-1=-\frac{g_s^2}{16\pi^2}R_\ep (\frac{2}{3}N_F-\frac{5}{3}C_A),\no
\de z_2=&Z_2-1=-\frac{g_s^2}{16\pi^2}R_\ep C_F,\no
\de z_{1F}=&Z_{1F}-1=-\frac{g_s^2}{16\pi^2}R_\ep (C_A+C_F),\no
\de z_{1}^{\ga}=&Z_{1}^{\ga}-1=-\frac{g_s^2}{16\pi^2}R_\ep C_F,
\end{align}
with $R_\ep=\frac{2}{\ep}-\ga_E+\ln 4\pi$. $N_F$ is the number of quark flavor. Note that $Z_1^\ga=Z_2$.

\section{Calculation of Fig.\ref{fig:hp1}(c) with transverse gluon}\label{sec:hp-c}
For hard pole, $k^+\neq 0$, we can use $G_\perp$ to calculate directly. With $G_\perp$ as the initial gluon, one of the hard parts 
of Fig.\ref{fig:hp1}(c) is
\begin{align}
H^{\mu\nu}_\rho=&\int\frac{d^n k_g}{(2\pi)^n}2\pi\de_+(k_g^2) \de^n(k_2+k_b-q-k_g)\no
&Tr[\ga^+(-ig_sT^b\ga^\tau)(\s{k}_1-\s{q})\ga^\mu
\ga^- T^a \ga^\nu \frac{-i}{\s{k}_2-\s{q}-i\ep}(ig_sT^c\ga_{\la'})]P^{\la\la'}(k_g)
(-g_s)f^{abc}\Gamma_{\rho\tau\la}(k,k_g-k,-k_g)\no
&\frac{-i}{(k_g-k)^2+i\ep}\pi\de((k_1-q)^2).
\end{align}
Then, we integrate out $k_g$ using the delta function for momentum conservation,
and get
\begin{align}
H^{\mu\nu}_\rho=&\frac{\pi}{(2\pi)^{n-1}}\de_+((k_2+k_b-q)^2)\no
&Tr[\ga^+(-ig_sT^b\ga^\tau)(\s{k}_1-\s{q})\ga^\mu
\ga^- T^a \ga^\nu \frac{-i}{\s{k}_2-\s{q}-i\ep}(ig_sT^c\ga_{\la'})]P^{\la\la'}(k_g)
(-g_s)f^{abc}\Gamma_{\rho\tau\la}(k,k_g-k,-k_g)\no
&\frac{-i}{(k_g-k)^2+i\ep}\de((k_1-q)^2)\Big|_{k_g\rightarrow k_2+k_b-q}.
\end{align}
For $G_\perp$ expansion, partons have no transverse momentum, so,
\begin{align}
k_2^\mu=x_2 p_a^\mu,\ k_1^\mu=x_1 p_a^\mu,\ k^\mu=x p_a^\mu,\ k_b^\mu=x_b p_b^\mu.
\end{align}
So, in $H^{\mu\nu}_\rho$ there is only one transverse momentum $q_\perp^\mu$.
Then,
\begin{align}
I\langle L\rangle =\int d^n q \de(q^2-Q^2)\int dk_b^- \bar{q}(x_b)
\int dk_2^+ dk_1^+
M^{(1)}_{\ga^+,\partial^+ G_\perp} \tilde{s}_\perp^\rho
\frac{i}{k^+}L_{\mu\nu}H^{\mu\nu}_{\rho}
\end{align}
Then, we integrate out $k_1^+$ with the pole condition $\de((k_1-q)^2)$, which
gives $x_1=x_1^*=Q^2/(2p_a\cdot q)$. In $L_{\mu\nu}$, $l^\mu$ is given by eq.(\ref{eq:lepton-momentum}).
This introduces another transverse momentum $l_{\perp,cs}^\mu$. In the simplifying
of the result, we use following replacement
\begin{align}
l_{\perp,cs}\cdot q_\perp q_\perp^\rho\rightarrow
\frac{l_{\perp,cs}^\rho q_\perp^2}{n-2}.
\end{align}
This is due to the symmetry of $q_\perp$ integration. After this, we get a
result with integrand depending on $q_\perp^2$ only, that is,
\begin{align}
I\langle L\rangle =& \tilde{s}_{\perp\rho}
\int d^n q \de(q^2-Q^2)\int dk_b^- \bar{q}(x_b)
\int dk_2^+ M^{(1)}_{\ga^+,\partial^+ G_\perp} l_{\perp,cs}^\rho
\bar{H}(q_\perp^2,p_a\cdot q,p_b\cdot q)\de((k_2+k_b-q)^2).
\end{align}
Then, we integrate out $q^-$ using $\de(q^2-Q^2)$ and then let $q^+=\xi p_a^+$.
$\de((k_2+k_b-q)^2)$ can be simplified as
\begin{align}
\de((k_2+k_b-q)^2)=\hat{x}_2 \de(\frac{Q^2}{\hat{x}_b}(1-\hat{x}_2)(1-\hat{x}_b)
-q_t^2).
\end{align}
This enables us to integrate out $q_\perp$ and get
\begin{align}
I\langle L\rangle
=& \tilde{s}_{\perp\rho}
\int \frac{d\xi}{2\xi} \int dk_b^- \bar{q}(x_b)
\int dk_2^+ M^{(1)}_{\ga^+,\partial^+ G_\perp} l_{\perp,cs}^\rho
\int d^{n-2}q_\perp \bar{H}(q_\perp^2,p_a\cdot q,p_b\cdot q)\de((k_2+k_b-q)^2)\no
=&\tilde{s}_{\perp\rho}
\int \frac{d\xi}{2\xi} \int dk_b^- \bar{q}(x_b)
\int dk_2^+ M^{(1)}_{\ga^+,\partial^+ G_\perp} l_{\perp,cs}^\rho
\frac{\Omega_{n-2}}{2}(q_t^2)^{-\ep/2} \bar{H}(q_\perp^2,p_a\cdot q,p_b\cdot q)\hat{x}_2.
\end{align}
Note that in $H^{\mu\nu}_\rho$, all
Lorentz vectors are defined in n-dim space, including $\rho$. $L_{\mu\nu}$ is
also defined in n-dim space. So, $l_{\perp,cs}^\rho$ is in n-dim space. After
contracting with $\tilde{s}_\perp$, it falls into 4-dim space. Anyhow, this
does not affect the calculation of $\bar{H}$. With some simplifications, we
have
\begin{align}
I\langle L\rangle
=&\tilde{s}_\perp\cdot l_{\perp,cs}\frac{\pi\Omega_{n-2}}{4(2\pi)^{n-1}}
\int \frac{d\xi}{\xi} \int \frac{x_b}{x_b}\bar{q}(x_b)
\int \frac{dx_2}{x_2} M^{(1)}_{\ga^+,\partial^+ G_\perp}
(q_t^2)^{-\ep/2}[\cdots],
\end{align}
with
\begin{align}
[\cdots]=& \frac{g_s^2 C_A}{16 N_c}
\frac{32\cos\theta \hat{x}_b}{(2-\ep)Q^2(1-\hat{x}_2)(1-\hat{x}_2)}
\Big[
(1-\ep)E_t Q\hat{x}_2 +Q^2[1+\hat{x}_2(\hat{x}_b-1)][
1+(2-\ep)\hat{x}_2(\hat{x}_b-1)]
\Big].
\end{align}
A nontrivial feature is the hard part is automatically proportional to $\cos\theta$. Writing the expression into standard form, we have
\begin{align}
I\langle L\rangle
=&\tilde{s}_\perp\cdot l_{\perp,cs}\cos\theta\frac{\al_s}{128\pi N_c^2}\bar{A}_\ep
\int_r \bar{q}(x_b)T_F(x_1^*,x_2)
\Big\{
C_A N_c \frac{64\hat{x}_b }{(2-\ep)Q^2(1-\hat{x}_2)(1-\hat{x}_b)}
\Big(\frac{Q^2}{q_t^2}\Big)^{\ep/2}[\cdots]
\Big\}.
\end{align}
Since there are singularities at $\hat{x}_2=1$ and $\hat{x}_b=1$, the expansion in
$\ep$ is realized by
\begin{align}
\frac{1}{(1-x)^{1+\ep/2}}=-\frac{2}{\ep}\de(1-x)
+\frac{1}{(1-x)_+}-\frac{\ep}{2}\Big(\frac{\ln(1-x)}{1-x}\Big)_+.
\end{align}
Then, we get
\begin{align}
\Big\{\cdots\Big\}
=&
C_A N_c \frac{64}{2-\ep}\frac{4}{\ep^2}\de(1-\hat{x}_2)\de(1-\hat{x}_b)[(1-\ep)+1]
+\cdots\no
=&C_A N_c\frac{256}{\ep^2}\de(1-\hat{x}_2)\de(1-\hat{x}_b)+\cdots.
\end{align}
$\cdots$ represents other terms which do not contain two delta functions.
This is the result given in eq.(\ref{eq:hp1}).
As can be seen, $(2-\ep)$ in denominator
is cancelled by the same factor in numerator. So, only $1/\ep^2$ is left.

\section{$\ga_5$ convention}\label{sec:ga5}
We use HVBM scheme, where $\ga_5$ is defined by $\ga_5=i\ga^0\ga^1\ga^2\ga^3$. In the trace it gives
$Tr(\ga_5\ga^\mu\ga^\nu\ga^\rho\ga^\tau)=-4i\ep^{\mu\nu\rho\tau}$, with $\ep^{0123}=+1$.
At the same time, spin vectors $s_\perp^\mu$ and $\tilde{s}_\perp^\mu$ are defined
as 4-dim vectors.
Following identity holds if all matrices are defined in 4-dim spacetime,
\begin{align}
\ga^\mu\ga^\rho \ga^\nu =& S^{\mu\rho\nu\sig}\ga_\sig-i\ep^{\mu\rho\nu\sig}
\ga_5\ga_\sig,\no
S^{\mu\rho\nu\sig}=&g^{\mu\rho}g^{\nu\sig}+g^{\mu\sig}g^{\nu\rho}
-g^{\mu\nu}g^{\rho\sig}.
\end{align}
With this identity, it is obvious that
\begin{align}
\ga^+\s{s}_\perp(i\ga_5\ga^-)=-i\ga_5 \s{s}_\perp -\s{\tilde{s}}_\perp,\
\ga^+\s{\tilde{s}}_\perp\ga^-=-\s{\tilde{s}}_\perp-i\ga_5\s{s}_\perp.
\end{align}
Note that $\tilde{s}_\perp^\mu =\ep_\perp^{\mu\nu}s_{\perp\nu}$,
$s_\perp^\mu=-\ep_\perp^{\mu\nu}\tilde{s}_{\perp\nu}$. This scheme is crucial
for getting correct virtual corrections.

\bibliography{ref2}

\begin{thebibliography}{44}%
\makeatletter
\providecommand \@ifxundefined [1]{%
 \@ifx{#1\undefined}
}%
\providecommand \@ifnum [1]{%
 \ifnum #1\expandafter \@firstoftwo
 \else \expandafter \@secondoftwo
 \fi
}%
\providecommand \@ifx [1]{%
 \ifx #1\expandafter \@firstoftwo
 \else \expandafter \@secondoftwo
 \fi
}%
\providecommand \natexlab [1]{#1}%
\providecommand \enquote  [1]{``#1''}%
\providecommand \bibnamefont  [1]{#1}%
\providecommand \bibfnamefont [1]{#1}%
\providecommand \citenamefont [1]{#1}%
\providecommand \href@noop [0]{\@secondoftwo}%
\providecommand \href [0]{\begingroup \@sanitize@url \@href}%
\providecommand \@href[1]{\@@startlink{#1}\@@href}%
\providecommand \@@href[1]{\endgroup#1\@@endlink}%
\providecommand \@sanitize@url [0]{\catcode `\\12\catcode `\$12\catcode
  `\&12\catcode `\#12\catcode `\^12\catcode `\_12\catcode `\%12\relax}%
\providecommand \@@startlink[1]{}%
\providecommand \@@endlink[0]{}%
\providecommand \url  [0]{\begingroup\@sanitize@url \@url }%
\providecommand \@url [1]{\endgroup\@href {#1}{\urlprefix }}%
\providecommand \urlprefix  [0]{URL }%
\providecommand \Eprint [0]{\href }%
\providecommand \doibase [0]{http://dx.doi.org/}%
\providecommand \selectlanguage [0]{\@gobble}%
\providecommand \bibinfo  [0]{\@secondoftwo}%
\providecommand \bibfield  [0]{\@secondoftwo}%
\providecommand \translation [1]{[#1]}%
\providecommand \BibitemOpen [0]{}%
\providecommand \bibitemStop [0]{}%
\providecommand \bibitemNoStop [0]{.\EOS\space}%
\providecommand \EOS [0]{\spacefactor3000\relax}%
\providecommand \BibitemShut  [1]{\csname bibitem#1\endcsname}%
\let\auto@bib@innerbib\@empty
\bibitem [{\citenamefont {Efremov}\ and\ \citenamefont
  {Teryaev}(1982)}]{Efremov:1981sh}%
  \BibitemOpen
  \bibfield  {author} {\bibinfo {author} {\bibfnamefont {A.}~\bibnamefont
  {Efremov}}\ and\ \bibinfo {author} {\bibfnamefont {O.}~\bibnamefont
  {Teryaev}},\ }\href@noop {} {\bibfield  {journal} {\bibinfo  {journal} {Sov.
  J. Nucl. Phys.}\ }\textbf {\bibinfo {volume} {36}},\ \bibinfo {pages} {140}
  (\bibinfo {year} {1982})}\BibitemShut {NoStop}%
\bibitem [{\citenamefont {Efremov}\ and\ \citenamefont
  {Teryaev}(1985)}]{Efremov:1984ip}%
  \BibitemOpen
  \bibfield  {author} {\bibinfo {author} {\bibfnamefont {A.}~\bibnamefont
  {Efremov}}\ and\ \bibinfo {author} {\bibfnamefont {O.}~\bibnamefont
  {Teryaev}},\ }\href {\doibase 10.1016/0370-2693(85)90999-2} {\bibfield
  {journal} {\bibinfo  {journal} {Phys. Lett. B}\ }\textbf {\bibinfo {volume}
  {150}},\ \bibinfo {pages} {383} (\bibinfo {year} {1985})}\BibitemShut
  {NoStop}%
\bibitem [{\citenamefont {Qiu}\ and\ \citenamefont
  {Sterman}(1991)}]{Qiu:1991pp}%
  \BibitemOpen
  \bibfield  {author} {\bibinfo {author} {\bibfnamefont {J.-w.}\ \bibnamefont
  {Qiu}}\ and\ \bibinfo {author} {\bibfnamefont {G.~F.}\ \bibnamefont
  {Sterman}},\ }\href {\doibase 10.1103/PhysRevLett.67.2264} {\bibfield
  {journal} {\bibinfo  {journal} {Phys. Rev. Lett.}\ }\textbf {\bibinfo
  {volume} {67}},\ \bibinfo {pages} {2264} (\bibinfo {year}
  {1991})}\BibitemShut {NoStop}%
\bibitem [{\citenamefont {Qiu}\ and\ \citenamefont
  {Sterman}(1999)}]{Qiu:1998ia}%
  \BibitemOpen
  \bibfield  {author} {\bibinfo {author} {\bibfnamefont {J.-w.}\ \bibnamefont
  {Qiu}}\ and\ \bibinfo {author} {\bibfnamefont {G.~F.}\ \bibnamefont
  {Sterman}},\ }\href {\doibase 10.1103/PhysRevD.59.014004} {\bibfield
  {journal} {\bibinfo  {journal} {Phys. Rev. D}\ }\textbf {\bibinfo {volume}
  {59}},\ \bibinfo {pages} {014004} (\bibinfo {year} {1999})},\ \Eprint
  {http://arxiv.org/abs/hep-ph/9806356} {arXiv:hep-ph/9806356} \BibitemShut
  {NoStop}%
\bibitem [{\citenamefont {Qiu}\ and\ \citenamefont
  {Sterman}(1992)}]{Qiu:1991wg}%
  \BibitemOpen
  \bibfield  {author} {\bibinfo {author} {\bibfnamefont {J.-w.}\ \bibnamefont
  {Qiu}}\ and\ \bibinfo {author} {\bibfnamefont {G.~F.}\ \bibnamefont
  {Sterman}},\ }\href {\doibase 10.1016/0550-3213(92)90003-T} {\bibfield
  {journal} {\bibinfo  {journal} {Nucl. Phys. B}\ }\textbf {\bibinfo {volume}
  {378}},\ \bibinfo {pages} {52} (\bibinfo {year} {1992})}\BibitemShut
  {NoStop}%
\bibitem [{\citenamefont {Eguchi}\ \emph {et~al.}(2006)\citenamefont {Eguchi},
  \citenamefont {Koike},\ and\ \citenamefont {Tanaka}}]{Eguchi:2006qz}%
  \BibitemOpen
  \bibfield  {author} {\bibinfo {author} {\bibfnamefont {H.}~\bibnamefont
  {Eguchi}}, \bibinfo {author} {\bibfnamefont {Y.}~\bibnamefont {Koike}}, \
  and\ \bibinfo {author} {\bibfnamefont {K.}~\bibnamefont {Tanaka}},\ }\href
  {\doibase 10.1016/j.nuclphysb.2006.05.036} {\bibfield  {journal} {\bibinfo
  {journal} {Nucl. Phys. B}\ }\textbf {\bibinfo {volume} {752}},\ \bibinfo
  {pages} {1} (\bibinfo {year} {2006})},\ \Eprint
  {http://arxiv.org/abs/hep-ph/0604003} {arXiv:hep-ph/0604003} \BibitemShut
  {NoStop}%
\bibitem [{\citenamefont {Eguchi}\ \emph {et~al.}(2007)\citenamefont {Eguchi},
  \citenamefont {Koike},\ and\ \citenamefont {Tanaka}}]{Eguchi:2006mc}%
  \BibitemOpen
  \bibfield  {author} {\bibinfo {author} {\bibfnamefont {H.}~\bibnamefont
  {Eguchi}}, \bibinfo {author} {\bibfnamefont {Y.}~\bibnamefont {Koike}}, \
  and\ \bibinfo {author} {\bibfnamefont {K.}~\bibnamefont {Tanaka}},\ }\href
  {\doibase 10.1016/j.nuclphysb.2006.11.016} {\bibfield  {journal} {\bibinfo
  {journal} {Nucl. Phys. B}\ }\textbf {\bibinfo {volume} {763}},\ \bibinfo
  {pages} {198} (\bibinfo {year} {2007})},\ \Eprint
  {http://arxiv.org/abs/hep-ph/0610314} {arXiv:hep-ph/0610314} \BibitemShut
  {NoStop}%
\bibitem [{\citenamefont {Ji}\ \emph {et~al.}(2006{\natexlab{a}})\citenamefont
  {Ji}, \citenamefont {Qiu}, \citenamefont {Vogelsang},\ and\ \citenamefont
  {Yuan}}]{Ji:2006br}%
  \BibitemOpen
  \bibfield  {author} {\bibinfo {author} {\bibfnamefont {X.}~\bibnamefont
  {Ji}}, \bibinfo {author} {\bibfnamefont {J.-W.}\ \bibnamefont {Qiu}},
  \bibinfo {author} {\bibfnamefont {W.}~\bibnamefont {Vogelsang}}, \ and\
  \bibinfo {author} {\bibfnamefont {F.}~\bibnamefont {Yuan}},\ }\href {\doibase
  10.1016/j.physletb.2006.05.044} {\bibfield  {journal} {\bibinfo  {journal}
  {Phys. Lett. B}\ }\textbf {\bibinfo {volume} {638}},\ \bibinfo {pages} {178}
  (\bibinfo {year} {2006}{\natexlab{a}})},\ \Eprint
  {http://arxiv.org/abs/hep-ph/0604128} {arXiv:hep-ph/0604128} \BibitemShut
  {NoStop}%
\bibitem [{\citenamefont {Ji}\ \emph {et~al.}(2006{\natexlab{b}})\citenamefont
  {Ji}, \citenamefont {Qiu}, \citenamefont {Vogelsang},\ and\ \citenamefont
  {Yuan}}]{Ji:2006ub}%
  \BibitemOpen
  \bibfield  {author} {\bibinfo {author} {\bibfnamefont {X.}~\bibnamefont
  {Ji}}, \bibinfo {author} {\bibfnamefont {J.-W.}\ \bibnamefont {Qiu}},
  \bibinfo {author} {\bibfnamefont {W.}~\bibnamefont {Vogelsang}}, \ and\
  \bibinfo {author} {\bibfnamefont {F.}~\bibnamefont {Yuan}},\ }\href {\doibase
  10.1103/PhysRevLett.97.082002} {\bibfield  {journal} {\bibinfo  {journal}
  {Phys. Rev. Lett.}\ }\textbf {\bibinfo {volume} {97}},\ \bibinfo {pages}
  {082002} (\bibinfo {year} {2006}{\natexlab{b}})},\ \Eprint
  {http://arxiv.org/abs/hep-ph/0602239} {arXiv:hep-ph/0602239} \BibitemShut
  {NoStop}%
\bibitem [{\citenamefont {Ji}\ \emph {et~al.}(2006{\natexlab{c}})\citenamefont
  {Ji}, \citenamefont {Qiu}, \citenamefont {Vogelsang},\ and\ \citenamefont
  {Yuan}}]{Ji:2006vf}%
  \BibitemOpen
  \bibfield  {author} {\bibinfo {author} {\bibfnamefont {X.}~\bibnamefont
  {Ji}}, \bibinfo {author} {\bibfnamefont {J.-W.}\ \bibnamefont {Qiu}},
  \bibinfo {author} {\bibfnamefont {W.}~\bibnamefont {Vogelsang}}, \ and\
  \bibinfo {author} {\bibfnamefont {F.}~\bibnamefont {Yuan}},\ }\href {\doibase
  10.1103/PhysRevD.73.094017} {\bibfield  {journal} {\bibinfo  {journal} {Phys.
  Rev. D}\ }\textbf {\bibinfo {volume} {73}},\ \bibinfo {pages} {094017}
  (\bibinfo {year} {2006}{\natexlab{c}})},\ \Eprint
  {http://arxiv.org/abs/hep-ph/0604023} {arXiv:hep-ph/0604023} \BibitemShut
  {NoStop}%
\bibitem [{\citenamefont {Vogelsang}\ and\ \citenamefont
  {Yuan}(2009)}]{Vogelsang:2009pj}%
  \BibitemOpen
  \bibfield  {author} {\bibinfo {author} {\bibfnamefont {W.}~\bibnamefont
  {Vogelsang}}\ and\ \bibinfo {author} {\bibfnamefont {F.}~\bibnamefont
  {Yuan}},\ }\href {\doibase 10.1103/PhysRevD.79.094010} {\bibfield  {journal}
  {\bibinfo  {journal} {Phys. Rev. D}\ }\textbf {\bibinfo {volume} {79}},\
  \bibinfo {pages} {094010} (\bibinfo {year} {2009})},\ \Eprint
  {http://arxiv.org/abs/0904.0410} {arXiv:0904.0410 [hep-ph]} \BibitemShut
  {NoStop}%
\bibitem [{\citenamefont {Kang}\ \emph {et~al.}(2013)\citenamefont {Kang},
  \citenamefont {Vitev},\ and\ \citenamefont {Xing}}]{Kang:2012ns}%
  \BibitemOpen
  \bibfield  {author} {\bibinfo {author} {\bibfnamefont {Z.-B.}\ \bibnamefont
  {Kang}}, \bibinfo {author} {\bibfnamefont {I.}~\bibnamefont {Vitev}}, \ and\
  \bibinfo {author} {\bibfnamefont {H.}~\bibnamefont {Xing}},\ }\href {\doibase
  10.1103/PhysRevD.87.034024} {\bibfield  {journal} {\bibinfo  {journal} {Phys.
  Rev. D}\ }\textbf {\bibinfo {volume} {87}},\ \bibinfo {pages} {034024}
  (\bibinfo {year} {2013})},\ \Eprint {http://arxiv.org/abs/1212.1221}
  {arXiv:1212.1221 [hep-ph]} \BibitemShut {NoStop}%
\bibitem [{\citenamefont {Dai}\ \emph {et~al.}(2015)\citenamefont {Dai},
  \citenamefont {Kang}, \citenamefont {Prokudin},\ and\ \citenamefont
  {Vitev}}]{Dai:2014ala}%
  \BibitemOpen
  \bibfield  {author} {\bibinfo {author} {\bibfnamefont {L.-Y.}\ \bibnamefont
  {Dai}}, \bibinfo {author} {\bibfnamefont {Z.-B.}\ \bibnamefont {Kang}},
  \bibinfo {author} {\bibfnamefont {A.}~\bibnamefont {Prokudin}}, \ and\
  \bibinfo {author} {\bibfnamefont {I.}~\bibnamefont {Vitev}},\ }\href
  {\doibase 10.1103/PhysRevD.92.114024} {\bibfield  {journal} {\bibinfo
  {journal} {Phys. Rev. D}\ }\textbf {\bibinfo {volume} {92}},\ \bibinfo
  {pages} {114024} (\bibinfo {year} {2015})},\ \Eprint
  {http://arxiv.org/abs/1409.5851} {arXiv:1409.5851 [hep-ph]} \BibitemShut
  {NoStop}%
\bibitem [{\citenamefont {Yoshida}(2016)}]{Yoshida:2016tfh}%
  \BibitemOpen
  \bibfield  {author} {\bibinfo {author} {\bibfnamefont {S.}~\bibnamefont
  {Yoshida}},\ }\href {\doibase 10.1103/PhysRevD.93.054048} {\bibfield
  {journal} {\bibinfo  {journal} {Phys. Rev. D}\ }\textbf {\bibinfo {volume}
  {93}},\ \bibinfo {pages} {054048} (\bibinfo {year} {2016})},\ \Eprint
  {http://arxiv.org/abs/1601.07737} {arXiv:1601.07737 [hep-ph]} \BibitemShut
  {NoStop}%
\bibitem [{\citenamefont {Chen}\ \emph {et~al.}(2017)\citenamefont {Chen},
  \citenamefont {Ma},\ and\ \citenamefont {Zhang}}]{Chen:2016dnp}%
  \BibitemOpen
  \bibfield  {author} {\bibinfo {author} {\bibfnamefont {A.~P.}\ \bibnamefont
  {Chen}}, \bibinfo {author} {\bibfnamefont {J.~P.}\ \bibnamefont {Ma}}, \ and\
  \bibinfo {author} {\bibfnamefont {G.~P.}\ \bibnamefont {Zhang}},\ }\href
  {\doibase 10.1103/PhysRevD.95.074005} {\bibfield  {journal} {\bibinfo
  {journal} {Phys. Rev. D}\ }\textbf {\bibinfo {volume} {95}},\ \bibinfo
  {pages} {074005} (\bibinfo {year} {2017})},\ \Eprint
  {http://arxiv.org/abs/1607.08676} {arXiv:1607.08676 [hep-ph]} \BibitemShut
  {NoStop}%
\bibitem [{\citenamefont {Chen}\ \emph {et~al.}(2018)\citenamefont {Chen},
  \citenamefont {Ma},\ and\ \citenamefont {Zhang}}]{Chen:2017lvx}%
  \BibitemOpen
  \bibfield  {author} {\bibinfo {author} {\bibfnamefont {A.~P.}\ \bibnamefont
  {Chen}}, \bibinfo {author} {\bibfnamefont {J.~P.}\ \bibnamefont {Ma}}, \ and\
  \bibinfo {author} {\bibfnamefont {G.~P.}\ \bibnamefont {Zhang}},\ }\href
  {\doibase 10.1103/PhysRevD.97.054003} {\bibfield  {journal} {\bibinfo
  {journal} {Phys. Rev. D}\ }\textbf {\bibinfo {volume} {97}},\ \bibinfo
  {pages} {054003} (\bibinfo {year} {2018})},\ \Eprint
  {http://arxiv.org/abs/1708.09091} {arXiv:1708.09091 [hep-ph]} \BibitemShut
  {NoStop}%
\bibitem [{\citenamefont {Ma}\ and\ \citenamefont {Zhang}(2015)}]{Ma:2014uma}%
  \BibitemOpen
  \bibfield  {author} {\bibinfo {author} {\bibfnamefont {J.~P.}\ \bibnamefont
  {Ma}}\ and\ \bibinfo {author} {\bibfnamefont {G.~P.}\ \bibnamefont {Zhang}},\
  }\href {\doibase 10.1007/JHEP02(2015)163} {\bibfield  {journal} {\bibinfo
  {journal} {JHEP}\ }\textbf {\bibinfo {volume} {02}},\ \bibinfo {pages} {163}
  (\bibinfo {year} {2015})},\ \Eprint {http://arxiv.org/abs/1409.2938}
  {arXiv:1409.2938 [hep-ph]} \BibitemShut {NoStop}%
\bibitem [{\citenamefont {Alexeev}\ \emph {et~al.}(2024)\citenamefont {Alexeev}
  \emph {et~al.}}]{COMPASS:2023vqt}%
  \BibitemOpen
  \bibfield  {author} {\bibinfo {author} {\bibfnamefont {G.~D.}\ \bibnamefont
  {Alexeev}} \emph {et~al.} (\bibinfo {collaboration} {COMPASS}),\ }\href
  {\doibase 10.1103/PhysRevLett.133.071902} {\bibfield  {journal} {\bibinfo
  {journal} {Phys. Rev. Lett.}\ }\textbf {\bibinfo {volume} {133}},\ \bibinfo
  {pages} {071902} (\bibinfo {year} {2024})},\ \Eprint
  {http://arxiv.org/abs/2312.17379} {arXiv:2312.17379 [hep-ex]} \BibitemShut
  {NoStop}%
\bibitem [{\citenamefont {Niemiec}(2025)}]{Niemiec:2024zou}%
  \BibitemOpen
  \bibfield  {author} {\bibinfo {author} {\bibfnamefont {M.}~\bibnamefont
  {Niemiec}},\ }\href {\doibase 10.22323/1.469.0229} {\bibfield  {journal}
  {\bibinfo  {journal} {PoS}\ }\textbf {\bibinfo {volume} {DIS2024}},\ \bibinfo
  {pages} {229} (\bibinfo {year} {2025})}\BibitemShut {NoStop}%
\bibitem [{\citenamefont {Rein}\ \emph
  {et~al.}(2025{\natexlab{a}})\citenamefont {Rein}, \citenamefont {Schlegel},
  \citenamefont {Tollk{\"u}hn},\ and\ \citenamefont {Vogelsang}}]{Rein:2025a}%
  \BibitemOpen
  \bibfield  {author} {\bibinfo {author} {\bibfnamefont {D.}~\bibnamefont
  {Rein}}, \bibinfo {author} {\bibfnamefont {M.}~\bibnamefont {Schlegel}},
  \bibinfo {author} {\bibfnamefont {P.}~\bibnamefont {Tollk{\"u}hn}}, \ and\
  \bibinfo {author} {\bibfnamefont {W.}~\bibnamefont {Vogelsang}},\ }\href
  {\doibase 10.1103/fhxq-njl2} {\bibfield  {journal} {\bibinfo  {journal}
  {Phys. Rev. Lett.}\ }\textbf {\bibinfo {volume} {135}},\ \bibinfo {pages}
  {251901} (\bibinfo {year} {2025}{\natexlab{a}})},\ \Eprint
  {http://arxiv.org/abs/2503.16097} {arXiv:2503.16097 [hep-ph]} \BibitemShut
  {NoStop}%
\bibitem [{\citenamefont {Rein}\ \emph
  {et~al.}(2025{\natexlab{b}})\citenamefont {Rein}, \citenamefont {Schlegel},
  \citenamefont {Tollk{\"u}hn},\ and\ \citenamefont {Vogelsang}}]{Rein:2025b}%
  \BibitemOpen
  \bibfield  {author} {\bibinfo {author} {\bibfnamefont {D.}~\bibnamefont
  {Rein}}, \bibinfo {author} {\bibfnamefont {M.}~\bibnamefont {Schlegel}},
  \bibinfo {author} {\bibfnamefont {P.}~\bibnamefont {Tollk{\"u}hn}}, \ and\
  \bibinfo {author} {\bibfnamefont {W.}~\bibnamefont {Vogelsang}},\ }\href
  {\doibase 10.1103/w5cv-ssdm} {\bibfield  {journal} {\bibinfo  {journal}
  {Phys. Rev. D}\ }\textbf {\bibinfo {volume} {112}},\ \bibinfo {pages}
  {114024} (\bibinfo {year} {2025}{\natexlab{b}})},\ \Eprint
  {http://arxiv.org/abs/2503.16119} {arXiv:2503.16119 [hep-ph]} \BibitemShut
  {NoStop}%
\bibitem [{\citenamefont {Kodaira}\ and\ \citenamefont
  {Tanaka}(1999)}]{Kodaira:1998jn}%
  \BibitemOpen
  \bibfield  {author} {\bibinfo {author} {\bibfnamefont {J.}~\bibnamefont
  {Kodaira}}\ and\ \bibinfo {author} {\bibfnamefont {K.}~\bibnamefont
  {Tanaka}},\ }\href {\doibase 10.1143/PTP.101.191} {\bibfield  {journal}
  {\bibinfo  {journal} {Prog. Theor. Phys.}\ }\textbf {\bibinfo {volume}
  {101}},\ \bibinfo {pages} {191} (\bibinfo {year} {1999})},\ \Eprint
  {http://arxiv.org/abs/hep-ph/9812449} {arXiv:hep-ph/9812449} \BibitemShut
  {NoStop}%
\bibitem [{\citenamefont {Collins}\ and\ \citenamefont
  {Soper}(1977)}]{Collins:1977iv}%
  \BibitemOpen
  \bibfield  {author} {\bibinfo {author} {\bibfnamefont {J.~C.}\ \bibnamefont
  {Collins}}\ and\ \bibinfo {author} {\bibfnamefont {D.~E.}\ \bibnamefont
  {Soper}},\ }\href {\doibase 10.1103/PhysRevD.16.2219} {\bibfield  {journal}
  {\bibinfo  {journal} {Phys. Rev. D}\ }\textbf {\bibinfo {volume} {16}},\
  \bibinfo {pages} {2219} (\bibinfo {year} {1977})}\BibitemShut {NoStop}%
\bibitem [{\citenamefont {Boer}\ \emph {et~al.}(2003)\citenamefont {Boer},
  \citenamefont {Mulders},\ and\ \citenamefont {Pijlman}}]{Boer:2003cm}%
  \BibitemOpen
  \bibfield  {author} {\bibinfo {author} {\bibfnamefont {D.}~\bibnamefont
  {Boer}}, \bibinfo {author} {\bibfnamefont {P.}~\bibnamefont {Mulders}}, \
  and\ \bibinfo {author} {\bibfnamefont {F.}~\bibnamefont {Pijlman}},\ }\href
  {\doibase 10.1016/S0550-3213(03)00527-3} {\bibfield  {journal} {\bibinfo
  {journal} {Nucl. Phys. B}\ }\textbf {\bibinfo {volume} {667}},\ \bibinfo
  {pages} {201} (\bibinfo {year} {2003})},\ \Eprint
  {http://arxiv.org/abs/hep-ph/0303034} {arXiv:hep-ph/0303034} \BibitemShut
  {NoStop}%
\bibitem [{\citenamefont {Zhou}\ \emph {et~al.}(2010)\citenamefont {Zhou},
  \citenamefont {Yuan},\ and\ \citenamefont {Liang}}]{Zhou:2009jm}%
  \BibitemOpen
  \bibfield  {author} {\bibinfo {author} {\bibfnamefont {J.}~\bibnamefont
  {Zhou}}, \bibinfo {author} {\bibfnamefont {F.}~\bibnamefont {Yuan}}, \ and\
  \bibinfo {author} {\bibfnamefont {Z.-T.}\ \bibnamefont {Liang}},\ }\href
  {\doibase 10.1103/PhysRevD.81.054008} {\bibfield  {journal} {\bibinfo
  {journal} {Phys. Rev. D}\ }\textbf {\bibinfo {volume} {81}},\ \bibinfo
  {pages} {054008} (\bibinfo {year} {2010})},\ \Eprint
  {http://arxiv.org/abs/0909.2238} {arXiv:0909.2238 [hep-ph]} \BibitemShut
  {NoStop}%
\bibitem [{\citenamefont {Kanazawa}\ \emph {et~al.}(2016)\citenamefont
  {Kanazawa}, \citenamefont {Koike}, \citenamefont {Metz}, \citenamefont
  {Pitonyak},\ and\ \citenamefont {Schlegel}}]{Kanazawa:2015ajw}%
  \BibitemOpen
  \bibfield  {author} {\bibinfo {author} {\bibfnamefont {K.}~\bibnamefont
  {Kanazawa}}, \bibinfo {author} {\bibfnamefont {Y.}~\bibnamefont {Koike}},
  \bibinfo {author} {\bibfnamefont {A.}~\bibnamefont {Metz}}, \bibinfo {author}
  {\bibfnamefont {D.}~\bibnamefont {Pitonyak}}, \ and\ \bibinfo {author}
  {\bibfnamefont {M.}~\bibnamefont {Schlegel}},\ }\href {\doibase
  10.1103/PhysRevD.93.054024} {\bibfield  {journal} {\bibinfo  {journal} {Phys.
  Rev. D}\ }\textbf {\bibinfo {volume} {93}},\ \bibinfo {pages} {054024}
  (\bibinfo {year} {2016})},\ \Eprint {http://arxiv.org/abs/1512.07233}
  {arXiv:1512.07233 [hep-ph]} \BibitemShut {NoStop}%
\bibitem [{\citenamefont {'t~Hooft}\ and\ \citenamefont
  {Veltman}(1972)}]{tHooft:1972tcz}%
  \BibitemOpen
  \bibfield  {author} {\bibinfo {author} {\bibfnamefont {G.}~\bibnamefont
  {'t~Hooft}}\ and\ \bibinfo {author} {\bibfnamefont {M.~J.~G.}\ \bibnamefont
  {Veltman}},\ }\href {\doibase 10.1016/0550-3213(72)90279-9} {\bibfield
  {journal} {\bibinfo  {journal} {Nucl. Phys. B}\ }\textbf {\bibinfo {volume}
  {44}},\ \bibinfo {pages} {189} (\bibinfo {year} {1972})}\BibitemShut
  {NoStop}%
\bibitem [{\citenamefont {Breitenlohner}\ and\ \citenamefont
  {Maison}(1977)}]{Breitenlohner:1977hr}%
  \BibitemOpen
  \bibfield  {author} {\bibinfo {author} {\bibfnamefont {P.}~\bibnamefont
  {Breitenlohner}}\ and\ \bibinfo {author} {\bibfnamefont {D.}~\bibnamefont
  {Maison}},\ }\href {\doibase 10.1007/BF01609069} {\bibfield  {journal}
  {\bibinfo  {journal} {Commun. Math. Phys.}\ }\textbf {\bibinfo {volume}
  {52}},\ \bibinfo {pages} {11} (\bibinfo {year} {1977})}\BibitemShut {NoStop}%
\bibitem [{\citenamefont {Zhou}\ and\ \citenamefont
  {Metz}(2012)}]{Zhou:2010ui}%
  \BibitemOpen
  \bibfield  {author} {\bibinfo {author} {\bibfnamefont {J.}~\bibnamefont
  {Zhou}}\ and\ \bibinfo {author} {\bibfnamefont {A.}~\bibnamefont {Metz}},\
  }\href {\doibase 10.1103/PhysRevD.86.014001} {\bibfield  {journal} {\bibinfo
  {journal} {Phys. Rev. D}\ }\textbf {\bibinfo {volume} {86}},\ \bibinfo
  {pages} {014001} (\bibinfo {year} {2012})},\ \Eprint
  {http://arxiv.org/abs/1011.5871} {arXiv:1011.5871 [hep-ph]} \BibitemShut
  {NoStop}%
\bibitem [{\citenamefont {Boer}\ \emph {et~al.}(1998)\citenamefont {Boer},
  \citenamefont {Mulders},\ and\ \citenamefont {Teryaev}}]{Boer:1997bw}%
  \BibitemOpen
  \bibfield  {author} {\bibinfo {author} {\bibfnamefont {D.}~\bibnamefont
  {Boer}}, \bibinfo {author} {\bibfnamefont {P.~J.}\ \bibnamefont {Mulders}}, \
  and\ \bibinfo {author} {\bibfnamefont {O.~V.}\ \bibnamefont {Teryaev}},\
  }\href {\doibase 10.1103/PhysRevD.57.3057} {\bibfield  {journal} {\bibinfo
  {journal} {Phys. Rev. D}\ }\textbf {\bibinfo {volume} {57}},\ \bibinfo
  {pages} {3057} (\bibinfo {year} {1998})},\ \Eprint
  {http://arxiv.org/abs/hep-ph/9710223} {arXiv:hep-ph/9710223} \BibitemShut
  {NoStop}%
\bibitem [{\citenamefont {Boer}\ and\ \citenamefont
  {Mulders}(2000)}]{Boer:1999si}%
  \BibitemOpen
  \bibfield  {author} {\bibinfo {author} {\bibfnamefont {D.}~\bibnamefont
  {Boer}}\ and\ \bibinfo {author} {\bibfnamefont {P.~J.}\ \bibnamefont
  {Mulders}},\ }\href {\doibase 10.1016/S0550-3213(99)00719-1} {\bibfield
  {journal} {\bibinfo  {journal} {Nucl. Phys. B}\ }\textbf {\bibinfo {volume}
  {569}},\ \bibinfo {pages} {505} (\bibinfo {year} {2000})},\ \Eprint
  {http://arxiv.org/abs/hep-ph/9906223} {arXiv:hep-ph/9906223} \BibitemShut
  {NoStop}%
\bibitem [{\citenamefont {Lu}\ and\ \citenamefont {Schmidt}(2011)}]{Lu:2011th}%
  \BibitemOpen
  \bibfield  {author} {\bibinfo {author} {\bibfnamefont {Z.}~\bibnamefont
  {Lu}}\ and\ \bibinfo {author} {\bibfnamefont {I.}~\bibnamefont {Schmidt}},\
  }\href {\doibase 10.1103/PhysRevD.84.114004} {\bibfield  {journal} {\bibinfo
  {journal} {Phys. Rev. D}\ }\textbf {\bibinfo {volume} {84}},\ \bibinfo
  {pages} {114004} (\bibinfo {year} {2011})},\ \Eprint
  {http://arxiv.org/abs/1109.3232} {arXiv:1109.3232 [hep-ph]} \BibitemShut
  {NoStop}%
\bibitem [{\citenamefont {Boer}\ and\ \citenamefont {Qiu}(2002)}]{Boer:2001tx}%
  \BibitemOpen
  \bibfield  {author} {\bibinfo {author} {\bibfnamefont {D.}~\bibnamefont
  {Boer}}\ and\ \bibinfo {author} {\bibfnamefont {J.-w.}\ \bibnamefont {Qiu}},\
  }\href {\doibase 10.1103/PhysRevD.65.034008} {\bibfield  {journal} {\bibinfo
  {journal} {Phys. Rev. D}\ }\textbf {\bibinfo {volume} {65}},\ \bibinfo
  {pages} {034008} (\bibinfo {year} {2002})},\ \Eprint
  {http://arxiv.org/abs/hep-ph/0108179} {arXiv:hep-ph/0108179} \BibitemShut
  {NoStop}%
\bibitem [{\citenamefont {Smirnov}(2008)}]{Smirnov:2008iw}%
  \BibitemOpen
  \bibfield  {author} {\bibinfo {author} {\bibfnamefont {A.~V.}\ \bibnamefont
  {Smirnov}},\ }\href {\doibase 10.1088/1126-6708/2008/10/107} {\bibfield
  {journal} {\bibinfo  {journal} {JHEP}\ }\textbf {\bibinfo {volume} {10}},\
  \bibinfo {pages} {107} (\bibinfo {year} {2008})},\ \Eprint
  {http://arxiv.org/abs/0807.3243} {arXiv:0807.3243 [hep-ph]} \BibitemShut
  {NoStop}%
\bibitem [{\citenamefont {Arnold}\ \emph {et~al.}(2009)\citenamefont {Arnold},
  \citenamefont {Metz},\ and\ \citenamefont {Schlegel}}]{Arnold:2008kf}%
  \BibitemOpen
  \bibfield  {author} {\bibinfo {author} {\bibfnamefont {S.}~\bibnamefont
  {Arnold}}, \bibinfo {author} {\bibfnamefont {A.}~\bibnamefont {Metz}}, \ and\
  \bibinfo {author} {\bibfnamefont {M.}~\bibnamefont {Schlegel}},\ }\href
  {\doibase 10.1103/PhysRevD.79.034005} {\bibfield  {journal} {\bibinfo
  {journal} {Phys. Rev. D}\ }\textbf {\bibinfo {volume} {79}},\ \bibinfo
  {pages} {034005} (\bibinfo {year} {2009})},\ \Eprint
  {http://arxiv.org/abs/0809.2262} {arXiv:0809.2262 [hep-ph]} \BibitemShut
  {NoStop}%
\bibitem [{\citenamefont {Collins}\ and\ \citenamefont
  {Rogers}(2008)}]{Collins:2008sg}%
  \BibitemOpen
  \bibfield  {author} {\bibinfo {author} {\bibfnamefont {J.~C.}\ \bibnamefont
  {Collins}}\ and\ \bibinfo {author} {\bibfnamefont {T.~C.}\ \bibnamefont
  {Rogers}},\ }\href {\doibase 10.1103/PhysRevD.78.054012} {\bibfield
  {journal} {\bibinfo  {journal} {Phys. Rev. D}\ }\textbf {\bibinfo {volume}
  {78}},\ \bibinfo {pages} {054012} (\bibinfo {year} {2008})},\ \Eprint
  {http://arxiv.org/abs/0805.1752} {arXiv:0805.1752 [hep-ph]} \BibitemShut
  {NoStop}%
\bibitem [{\citenamefont {Kang}\ and\ \citenamefont {Qiu}(2009)}]{Kang:2008ey}%
  \BibitemOpen
  \bibfield  {author} {\bibinfo {author} {\bibfnamefont {Z.-B.}\ \bibnamefont
  {Kang}}\ and\ \bibinfo {author} {\bibfnamefont {J.-W.}\ \bibnamefont {Qiu}},\
  }\href {\doibase 10.1103/PhysRevD.79.016003} {\bibfield  {journal} {\bibinfo
  {journal} {Phys. Rev. D}\ }\textbf {\bibinfo {volume} {79}},\ \bibinfo
  {pages} {016003} (\bibinfo {year} {2009})},\ \Eprint
  {http://arxiv.org/abs/0811.3101} {arXiv:0811.3101 [hep-ph]} \BibitemShut
  {NoStop}%
\bibitem [{\citenamefont {Braun}\ \emph {et~al.}(2009)\citenamefont {Braun},
  \citenamefont {Manashov},\ and\ \citenamefont {Pirnay}}]{Braun:2009mi}%
  \BibitemOpen
  \bibfield  {author} {\bibinfo {author} {\bibfnamefont {V.~M.}\ \bibnamefont
  {Braun}}, \bibinfo {author} {\bibfnamefont {A.~N.}\ \bibnamefont {Manashov}},
  \ and\ \bibinfo {author} {\bibfnamefont {B.}~\bibnamefont {Pirnay}},\ }\href
  {\doibase 10.1103/PhysRevD.80.114002} {\bibfield  {journal} {\bibinfo
  {journal} {Phys. Rev. D}\ }\textbf {\bibinfo {volume} {80}},\ \bibinfo
  {pages} {114002} (\bibinfo {year} {2009})},\ \bibinfo {note} {[Erratum:
  Phys.Rev.D 86, 119902 (2012)]},\ \Eprint {http://arxiv.org/abs/0909.3410}
  {arXiv:0909.3410 [hep-ph]} \BibitemShut {NoStop}%
\bibitem [{\citenamefont {Ma}\ and\ \citenamefont {Wang}(2012)}]{Ma:2012xn}%
  \BibitemOpen
  \bibfield  {author} {\bibinfo {author} {\bibfnamefont {J.~P.}\ \bibnamefont
  {Ma}}\ and\ \bibinfo {author} {\bibfnamefont {Q.}~\bibnamefont {Wang}},\
  }\href {\doibase 10.1016/j.physletb.2012.07.036} {\bibfield  {journal}
  {\bibinfo  {journal} {Phys. Lett. B}\ }\textbf {\bibinfo {volume} {715}},\
  \bibinfo {pages} {157} (\bibinfo {year} {2012})},\ \Eprint
  {http://arxiv.org/abs/1205.0611} {arXiv:1205.0611 [hep-ph]} \BibitemShut
  {NoStop}%
\bibitem [{\citenamefont {Schafer}\ and\ \citenamefont
  {Zhou}(2012)}]{Schafer:2012ra}%
  \BibitemOpen
  \bibfield  {author} {\bibinfo {author} {\bibfnamefont {A.}~\bibnamefont
  {Schafer}}\ and\ \bibinfo {author} {\bibfnamefont {J.}~\bibnamefont {Zhou}},\
  }\href {\doibase 10.1103/PhysRevD.85.117501} {\bibfield  {journal} {\bibinfo
  {journal} {Phys. Rev. D}\ }\textbf {\bibinfo {volume} {85}},\ \bibinfo
  {pages} {117501} (\bibinfo {year} {2012})},\ \Eprint
  {http://arxiv.org/abs/1203.5293} {arXiv:1203.5293 [hep-ph]} \BibitemShut
  {NoStop}%
\bibitem [{\citenamefont {Kang}\ and\ \citenamefont {Qiu}(2012)}]{Kang:2012em}%
  \BibitemOpen
  \bibfield  {author} {\bibinfo {author} {\bibfnamefont {Z.-B.}\ \bibnamefont
  {Kang}}\ and\ \bibinfo {author} {\bibfnamefont {J.-W.}\ \bibnamefont {Qiu}},\
  }\href {\doibase 10.1016/j.physletb.2012.06.021} {\bibfield  {journal}
  {\bibinfo  {journal} {Phys. Lett. B}\ }\textbf {\bibinfo {volume} {713}},\
  \bibinfo {pages} {273} (\bibinfo {year} {2012})},\ \Eprint
  {http://arxiv.org/abs/1205.1019} {arXiv:1205.1019 [hep-ph]} \BibitemShut
  {NoStop}%
\bibitem [{\citenamefont {Scimemi}\ \emph {et~al.}(2019)\citenamefont
  {Scimemi}, \citenamefont {Tarasov},\ and\ \citenamefont
  {Vladimirov}}]{Scimemi:2019gge}%
  \BibitemOpen
  \bibfield  {author} {\bibinfo {author} {\bibfnamefont {I.}~\bibnamefont
  {Scimemi}}, \bibinfo {author} {\bibfnamefont {A.}~\bibnamefont {Tarasov}}, \
  and\ \bibinfo {author} {\bibfnamefont {A.}~\bibnamefont {Vladimirov}},\
  }\href {\doibase 10.1007/JHEP05(2019)125} {\bibfield  {journal} {\bibinfo
  {journal} {JHEP}\ }\textbf {\bibinfo {volume} {05}},\ \bibinfo {pages} {125}
  (\bibinfo {year} {2019})},\ \Eprint {http://arxiv.org/abs/1901.04519}
  {arXiv:1901.04519 [hep-ph]} \BibitemShut {NoStop}%
\bibitem [{\citenamefont {Vladimirov}\ \emph {et~al.}(2022)\citenamefont
  {Vladimirov}, \citenamefont {Moos},\ and\ \citenamefont
  {Scimemi}}]{Vladimirov:2021hdn}%
  \BibitemOpen
  \bibfield  {author} {\bibinfo {author} {\bibfnamefont {A.}~\bibnamefont
  {Vladimirov}}, \bibinfo {author} {\bibfnamefont {V.}~\bibnamefont {Moos}}, \
  and\ \bibinfo {author} {\bibfnamefont {I.}~\bibnamefont {Scimemi}},\ }\href
  {\doibase 10.1007/JHEP01(2022)110} {\bibfield  {journal} {\bibinfo  {journal}
  {JHEP}\ }\textbf {\bibinfo {volume} {01}},\ \bibinfo {pages} {110} (\bibinfo
  {year} {2022})},\ \Eprint {http://arxiv.org/abs/2109.09771} {arXiv:2109.09771
  [hep-ph]} \BibitemShut {NoStop}%
\bibitem [{\citenamefont {Muta}(1998)}]{Muta:1998vi}%
  \BibitemOpen
  \bibfield  {author} {\bibinfo {author} {\bibfnamefont {T.}~\bibnamefont
  {Muta}},\ }\href@noop {} {\emph {\bibinfo {title} {{Foundations of quantum
  chromodynamics. Second edition}}}},\ Vol.~\bibinfo {volume} {57}\ (\bibinfo
  {year} {1998})\BibitemShut {NoStop}%
\end{thebibliography}%

\end{document}